\documentclass[twocolumn, superscriptaddress, secnumarabic, nobibnotes, aps, prx, amsmath, amssymb, floatfix]{revtex4-2}
\usepackage{enumitem}
\usepackage{graphicx}
\usepackage{dcolumn}
\usepackage{bm}
\usepackage[english]{babel}

\usepackage{slashed}
\usepackage[separate-uncertainty]{siunitx}
\usepackage[breaklinks=true]{hyperref}
\hypersetup{
	colorlinks=true,
	linkcolor=blue,
	citecolor=blue,
	urlcolor=blue
}

\newcommand{\Ai}{\,\text{Ai}}

\newcommand{\atan}{\text{atan2}}

\allowdisplaybreaks

\begin{document}

\title{Intrinsic Nonlocality of Spin- and Polarization-Resolved Probabilities\\in Strong-Field Quantum Electrodynamics}

\author{Samuele~Montefiori}
\affiliation{Max-Planck-Institut f\"{u}r Kernphysik, Saupfercheckweg 1, 69117 Heidelberg, Germany}

\author{Antonino~Di~Piazza}
\affiliation{Department of Physics and Astronomy, University of Rochester, Rochester, New York 14627, USA}
\affiliation{Laboratory for Laser Energetics, University of Rochester, Rochester, New York 14623, USA}

\author{Tobias~Podszus}
\affiliation{Max-Planck-Institut f\"{u}r Kernphysik, Saupfercheckweg 1, 69117 Heidelberg, Germany}

\author{Christoph~H.~Keitel}
\affiliation{Max-Planck-Institut f\"{u}r Kernphysik, Saupfercheckweg 1, 69117 Heidelberg, Germany}

\author{Matteo~Tamburini}
\email{matteo.tamburini@mpi-hd.mpg.de}
\affiliation{Max-Planck-Institut f\"{u}r Kernphysik, Saupfercheckweg 1, 69117 Heidelberg, Germany}

\date{\today}


\begin{abstract}
Spin and polarization are central to precision tests of fundamental physics and for interpreting radiation from astrophysical sources and ultraintense laser-matter experiments. 
Here, focusing on the fundamental process of nonlinear Compton scattering, we demonstrate that a key assumption underlying current strong-field quantum electrodynamics (SFQED) models, i.e., that emission can be treated as an instantaneous random event sampled from a local differential rate, is inconsistent once emission angles, electron spin, and/or photon polarization are resolved. Namely, \emph{even in strictly constant and uniform fields}, the resulting fully differential distribution is sign-indefinite, yielding negative inferred probabilities. The physical reason is that the photon emission probability builds up over a finite length of the electron trajectory, the formation region, during which the electron direction changes by roughly the same small angle that defines the radiation cone. Therefore, we put forward a new method where we integrate over this formation region analytically to obtain a physically consistent electron spin and photon polarization model. We show that the implementation of our model is compatible with existing Monte Carlo (MC) and particle-in-cell (PIC) workflows. Simulations of a GeV-class electron--laser collision accessible at current petawatt facilities and of emission in a pulsar-like magnetic field are shown to reveal spin and polarization patterns that differ even qualitatively from state-of-the-art local models. In particular, our new model predicts substantial angle-dependent circular photon polarization where the well-known collinear-emission approach yields none, and a pronounced helicity bias in the recoiling electrons absent from current predictions. These findings have direct implications for upcoming strong-field QED experiments and for interpreting polarized radiation from extreme astrophysical environments.
\end{abstract}

\maketitle


\section{\label{sec:intro}Introduction}

Lepton spin and photon polarization are fundamental degrees of freedom that determine scattering amplitudes and observable asymmetries in physical processes, rendering them invaluable probes across many areas of physics. In high-energy physics, particle beams with controlled polarization enable measurements of parity-violating interactions~\cite{anthonyPRL04, anthonyPRL05}, and searches for physics beyond the Standard Model~\cite{androicN18}. In materials science, spin-polarized electron and positron beams enable materials characterization~\cite{tolhoekRMP56, gidleyPRL82, jungwirthRMP14, danielsonRMP15} and the study of magnetic order~\cite{zuticRMP04}. In laboratory plasmas, photon polarization and lepton spin trace ultrafast field structures, enabling femtosecond diagnostics of transient magnetic fields~\cite{gongPRL21, gongPRR22}, plasma instabilities~\cite{gongPRL23a}, shock-driven electron acceleration~\cite{gongPRL23b}, and magnetic reconnection~\cite{gongPRL25}. In nuclear and hadronic physics, polarized gamma-ray beams provide a sensitive probe of nuclear structure~\cite{abePRL, schlimmPRL2013}, enable selective excitation of polarization-dependent nuclear resonances~\cite{schaerfPT05, guntherNC22}, and facilitate investigations of meson photoproduction~\cite{akbarPRC17}. Photon polarization also serves as a unique probe of vacuum birefringence~\cite{nakamiyaPRD17, braginPRL17}, an intrinsically quantum-field effect of central importance in quantum electrodynamics (QED)~\cite{dipiazzaRMP12, gonoskovRMP22, fedotovPR23} and cosmology~\cite{minamiPRL20}. In astrophysics, polarization of photons provides clues on their source and radiation mechanism~\cite{uzdenskyRPP2014}. Prominent examples include pulsars~\cite{deanS08, songMNRAS24}, black holes~\cite{laurentS11}, fast radio bursts~\cite{zhangRMP23}, and gamma-ray bursts~\cite{zhangNA19}.

Motivated by the wide range of applications of polarized lepton and gamma-ray beams, the past decade has seen rapid progress in developing new concepts for polarized lepton~\cite{Del-SorboPRA17, wenPRL19, liPRL19, chenPRL19,  Song:2019aa, Seipt:2019aa, liPRL20b, guoPRR20, Wan:2020aa,  Nie:2021aa, songNJP21, Tang:2021aa, xueFR22, liPRL22, xuePRL23,  gongMRE23, bohlenPRR23, Li:2023aa, qianPoP23, Yin:2024aa} and gamma-ray sources~\cite{omoriPRL06, liPRL20a, xueMRE20, wanPRR20, daiMRE21, WangPoP24, JiangPRL25}. On the theoretical side, these concepts are supported by simulations that incorporate analytical calculations of polarization-dependent nonlinear Compton scattering (NCS) and nonlinear Breit-Wheeler pair production (NBW)~\cite{Baier-book, seiptPRA18, Wistisen:2019aa, torgrimssonNJP21, seiptPRA20, podszusPRD22, chenPRD22, liPRD23, daiPRD23,  montefioriPhD25}. On the experimental side, advances in accelerator technology~\cite{yakimenkoPRAB19} and high-power laser technology~\cite{dansonHPLSE19} have opened access to the strong-field quantum electrodynamics (SFQED) regime~\cite{ritusJSLR85, Baier-book, dipiazzaRMP12, gonoskovRMP22, fedotovPR23}, with flagship experiments providing evidence of NCS~\cite{mirzaieNP24}, of radiation reaction~\cite{colePRX18, poderPRX18}, and of its quantum nature~\cite{poderPRX18, losNC26}. In these beam--laser collision experiments, electrons in the beam can experience electromagnetic fields approaching the QED critical field $F_{\mathrm{cr}} \equiv m^{2}/|e| \approx 1.3\times10^{18}\,\mathrm{V/m}$ ($\approx 4.4\times10^{9}\,\mathrm{T}$) in their instantaneous rest frame, leading to qualitatively new dynamics including stochastic photon emission with significant quantum recoil, and radiative spin polarization of the beam, phenomena absent in conventional low-intensity laser-matter experiments. Here $m$ and $e<0$ denote the electron mass and charge, respectively, and units with $\hbar = c = \epsilon_0 = 1$ (fine-structure constant $\alpha = e^2/4\pi \approx 1/137$) are employed.

A central objective of the ongoing and upcoming experimental effort is the accurate measurement of particle distributions in NCS and NBW, a critical step in validating the theoretical models and simulation codes upon which experimental interpretation relies~\cite{E320FACET, abramowiczEPJSP21, mirzaieNP24}. These models underpin both laboratory and astrophysical observations, but rest on two approximations whose validity for spin- and polarization-resolved observables has not been fully established: the \emph{collinear approximation} and, within the locally constant field approximation (LCFA), that emission events can be sampled from an \emph{instantaneous} differential rate. The collinear approximation assumes that produced particles initially propagate strictly along the momentum direction of the parent particle~\cite{dipiazzaRMP12, gonoskovRMP22, fedotovPR23}. This is supported by kinematics: at ultrarelativistic energies, i.e., $\gamma \gg 1$ for a lepton and $\omega \gg m$ for a photon, where $\gamma$ is the Lorentz factor of the lepton and $\omega$ the energy of the photon, the emission products are collimated within a narrow forward cone of half-angle ${\sim}1/\gamma$ (or ${\sim}m/\omega$ for high-energy photons)~\footnote{An important exception is Ref.~\cite{blackburnPRA20}, where NCS is studied beyond the collinear approximation while still averaging/summing over spin and polarization.}. However, spin and polarization are defined relative to the particle's momentum: the spin density matrix of the outgoing lepton and the polarization density matrix of the emitted photon both depend on the propagation direction. Resolving angles is therefore not optional for spin- or polarization-sensitive observables. This motivates the construction of fully differential, angle-resolved distributions, and raises the question, addressed in this work, of whether such distributions can be derived consistently from local, instantaneous models. The second assumption is emission sampling from instantaneous rates: within the LCFA every emission event is treated as instantaneous, determined solely by the particle's state at a single point along its trajectory, with the background field locally replaced by a constant crossed field (CCF) matching the instantaneous field invariants. For a laser-pulse background, the LCFA is known to be accurate when $a_0 \equiv |e|E_0/m\omega_0 \gg 1$, where $E_0$ is the laser field amplitude and $\omega_0$ its central angular frequency, while it breaks down for $a_0\lesssim1$ because the background field then varies on the scale of the photon formation length~\cite{ritusJSLR85, dipiazzaRMP12, gonoskovRMP22, fedotovPR23}.

In this article we identify a conceptually distinct and more fundamental limitation. The known failure of the LCFA at $a_0\lesssim1$ is a \emph{regime} failure: the background field varies too rapidly for a local-rate description to be accurate. The problem we expose here is \emph{structural}: even in a strictly uniform CCF, where the LCFA holds \emph{exactly} for spin- and polarization-summed quantities, the angle- and spin/polarization-resolved local differential rate is not positive definite and therefore does not admit a probabilistic interpretation. The physical origin is the finite photon formation region: emission probability builds up coherently over a finite segment of the electron trajectory during which the electron momentum rotates by ${\sim}1/\gamma$, an angle comparable to the radiation cone, so that spin and polarization cannot be self-consistently assigned from the emitter's instantaneous local state. Specifically, we demonstrate that angle- and spin/polarization-resolved distributions constructed from the local lepton state can yield spin $\bm{\eta}$ and polarization $\bm{\xi}$ vectors with magnitude exceeding unity even in a uniform CCF, implying unphysical negative probabilities. When the full photon formation length is accounted for, these pathologies disappear. This result is not a special feature of crossed fields: for high-energy particles the CCF is precisely the local model underlying the LCFA for \emph{any} slowly varying background ($a_0\gg1$), because the field experienced over the formation region is locally a CCF in the particle's instantaneous rest frame. We stress that the sign-indefiniteness of the angle- and spin/polarization-resolved rate 1) occurs in a  CCF and is therefore unrelated to the validity of the LCFA and 2) propagates directly into every LCFA-based Monte Carlo (MC) and PIC code, regardless of the field configuration being simulated.

We introduce a new LCFA model for angle- and spin/polarization-resolved particle distributions that accounts for the full photon formation region, and compare its predictions against the standard collinear approximation in two paradigmatic configurations. The first, a GeV-class electron bunch colliding head-on with a petawatt-class laser pulse, is directly accessible at current high-power laser facilities, and reveals angle-dependent circular photon polarization absent from the collinear local models that form the basis of current LCFA-based PIC simulations. The second, emission in a pulsar-like uniform magnetic field representative of conditions in the polar cap of an ordinary radio pulsar, reveals a pronounced helicity bias in the recoiling electrons that the collinear model predicts to vanish identically. In both cases the differences are qualitative, not merely quantitative, underscoring that the new model is a fundamental modification to existing frameworks.

Since NCS and NBW are the elementary building blocks of all SFQED plasma simulations, these findings have direct implications for upcoming strong-field QED experiments at high-power laser facilities, and for the quantitative interpretation of polarized high-energy radiation from pulsars, magnetars, and other extreme astrophysical environments.


\section{\label{sec:break_prob}Spin- and polarization-resolved Nonlinear Compton scattering}

We first collect the covariant descriptions of lepton spin and photon polarization used throughout. For a spin-$1/2$ particle with four-momentum $p^\mu=(\varepsilon,\bm{p})$, the covariant spin density matrix is~\cite{Berestetskii-book}
\begin{equation}
\rho(p) = \frac{1}{2}\,(\slashed{p} + m)\,\bigl(1 + \gamma^5\,\slashed{\zeta}\bigr),
\label{spin_dens_mtrx}
\end{equation}
where the spin four-vector $\zeta^\mu \equiv \zeta^\mu(p)$ satisfies $(\zeta p) = 0$ and $\zeta^2 = -\bm{\eta}^2$. Here $\bm{\eta}$ is the rest-frame spin-polarization vector with $|\bm{\eta}|\le 1$ (for a pure state $|\bm{\eta}|=1$ and $\bm{\eta}$ equals twice the mean spin vector in the rest frame). In this article we employ the notation $\slashed{p} \equiv \gamma^\mu p_\mu$, where $\gamma^\mu$ are the Dirac matrices with metric $g_{\mu\nu}=\mathrm{diag}(+1,-1,-1,-1)$, $\gamma^5 \equiv i\gamma^0\gamma^1\gamma^2\gamma^3$, Levi-Civita tensor $\epsilon^{0123}=+1$, and $(\zeta p) \equiv \zeta^\mu p_\mu$. Because $\zeta^\mu$ depends explicitly on $p^\mu$, the spin state of a lepton cannot be specified independently of its momentum in the relativistic regime. Photon polarization is likewise tied to momentum. For a photon with four-momentum $k^\mu=(\omega,\bm{k})$, one fixes a transverse polarization basis $\{e^\mu_{(1)},e^\mu_{(2)}\}$ such that
\begin{equation}
(e_{(i)} e_{(j)}) = -\delta_{ij}, \qquad (e_{(i)} k) = 0, \qquad i,j=1,2.
\label{pola_constraint}
\end{equation}
Up to gauge transformations, the photon polarization density matrix is~\cite{Berestetskii-book}
\begin{equation}
\rho^{\mu\nu}(k) = \frac{1}{2} (1 + \bm{\xi}\cdot\bm{\sigma})_{(i,j)} \, e^\mu_{(i)}\,e^\nu_{(j)}, \quad \bm{\xi} \equiv (\xi^1,\xi^2,\xi^3),
\label{phtn_dens_mtrx}
\end{equation}
where $1_{(i,j)}$ and $\bm{\sigma}_{(i,j)}$ are the $(i,j)$-element of the $2\times 2$ unity matrix and of the Pauli matrices, respectively, and Einstein's summation convention over repeated indices is employed throughout. In Eq.~(\ref{phtn_dens_mtrx}) the Stokes parameters $\xi^1, \xi^2, \xi^3$ are defined with respect to a basis transverse to $k^\mu$ [see the constraints in Eq.~(\ref{pola_constraint})] and satisfy $|\bm{\xi}| \le 1$, where, analogously to the lepton case, the equal sign corresponds to a pure polarization state. Consequently, a photon's polarization is necessarily tied to its propagation direction: integrating over the emission solid angle mixes Stokes vectors defined in different, incompatible transverse bases, so spin- and polarization-resolved observables require angle-resolved differential distributions (see also Ref.~\cite{podszusPRD22}).

We begin by considering the differential probability $dP^{(\zeta,\zeta^\prime,j)}_{\text{NCS}}$ for the emission of a photon with four-momentum $k^\mu$ by an electron with incoming four-momentum $p^\mu$ and spin four-vector $\zeta^\mu$, interacting with a plane-wave background field characterized by the four-potential $A^\mu$ and propagating along $\bm{d}$ ($\bm{d}^2=1$). Here and in what follows, for an arbitrary four-vector $v^\mu=(v^0,\bm{v})$ we use the subscripts $\parallel$ and $\perp$ to denote the spatial components parallel and transverse to the propagation direction $\bm{d}$, i.e., $v_\parallel=\bm{v}\cdot\bm{d}$ and $\bm{v}_\perp=\bm{v}-v_\parallel \bm{d}$. It is useful to introduce the quantities $d^\mu=(1,\bm{d})$ and $\tilde d_\mu = (1,-\bm{d})/2$ and the light-cone coordinates $v_+ = (\tilde d v)$ and $v_- = (d v)$. We also introduce two mutually orthogonal four-vectors $a_i^\mu$ ($i=1,2$), satisfying $(a_i a_l)=-\delta_{il}$ and orthogonal also to $d^\mu$, i.e., $(d a_i)=0$. For an arbitrary plane wave, the four-potential depends on space $\bm{x}$ and time $t$ [with $x^\mu=(t,\bm{x})$] only through the phase $\phi=(x d)= x_-$, namely $A^\mu = A^\mu(\phi)$ (the quantity $\phi$ is more precisely the light-cone time but we will refer to it as a ``phase'' for simplicity). Choosing the Lorenz gauge with the additional condition $A^0=0$, the potential can be written as a linear combination of the transverse basis vectors $a_i^\mu=(0,\bm{a}_i)$. No specific choice of the emitted-photon polarization basis $e^\mu_j$ ($j=1,2$) is made. Within the $S$-matrix formalism, the probability reads
\begin{align}
dP_{\text{NCS}}^{(\zeta,\zeta^\prime,j)}
&= V \frac{d^3 k}{(2 \pi)^3} V \int \frac{d^3 p^\prime}{(2 \pi)^3} \, |S_{fi}|^2,
\label{prob_tob_pre}
\end{align}
where $V$ is the quantization volume, $p^{\prime\mu}=(\varepsilon^\prime,\bm{p}^{\,\prime})$ and $\zeta^{\prime\mu}$ denote the outgoing electron four-momentum and spin four-vector, respectively, and
\begin{align}
S_{fi} &=
(2 \pi)^3 \delta^{(2)} \!\left( \bm{p}^{\,\prime}_\perp + \bm{k}_\perp - \bm{p}_\perp \right)
\delta\!\left(p^\prime_- + k_- - p_- \right)\, i\, \mathcal{M}_{fi},
\label{smatrix}
\end{align}
is the transition-matrix element. When computing $|S_{fi}|^2$, the square of the $\delta$-function $\delta(p'_- + k_- - p_-)$ is evaluated after expressing it in terms of the longitudinal momentum: $\delta(p^\prime_- + k_- - p_-) = (\varepsilon/p_-) \delta(p_\parallel - p^*_\parallel)$, with
\begin{equation}
p_\parallel^{*}=\frac{m^{2}+\bm{p}_{\perp}^{2}-\left(p'_-+k_-\right)^{2}}{2\left(p'_-+k_-\right)}\,.
\end{equation}
This change of variables then allows the standard application of periodic boundary conditions along $\bm{d}$, simplifying the subsequent integration. At leading order in the Furry picture, the invariant matrix element $\mathcal{M}_{fi}$ takes the form (see, e.g., Ref.~\cite{dipiazzaPRA18})
\begin{widetext}
\begin{align}
\mathcal{M}_{fi}
&= - \frac{e}{\sqrt{8 V^3 \varepsilon \varepsilon^\prime \omega}}
\int d\phi \,
\Biggl[
\bar{u}_{p^\prime,\zeta^\prime}
\Bigl( 1 - \frac{\slashed{d}\slashed{\mathcal{A}}(\phi)}{2 p^\prime_-} \Bigr)
\slashed{e}^{\,*}_j
\Bigl( 1 + \frac{\slashed{d}\slashed{\mathcal{A}}(\phi)}{2 p_-} \Bigr)
u_{p,\zeta}
\Biggr]
\exp\!\bigl\{ i \Phi(\phi) \bigr\},
\label{matrix_element}
\end{align}
where the function in the exponential reads
\begin{align}
\Phi(\phi)
&= (p^\prime_+ + k_+ - p_+)\phi
+ \int_{0}^{\phi} d\phi^\prime
\Biggl[
\frac{p^\prime_\mu \mathcal{A}^\mu(\phi^\prime)}{p^\prime_-}
- \frac{p_\mu \mathcal{A}^\mu(\phi^\prime)}{p_-}
- \frac{\mathcal{A}^2(\phi^\prime)}{2}
\Bigl( \frac{1}{p^\prime_-} - \frac{1}{p_-} \Bigr)
\Biggr],
\label{exponential}
\end{align}
\end{widetext}
with $u_{p,\zeta}$ ($u_{p^\prime,\zeta^\prime}$) being the free Dirac spinor for the incoming (outgoing) electron, and $\mathcal{A}^\mu \equiv e A^\mu$. Squaring the modulus of the matrix element and integrating over the momentum of the outgoing electron as in Eq.~\eqref{prob_tob_pre}, one obtains
\begin{widetext}
\begin{align}
\mkern-18mu \mkern-18mu \frac{dP^{(\zeta,\zeta^\prime,j)}_{NCS}}{d^3 k} = & \frac{\alpha}{16 \pi^2} {\frac{1}{p_- p^\prime_- \omega}} \int d\phi \int d\phi^\prime \exp{ \Bigl\{i [ \Phi(\phi) - \Phi(\phi^\prime) ] \Bigr\} } \times \notag \\
& \qquad \qquad \times \frac{1}{4} \text{tr} \Bigl\{ \Bigl[ \slashed{e}^*_j +\frac{1}{2}\Bigl( \frac{\slashed{\mathcal{A}}(\phi) \slashed{d} \slashed{e}^*_j}{p^\prime_-} + \frac{\slashed{e}^*_j \slashed{d} \slashed{\mathcal{A}}(\phi)}{p_-}\Bigr) -\frac{\mathcal{A}^2(\phi) \slashed{d} e^{*}_{j-} }{2 p^\prime_- p_-} \Bigr](\slashed{p} + m)(1+ \gamma^5 \slashed{\zeta}) \times \notag \\
& \qquad \qquad \qquad \quad \times \Bigl[ \slashed{e}_j +\frac{1}{2}\Bigl( \frac{\slashed{\mathcal{A}}(\phi^\prime) \slashed{d} \slashed{e}_j}{p_-} + \frac{\slashed{e}_j \slashed{d} \slashed{\mathcal{A}}(\phi^\prime)}{p^\prime_-}\Bigr) -\frac{\mathcal{A}^2(\phi^\prime) \slashed{d} e_{j-} }{2 p^\prime_- p_-} \Bigr] (\slashed{p}^\prime + m)(1+ \gamma^5 \slashed{\zeta}^\prime) \Bigr\}.
\label{squared}
\end{align}
\end{widetext}
Having in mind current models based on the LCFA, in Eq.~\eqref{squared} it is convenient to change to the variables $\phi_+ = (\phi + \phi^\prime)/2$ and $\phi_- = \phi - \phi^\prime$. The differential probability can then be written as $dP_{\mathrm{NCS}} / d^3k = \int d\phi_+ \, dP_{\mathrm{NCS}} / d^3k\, d\phi_+$. 

For $a_0\lesssim 1$ the probability is not localized, $dP_{\mathrm{NCS}}/(d^3k\, d\phi_+)$ takes on both positive and negative values, and only the integration over both $\phi_-$ and $\phi_+$ yields a physically meaningful result. By contrast, it is known that, in the limit $a_0\gg1$ and by averaging (summing) over the initial (final) spin and polarization degrees of freedom [$dP_{\mathrm{NCS}} = \sum_{(\zeta,\zeta^\prime,j)}{dP^{(\zeta,\zeta^\prime,j)}_{\mathrm{NCS}}}/2$], the total probability in the plane-wave field can be obtained by summing the contributions of probabilities calculated at each $\phi_+$ in a locally-matched CCF \cite{ritusJSLR85}. This implies that, for $a_0\gg1$, $dP_{\mathrm{NCS}} / d^3k\, d\phi_+$ can be interpreted as a probability per unit $\bm{k}$ and per unit laser phase, and $dP_{\mathrm{NCS}} / d^3k\, d\phi_+ \approx dP_{\mathrm{CCF}} / d^3k\, d\phi_+$, where $dP_{\mathrm{CCF}} / d^3k\, d\phi_+$ denotes the differential probability in a CCF matching the local value of the plane-wave field. 

Physically, this reflects the fact that, for $a_0\gg 1$, at each $\phi_+$ only a small formation region around $\phi_-=0$, where the field is well approximated by a CCF, contributes significantly to the total probability. Analytically, after averaging (summing) over the initial (final) spin and polarization degrees of freedom, this result follows from the integrand in Eq.~\eqref{squared} by changing variables to $(\phi_+,\phi_-)$, expanding the pre-exponent and the phase of the integrand in $\phi_-$ about $\phi_- = 0$ each to the appropriate order, and integrating the resulting expression over $\phi_-$ (see, e.g., Refs.~\cite{dipiazzaPRA18, dipiazzaPRA19} for details). The resulting matched-CCF distribution $dP_{\mathrm{CCF}} / d^3k\, d\phi_+$ is the quantity actually sampled in LCFA-based MC event generators. As we prove below, for spin-/polarization-resolved and angle-resolved NCS this quantity can take both positive and negative values, and therefore cannot be interpreted as a probability density per unit time/phase, contrary to the assumption at the core of current LCFA-models.

Notably, the size of the formation region is determined by the phase function $\Phi(\phi)$ in the exponent of Eq.~\eqref{matrix_element}, and $\Phi(\phi)$ is \emph{independent} of the spin and polarization degrees of freedom [see Eq.~\eqref{exponential}]. Corrections to the LCFA due to the space-time structure of the background field are therefore expected to be suppressed by $1/a_0^2$ (see Sec.~IV of Ref.~\cite{dipiazzaPRA19}) for both spin-averaged/summed and spin-resolved probabilities.


\subsection{\label{subsec:sign}Structure of the angle- and spin-resolved differential distribution in a uniform CCF}

One may intuitively expect that the considerations above, obtained for spin- and polarization-averaged/summed quantities, also apply to the spin-/polarization-resolved case. In particular, one might expect that for $a_0\gg 1$ the integrand $dP^{(\zeta,\zeta^\prime,j)}_{\mathrm{NCS}}/(d^3 k\, d\phi_+)$ is non-negative and therefore admits an interpretation as a probability density. In the following, however, we show that even in a CCF (which is strictly uniform and constant, i.e., $a_0 \to \infty$), $dP^{(\zeta,\zeta^\prime,j)}_{\mathrm{NCS}}/(d^3 k\, d\phi_+)$ can take both positive and negative values and thus does not admit a probabilistic interpretation. In fact, we will demonstrate that this occurs also if one sums over the photon-polarization states~\footnote{Although not demonstrated analytically, panel~\textbf{b} and \textbf{d} of Fig.~\ref{fig:scatter} show that this occurs also if one sums over the final electron spin states.}. 

For definiteness, we set the vector potential to $A^\mu(\phi)=-|\bm{E}| \phi\, a_1^\mu$, with $a_1^\mu=(0,\bm{a}_1)$. This choice corresponds to mutually orthogonal uniform and constant electric $\bm{E}$ and magnetic $\bm{B}$ fields satisfying $|\bm{E}|=|\bm{B}|$ and $(\bm{E}\times \bm{B})/|\bm{E}|^2=\bm{d}$. Notice that the electric field $\bm{E}=-d\bm{A}/d\phi$ is aligned with $\bm{a}_1$. In a CCF, the quantum nonlinearity parameter $\chi_e = \sqrt{|(F^{\mu \nu} p_\nu)^2|}/mF_{\text{cr}}$, where $F^{\mu\nu} = (\bm{E}, \bm{B})$ is the electromagnetic field tensor, is constant and can be written as $\chi_e = (p_-/m) |\bm{E}| /F_{\text{cr}}$. Without loss of generality (apart from assuming that $\bm{k}$ is not parallel to $\bm{d}$), we choose a gauge such that
\begin{align}
\label{gauge_fixing_0}
k_\mu e^\mu_j &= 0, \\
d_\mu e^\mu_j &\equiv e_{j-} = 0,
\label{gauge_fixing}
\end{align}
and
\begin{align}
\sum_{j=1,2} e^{*}_{j,\mu} e_{j,\nu}
=
- g_{\mu\nu} + \frac{k_\mu d_\nu + d_\mu k_\nu}{k_-}.
\label{completeness}
\end{align}
To simplify the discussion, we consider $\bm{p}/|\bm{p}|=-\bm{d}$ and focus only on the lepton spin, summing over the photon polarization, i.e., $dP^{(\zeta,\zeta^\prime)}_{\mathrm{NCS}} = \sum_{j=1,2} dP^{(\zeta,\zeta^\prime,j)}_{\mathrm{NCS}}$. With this choice, Eq.~\eqref{squared} reduces to
\begin{equation}
dP^{(\zeta,\zeta^\prime)}_{\mathrm{NCS}} = \frac{d^3 k}{8 \pi}\, \frac{\alpha}{p_- p^\prime_- \omega} \int d\phi_+\, \Bigl[ \bar I(\phi_+) + \bigl(\zeta^\prime \bar l(\phi_+)\bigr) \Bigr],
\label{matrix_element_squared_final}
\end{equation}
where details of the derivation and the explicit expressions for $\bar I(\phi_+)$ and $\bar l^\mu(\phi_+)$ are given in the first subsection of Appendix~\ref{CCFint}.

Since the choice of the spin quantization axis (SQA) for the incoming and outgoing lepton is arbitrary, we follow Ref.~\cite{podszusPRD22} and choose, for convenience,
\begin{align}
\zeta^\mu &= - \frac{(d^\mu a_2^\nu - d^\nu a_2^\mu)p_\nu}{p_-},
\label{sub_SQA_i} \\
\zeta^{\prime\mu} &= - \frac{(d^\mu a_2^\nu - d^\nu a_2^\mu)p^\prime_\nu}{p^\prime_-},
\label{sub_SQA_f}
\end{align}
which corresponds to the three-dimensional spin vector $\bm{\eta}$ ($\bm{\eta}^\prime$) pointing along the direction of the magnetic field in the instantaneous rest frame of the incoming (outgoing) lepton. This allows us to label the initial and final spin states by the quantum numbers $\sigma,\sigma^\prime=\pm 1$, defined with respect to the SQAs in Eqs.~\eqref{sub_SQA_i}--\eqref{sub_SQA_f}. With these choices and definitions, Eq.~\eqref{matrix_element_squared_final} can be rewritten as
\begin{align}
\frac{dP^{(\sigma,\sigma^\prime)}_{\mathrm{NCS}}}{d\omega d\theta d\varphi} &=\int d\phi_+\frac{dP^{(\sigma,\sigma^\prime)}_{\mathrm{NCS}}}{d\phi_+ d\omega d\theta d\varphi},\\
\frac{dP^{(\sigma,\sigma^\prime)}_{\mathrm{NCS}}}{d\phi_+ d\omega d\theta d\varphi}& = \frac{\alpha}{4 \pi} \frac{\omega \sin\theta}{p_- p^\prime_-} \bar{T}^{\sigma,\sigma^\prime}(\phi_+),
\label{prob_tob_3sum_noint}
\end{align}
where $d^3 k = \omega^2 d\omega \sin\theta\, d\theta\, d\varphi$, and $\theta$ and $\varphi$ are the polar and azimuthal angles, respectively. For the sake of comparison with Baier's quasiclassical method (see Sec.~\ref{subsec:QM}), we define the polar angle with respect to $\bm{p}$ as polar axis, and the azimuthal angle in the reference plane orthogonal to $\bm{p}$ (see Sec.~\ref{subsec:QM}). The function $\bar{T}^{\sigma,\sigma^\prime}(\phi_+)$ is
\begin{widetext}
\begin{align}
\bar{T}^{\sigma,\sigma^\prime}(\phi_+) & = - 2(1 + \sigma \sigma^\prime) m^2 \Bigl( \frac{p_- p^\prime_-}{k_- \mathcal{E}^2} \Bigr)^\frac{1}{3} \Ai(z) - 2 m \mathcal{E} \frac{k_-}{p_-} \Bigl( \sigma + \sigma^\prime + \frac{k_-}{p^\prime_-} \sigma^\prime \Bigr) \Bigl( \frac{p_- p^\prime_-}{k_- \mathcal{E}^2} \Bigr)^\frac{2}{3} \Ai^\prime(z)+ \notag \\
& \quad + 2 \Bigl[ 2 (1 + \sigma \sigma^\prime) + \frac{k^2_-}{p_- p^\prime_-} \Bigr] \Bigl( \frac{p_- p^\prime_-}{k_-} \Bigr) z \Ai(z),
\label{bar tsum}
\end{align}
\end{widetext}
where we have introduced $\mathcal{E} = e |\bm{E}|$ ($\mathcal{E} <0$ for an electron~\footnote{Throughout this work, for any real number $a$ the cube root $a^{1/3}$ is defined as the unique real number $b$ satisfying $b^3 = a$.}) and
\begin{align}
z(\phi_+) &= C_3^2 [1 + \bm{\pi}^2_\perp(\phi_+)], \label{z integral tobs} \\
C_3 &= \Bigl( \frac{k_- m^3}{ p_- p^{\prime}_- \mathcal{E}} \Bigr)^\frac{1}{3}, \label{C3_parameter_def} \\
\bm{\pi}_\perp(\phi_+) &= \frac{1}{m} \Bigl( \bm{p}_\perp - \bm{\mathcal{A}}_\perp(\phi_+)  - \frac{p_-}{k_-} \bm{k}_\perp \Bigr).
\label{pi}
\end{align}
Equation~\eqref{bar tsum} involves the Airy function $\Ai$ and its derivative $\Ai^\prime$ (see Appendix~\ref{sec:airy}) and coincides with the integral over $\phi_-$ of the sum of Eqs.~(8) and (9) in Ref.~\cite{podszusPRD22}.

We now show that Eq.~\eqref{prob_tob_3sum_noint} can become negative over a finite range of $\phi_+$. Since the prefactor in Eq.~\eqref{prob_tob_3sum_noint} is non-negative, the sign of $d P^{(\sigma,\sigma^\prime)}_{\mathrm{NCS}}/(d\phi_+\, d\omega\, d\theta\, d\varphi)$ is entirely determined by $\bar{T}^{\sigma,\sigma^\prime}(\phi_+)$. We therefore study the condition
\begin{equation}
\bar{T}^{\sigma,\sigma^\prime}(\phi_+) \geq 0,
\label{diseq}
\end{equation}
considering separately (i) $\sigma=-\sigma^\prime$ (spin flip upon emission) and (ii) $\sigma=\sigma^\prime$ (no spin flip). In case (i), Eq.~\eqref{diseq} reduces to
\begin{align}
\Bigl( \frac{p_- p^\prime_-}{k_-} \Bigr) z \Ai(z) & \geq \sigma^\prime m \mathcal{E} \Bigl( \frac{p_- p^\prime_-}{k_- \mathcal{E}^2} \Bigr)^\frac{2}{3} \Ai^\prime(z).
\label{diseq_2}
\end{align}
Since $z(\phi_+)>0$, and for each $z>0$ the Airy function and its derivative satisfy $\Ai(z) > 0$ and $\Ai^\prime(z) < 0$, respectively, then Eq.~\eqref{diseq_2} can be rewritten as
\begin{align}
z \frac{\Ai(z)}{\Ai^\prime(z)} \leq \sigma^\prime C_3.
\label{diseq_3}
\end{align}
\begin{figure}[t]
\centering
\includegraphics[width=1\linewidth]{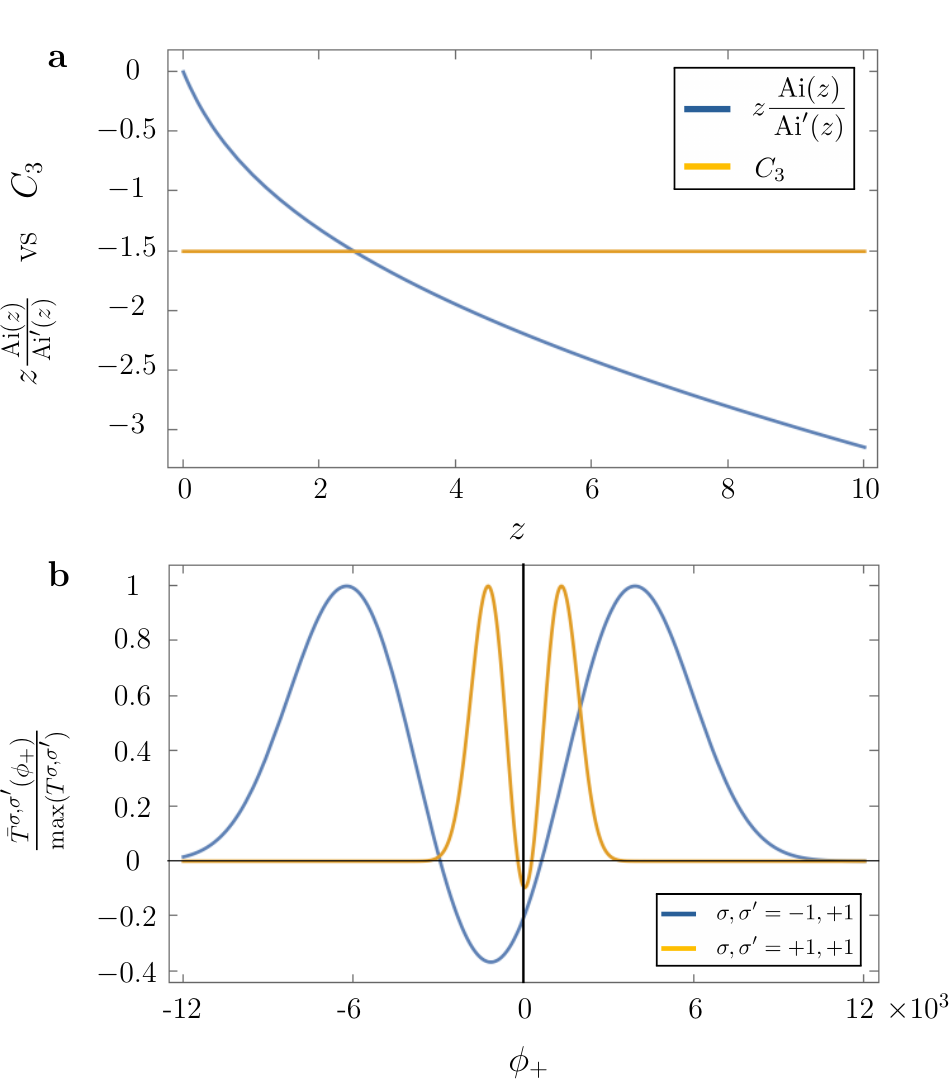}
\caption{Panel~\textbf{a}: Graphical illustration of the inequality in Eq.~\eqref{diseq_3}. The blue curve shows $z\,\Ai(z)/\Ai^\prime(z)$, while the orange horizontal line indicates a representative value of the constant $\sigma^\prime C_3$. The condition in Eq.~\eqref{diseq_3} is satisfied where the blue curve lies below the orange line. Panel~\textbf{b}: $\bar{T}^{-1,+1}(\phi_+)$ (blue) and $\bar{T}^{+1,+1}(\phi_+)$ (orange) as functions of $\phi_+$, each normalized to its respective maximum. Both functions become negative over finite intervals of $\phi_+$, illustrating that $d P^{(\sigma,\sigma^\prime)}_{\mathrm{NCS}}/(d\phi_+\, d\omega\, d\theta\, d\varphi)$ is not positive definite. See the text for details.}
\label{graphical}
\end{figure}
For any choice of initial and final states, the right-hand side of Eq.~\eqref{diseq_3} is a constant. To gain intuition about the meaning of the parameters involved, it is useful to consider the important case of an ultrarelativistic electron counterpropagating with the CCF. To leading order in $1/\gamma$ (with $\theta\sim 1/\gamma$), one has $p_- \approx 2\varepsilon$ and $k_- \approx 2\omega$, and Eq.~\eqref{C3_parameter_def} reduces to 
\begin{align}
C_3 \approx - \Bigl[ \frac{ \omega/\varepsilon }{ (1 - \omega/\varepsilon) \chi_e} \Bigr]^{\frac{1}{3}},
\label{C3_parameter}
\end{align}
which depends only on the ratio $\omega/\varepsilon$ and on $\chi_e$. The general solution of Eq.~\eqref{diseq_3} can be written as
\begin{equation}
z(\phi_+) \geq \tilde{z},
\label{final_dis_z}
\end{equation}
where $\tilde{z}$ is the value of $z$ at which the two sides of Eq.~\eqref{diseq_3} are equal, i.e., the intersection point of the two curves in Fig.~\ref{graphical}(a), representing the left- (blue) and right-hand (orange) sides of Eq.~\eqref{diseq_3}. The example shown in Fig.~\ref{graphical}(a) corresponds to $\chi_e = 1.2$, emission angle $\theta = 2 \times 10^{-5}$ rad, and photon energy fraction $\omega/\varepsilon = 0.8$. Substituting Eqs.~\eqref{z integral tobs} and \eqref{pi} into Eq.~\eqref{final_dis_z} reduces the latter to a quadratic inequality in $\phi_+$
\begin{equation}
\bar{a}\phi_+^2 + \bar{b}\phi_+ + \bar{d} \geq 0,
\label{solutions_pre}
\end{equation}
with $\bar{a}$, $\bar{b}$, and $\bar{d}$ depending on the relative orientation of the incoming electron momentum $\bm{p}$ and the direction $\bm{d}$ of the CCF Poynting vector. For a counterpropagating geometry one has $\bar{a} \geq 0$, and Eq.~\eqref{solutions_pre} yields the two-interval solution
\begin{equation}
\phi_+ \leq \frac{ - \bar{b} - \sqrt{\bar{b}^2 - 4 \bar a \bar d}}{2\bar a} \quad \text{or} \quad \phi_+ \geq \frac{ - \bar{b} + \sqrt{\bar{b}^2 - 4 \bar a \bar d}}{2\bar a},
\label{solutions}
\end{equation}
This is consistent with the plot of $\bar{T}^{-1,+1}(\phi_+)$ shown in Fig.~\ref{graphical}(b) for $\sigma^\prime=-\sigma=1$, $\chi_e=2$, $\omega/\varepsilon=1/10$, and a fixed photon emission direction $\bm{n}$ making an angle $\theta = 2 \times 10^{-5}$ rad with $\bm{p}$, where $\bar{T}^{-1,+1}(\phi_+)$ becomes negative over an extended range of $\phi_+$. Owing to the factor $\sigma^\prime$ in Eq.~\eqref{diseq_3}, only the spin-flip case with $\sigma^\prime=+1$ (i.e., $\bar T^{-1,+1}$) exhibits the two-interval structure in Eq.~\eqref{solutions}, whereas $\bar T^{+1,-1}$ remains positive for all $\phi_+$. For positrons, the situation is reversed.

An analogous analysis applies to the case $\sigma=\sigma^\prime$, for which the condition in Eq.~\eqref{diseq} becomes
\begin{align}
&\Bigl[- 2 m^2 \Bigl( \frac{p_- p^\prime_-}{k_- \mathcal{E}^2(\phi_+)} \Bigr)^\frac{1}{3} + \Bigl(4 + \frac{k_-}{p_-} \frac{k_-}{p'_-} \Bigr) \Bigl( \frac{p_- p^\prime_-}{k_-} \Bigr) z \Bigr] \Ai(z) \geq \notag \\
&\sigma^\prime m \mathcal{E}(\phi_+) \frac{k_-}{p_-} \Bigl( 2 + \frac{k_-}{p'_-} \Bigr) \Bigl( \frac{p_- p^\prime_-}{k_- \mathcal{E}^2(\phi_+)} \Bigr)^\frac{2}{3} \Ai^\prime(z).
\label{diseq_4}
\end{align}
After straightforward algebra, this inequality can be recast as
\begin{equation}
\bigl[- 2 C_3^2 C_4 + ( 4 C_4 + 1 ) z \bigr] \frac{\Ai(z)}{\Ai^\prime(z)} \leq \sigma^\prime C_3 \bigl( 2 C_5 + 1 \bigr),
\label{diseq_5}
\end{equation}
with $C_4 = p_- p_-^\prime / k_-^2$ and $C_5 = p_-^\prime / k_-$. Specializing, as above, to an ultrarelativistic electron counterpropagating with the CCF, these parameters reduce to functions of the energy ratio $\omega/\varepsilon$:
\begin{align}
C_4 & \approx \frac{1-\omega/\varepsilon}{\bigl( \omega/\varepsilon \bigr)^2}, \\
C_5 & \approx \frac{1-\omega/\varepsilon}{ \omega/\varepsilon }.
\label{parameters list_2}
\end{align}
Although a graphical illustration is less convenient in this case, considering limiting regimes is still instructive. For $\omega/\varepsilon \to 0$, the term proportional to $C_4 z$, which scales with the highest negative power of $\omega/\varepsilon$, dominates Eq.~\eqref{diseq_5}, and the inequality reduces to
\begin{equation}
4 C_4 z \frac{\Ai(z)}{\Ai^\prime(z)} \leq 0,
\label{diseq_6}
\end{equation}
which is satisfied for all $z>0$ (and hence for all $\phi_+$). Conversely, in the limit $\omega/\varepsilon \to 1$ most terms vanish, and Eq.~\eqref{diseq_5} reduces to the same form as Eq.~\eqref{diseq_3}. Indeed, Fig.~\ref{graphical}(b) shows that already at $\omega/\varepsilon=0.8$, for a counterpropagating electron--CCF configuration, $\bar{T}^{+1,+1}(\phi_+)$ (orange) exhibits the same qualitative behavior as $\bar{T}^{-1,+1}(\phi_+)$ (blue). Thus, as $\omega/\varepsilon$ increases, $\bar{T}^{+1,+1}(\phi_+)$ ceases to be non-negative.

In summary, we have demonstrated that, in contrast to the spin-summed case, the angle- and spin-resolved distribution in Eq.~\eqref{prob_tob_3sum_noint} is not positive definite and therefore does not admit a probabilistic interpretation, even in a strictly constant and uniform crossed field.


\subsection{\label{subsec:QM}Spin- and polarization-resolved differential distribution within the quasiclassical method}

Inspired by Schwinger’s seminal work \cite{schwingerPNAS54}, the quasiclassical method (QM) developed by Baier and Katkov provides a powerful approach to describe radiation from ultrarelativistic charges and pair production by high-energy photons in strong external electromagnetic fields \cite{Baier-book, baierPLA67}. Their key observation is that, at ultrarelativistic energies, the noncommutativity of the operators associated with the particle dynamical variables (such as position and momentum) can be neglected, whereas commutators involving the photon field operators, which are ultimately related to the importance of photon recoil, must be retained \cite{Baier-book}. A distinctive feature of the QM is that it evaluates quantum transition amplitudes along the classical trajectory determined by the Lorentz force, while the spin evolution between emissions is described by the Bargmann-Michel-Telegdi (BMT) equation. Compared with the Furry-picture formulation, the QM often yields more compact analytic expressions for transition amplitudes and distributions, while reproducing the corresponding strong-field QED results in the WKB limit at high particle energies \cite{dipiazzaPRD21}. In fact, after taking the ultrarelativistic limit, QM expressions can be obtained from the corresponding Furry-picture results written in light-front coordinates and by the substitutions $p_- \approx 2\varepsilon$, $k_- \approx 2\omega$, and $\phi \approx 2t$ (see the second subsection of Appendix~\ref{CCFint} for an explicit example). 
\begin{figure}[tb]
\centering
\includegraphics[width=0.5\linewidth]{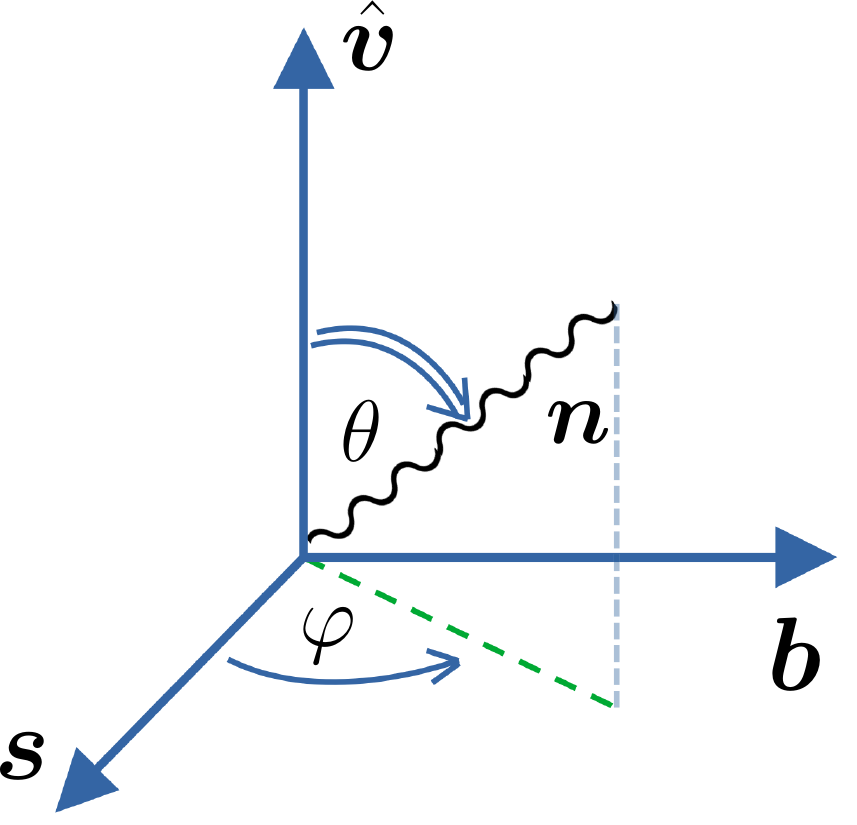}
\caption{Angular coordinate system used to specify the photon emission direction $\bm{n}$ with respect to the electron velocity unit vector $\hat{\bm{v}}$. The polar angle $\theta$ is defined as the angle between $\bm{n}$ and $\hat{\bm{v}}$. Let $\bm{s}$ be the unit vector along the electron’s transverse (with respect to $\hat{\bm{v}}$) acceleration, and define $\bm{b}=\hat{\bm{v}}\times\bm{s}$. The azimuthal angle $\varphi$ is measured in the plane orthogonal to $\hat{\bm{v}}$, spanned by $(\bm{s},\bm{b})$, from $\bm{s}$ to the projection of $\bm{n}$ onto that plane.}
\label{angles_pic}
\end{figure}
In general, since the squared matrix element $|\mathcal{M}_{fi}|^2$ in Eq.~\eqref{prob_tob_pre} must be linear in $\bm{\xi}$ and $\bm{\eta}$, $\bm{\eta}^\prime$ (see, e.g., Ref.~\cite{Berestetskii-book}), then $dP^{(\eta\eta^\prime\xi)}_{NCS} / (d t d\omega d\theta d\varphi)$ can be written in the form 
\begin{equation}
\frac{dP^{(\eta\eta^\prime\xi)}_{\mathrm{NCS}}}{d t\, d\omega\, d\theta\, d\varphi} = C_{\mathrm{NCS}} \left(w + \xi^i h^i + \eta^{\prime j} S^j + \xi^i \eta^{\prime j} P^{ij} \right),
\label{spa_res_2}
\end{equation}
where Einstein summation over $i,j=1,2,3$ is implied. In Eq.~\eqref{spa_res_2} we have collected terms depending on $\bm{\eta}^{\prime}$ and $\bm{\xi}$ to make their dependence explicit. Here, $h^i$ and $S^j$ determine the mean polarization- and spin- vectors, respectively (see below), and $P^{ij}$ is the spin-polarization correlation matrix. Note that $w$, $h^i$, $S^j$, $P^{ij}$ are functions of the incoming and outgoing particles parameters, and depend also on $\bm{\eta}$. Their explicit expressions as well as the expression of the prefactor $C_{\mathrm{NCS}}$ as obtained within the QM for NCS in the LCFA, are given in Appendix~\ref{sec:yy-comp}. 

The quantity $dP^{(\eta\eta^\prime\xi)}_{NCS} / (d t d\omega d\theta d\varphi)$ in Eq.~\eqref{spa_res_2} is differential in the time $t$, photon energy $\omega$ and polar $\theta$ and azimuthal $\varphi$ angles. As in the corresponding Furry-picture derivation, it is obtained analytically from the QM matrix element by introducing the variables $t_+ = (t_1+t_2)/2$ and $t_- = t_1-t_2$ (see Ref. \cite{Baier-book} for this notation), expanding the integrand to leading order in $t_-$ about $t_-=0$, integrating the resulting expression over $t_-$, and identifying $t_+$ with the local time along the particle trajectory, i.e., $t=t_+$ (see Refs.~\cite{daiPRD23, montefioriPhD25}). The polar angle $\theta$ is defined as the angle between the unit vectors $\bm{n}$ and $\hat{\bm{v}}$ (to be distinguished from the particle's velocity $\bm{v}=|\bm{v}| \hat{\bm{v}}$). Let $\bm{s}$ be the unit vector along the electron’s transverse (with respect to $\hat{\bm{v}}$) acceleration, and define $\bm{b}=\hat{\bm{v}}\times\bm{s}$. The azimuthal angle $\varphi$ is measured in the plane orthogonal to $\hat{\bm{v}}$, spanned by $(\bm{s},\bm{b})$, from $\bm{s}$ to the projection of $\bm{n}$ onto that plane (see Fig. \ref{angles_pic}). This definition of angles is consistent with that given above for the Furry-picture calculation in a CCF for $\bm{d}$ being in the opposite direction of the local electron kinetic momentum, and it corresponds to the standard definition of spherical angular coordinates with respect to a system where $\bm{s}$ is the $x$ axis, $\bm{b}$ is the $y$ axis and $\hat{\bm{v}}$ is the $z$ axis.

Owing to their simplicity, commonly used local, collinear-emission models are based on angle-integrated expressions, where the solid-angle integration is carried out analytically within the QM (see, e.g., Ref.~\cite{chenPRD22} and references therein). In practice, starting from Eq.~\eqref{spa_res_2}, these models determine spin and polarization from
\begin{align}
\frac{dP^{(\eta\eta^\prime\xi)}_{\mathrm{NCS}}}{d t\, d\omega} &= \int_0^{\pi} d\theta \int_0^{2\pi} d\varphi\,\frac{dP^{(\eta\eta^\prime\xi)}_{\mathrm{NCS}}}{d t\, d\omega\, d\theta\, d\varphi} \nonumber\\
& =
\underline{C}_{\mathrm{NCS}} \left(\underline{w} + \xi^i \underline{h}^i + \eta^{\prime j} \underline{S}^j + \xi^i \eta^{\prime j} \underline{P}^{ij}\right).
\label{sp_res}
\end{align}
Explicit expressions for the angle-integrated functions $\underline{w}$, $\underline{h}^i$, $\underline{S}^j$, correlation matrix $\underline{P}^{ij}$, and for the prefactor $\underline{C}_{\mathrm{NCS}}$ are given in Appendix~\ref{sec:and_integrate}.

Note that, in both Eq.~\eqref{spa_res_2} and Eq.~\eqref{sp_res}, $\bm{\eta}^\prime$ and $\bm{\xi}$ are the spin and the polarization vectors characterizing the spin and polarization detector, respectively, and can be chosen arbitrarily. In the spin case, $\bm{\eta}^\prime$ specifies the axis with respect to which the outgoing lepton spin is measured, and the corresponding probabilities refer to spin aligned ($+|\bm{\eta}^\prime|$) or antialigned ($-|\bm{\eta}^\prime|$) with that axis. Likewise, in the polarization case, the Stokes vector $\bm{\xi}$ characterizes the photon polarization detector, and corresponds to selecting photon polarization states with $\bm{\xi}=\pm|\bm{\xi}|$ (see Sec.~65 of Ref.~\cite{Berestetskii-book}).

In kinetic descriptions, the relevant quantities are spin- and polarization-resolved distribution functions. The spin-resolved differential distribution is therefore obtained by summing over the two polarization states, yielding
\begin{align}
\frac{dP^{(\eta\eta^\prime)}_{\mathrm{NCS}}}{d t\, d\omega\, d\theta\, d\varphi} &= 2 C_{\mathrm{NCS}} \left(w + \bm{\eta}^{\,\prime}\cdot \bm{S} \right), \; \text{[cfr. Eq.~\eqref{spa_res_2}],}
\label{spa_res_2_sum_pola} \\
\frac{dP^{(\eta\eta^\prime)}_{\mathrm{NCS}}}{d t\, d\omega} &= 2 \underline{C}_{\mathrm{NCS}} \left(\underline{w} + \bm{\eta}^{\,\prime}\cdot \underline{\bm{S}} \right), \; \text{[cfr. Eq.~\eqref{sp_res}].}
\label{spa_res_2_sum_pola_u}
\end{align}
From these expressions, the mean outgoing spin vector (in the electron rest frame) produced by the scattering process follows as (see Sec.~65 of Ref.~\cite{Berestetskii-book} and, e.g., Refs.~\cite{chenPRD22, daiPRD23} and references therein)
\begin{align}
\langle \bm{\eta}^{\,\prime} \rangle &= \frac{\bm{S}(t)}{w(t)}, 
\label{final_spin_angles_def} \quad \text{(angle-resolved),}\\
\langle \bm{\eta}^{\,\prime} \rangle &= \frac{\underline{\bm{S}}(t)}{\underline{w}(t)},
\quad \text{(angle-integrated),}
\label{final_spin_angles_def_u}
\end{align}
where $w$ ($\underline{w}$) and $\bm{S}$ ($\underline{\bm{S}}$) are linear functions of the incoming spin vector $\bm{\eta}$, and depend on time $t$ and on the specific incoming lepton $\bm{p}$ and outgoing photon $\bm{k}$ ($\omega$) momentum (energy). Similarly, summing Eq.~\eqref{spa_res_2} over the two spin states of the outgoing electron yields the polarization-resolved distribution
\begin{align}
\frac{dP^{(\eta\xi)}_{\mathrm{NCS}}}{d t\, d\omega\, d\theta\, d\varphi} &= 2 C_{\mathrm{NCS}} \left(w + \bm{\xi}\cdot \bm{h}
\right), \; \text{[cfr. Eq.~\eqref{spa_res_2}],}
\label{spa_res_2_sum_spin} \\
\frac{dP^{(\eta\xi)}_{\mathrm{NCS}}}{d t\, d\omega} &= 2 \underline{C}_{\mathrm{NCS}} \left(\underline{w} + \bm{\xi}\cdot \underline{\bm{h}}
\right), \; \text{[cfr. Eq.~\eqref{sp_res}],}
\label{spa_res_2_sum_spin_u}
\end{align}
from which the mean photon Stokes vector describing the photon polarization state is obtained as
\begin{align}
\langle \bm{\xi} \rangle &= \frac{\bm{h}(t)}{w(t)}, \quad \text{(angle-resolved),}
\label{final_polarization_angles_def} \\
\langle \bm{\xi} \rangle &= \frac{\underline{\bm{h}}(t)}{\underline{w}(t)}, 
\quad \text{(angle-integrated),}
\label{final_polarization_angles_def_u}
\end{align}
in the chosen polarization basis. For the angle-integrated collinear model based on Eq.~\eqref{sp_res}, the same reasoning has led to analogous expressions with $w$, $\bm{h}$, and $\bm{S}$ replaced by their angle-integrated counterparts $\underline{w}$, $\underline{\bm{h}}$, and $\underline{\bm{S}}$, respectively. Note that, in both angle-resolved and angle-integrated descriptions, the spin-polarization correlation matrix $P^{ij}$ ($\underline{P}^{ij}$) encodes joint correlations between the outgoing electron spin and the emitted photon polarization. Accessing these correlations requires event-by-event coincidence information, i.e., the ability to correlate each photon with its parent electron and to measure both the photon polarization and the recoiling-electron spin in the same event. In a kinetic (plasma) description one evolves single-particle distribution functions, which are fully characterized by $w$ together with $\bm{S}$ (electron) or $\bm{h}$ (photon) neglecting higher moments and electron-photon correlations.

\begin{figure}[tb]
\centering
\includegraphics[width=1\linewidth]{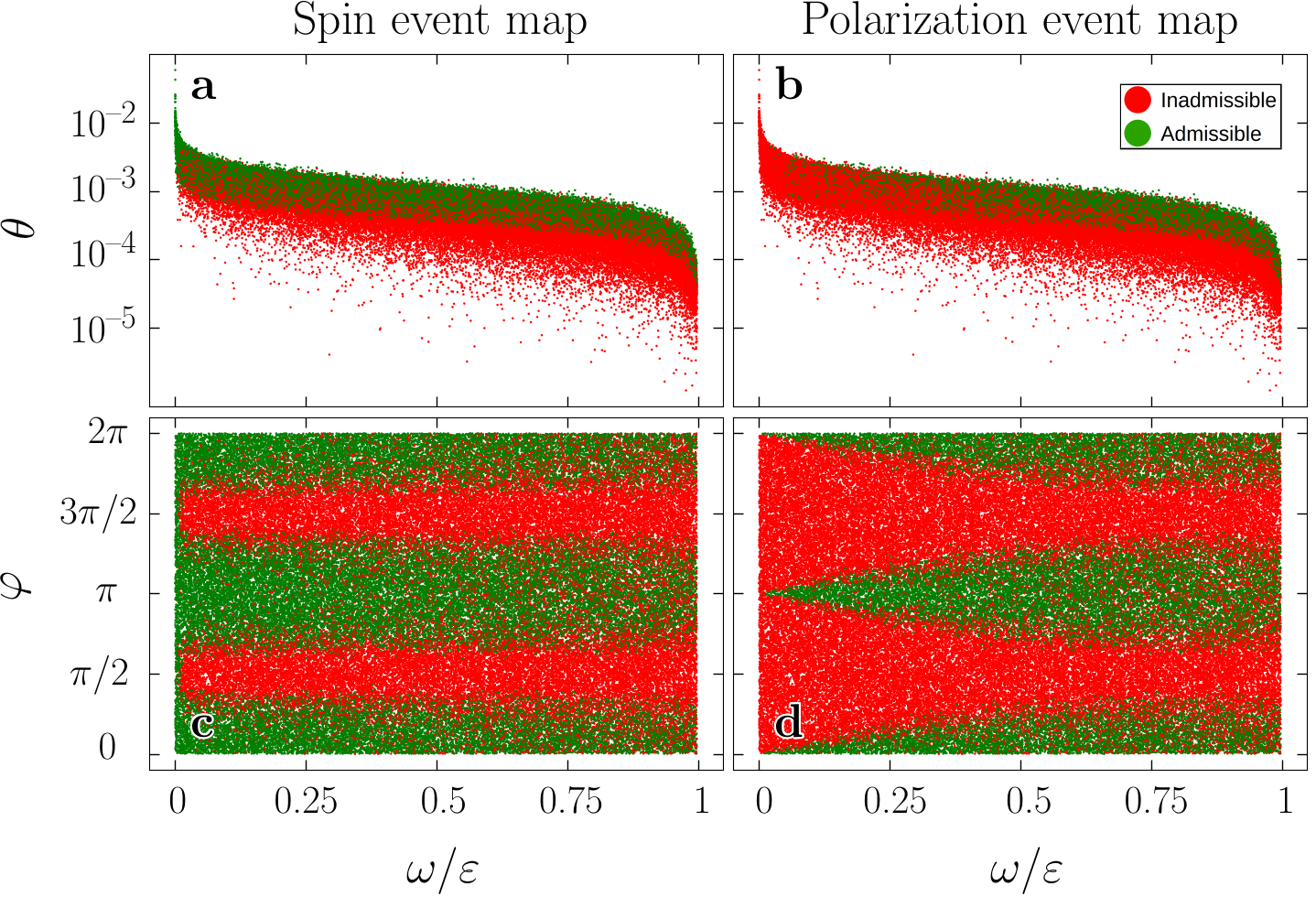}
\caption{Distribution of admissible (green) and inadmissible (red) absolute values of the mean outgoing electron spin $\langle\bm{\eta}^{\,\prime}\rangle$ (panels \textbf{a}, \textbf{c}) and of the photon Stokes vector $\langle\bm{\xi}\rangle$ (panels \textbf{b}, \textbf{d}) versus the photon-to-electron energy ratio $\omega/\varepsilon$ and the photon emission angles $\theta$ and $\varphi$. An incoming electron has energy $\varepsilon=1~\mathrm{GeV}$  in a magnetic field $\bm{B}$ with velocity perpendicular to $\bm{B}$ and with spin vector $\bm{\eta}$ anti-aligned with $\bm{B}$ and quantum parameter $\chi_e=2$. We simulate $10^7$ events by sampling $\omega$ uniformly in $(0,\varepsilon)$ and $(\theta,\varphi)$ from the distribution $dP^{(\eta)}_{\mathrm{NCS}}/(dt\,d\omega\,d\theta\,d\varphi)$ obtained from Eq.~\eqref{spa_res_2} after summing over the final spin and polarization states. For each event, $\langle\bm{\eta}^{\,\prime}\rangle$ and $\langle\bm{\xi}\rangle$ are computed from Eqs.~\eqref{final_spin_angles_def} and \eqref{final_polarization_angles_def}. Markers are green when $|\langle\bm{\eta}^{\,\prime}\rangle|\le 1$ (panels \textbf{a}, \textbf{c}) or $|\langle\bm{\xi}\rangle|\le 1$ (panels \textbf{b}, \textbf{d}), and red otherwise; red markers thus indicate events for which the local model implies a negative inferred probability, in direct violation of its probabilistic interpretation.}
\label{fig:scatter}
\end{figure}
\begin{figure}[tb]
\centering
\includegraphics[width=1\linewidth]{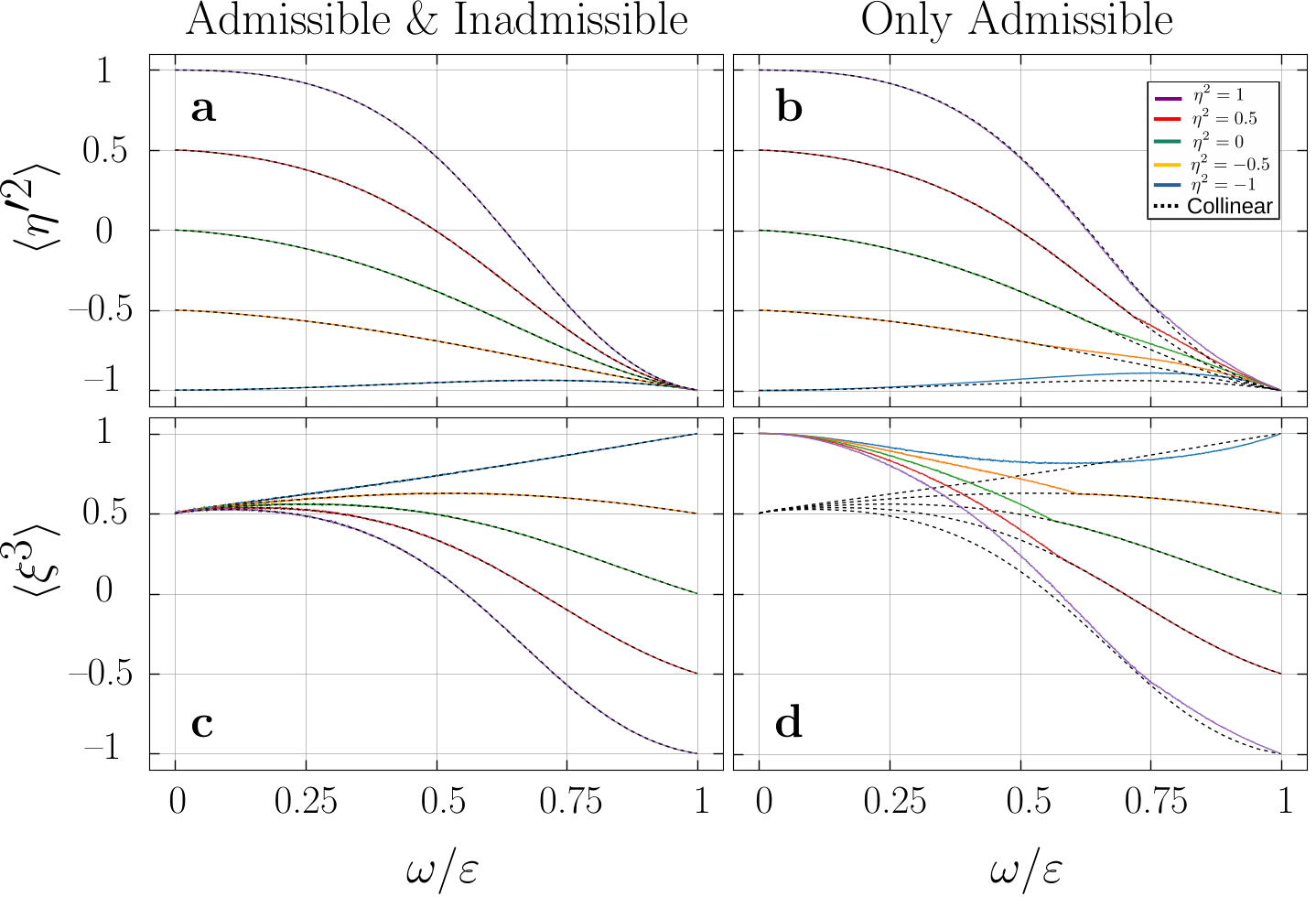}
\caption{Spin and polarization curves for NCS of an ultrarelativistic electron ($\varepsilon = 1$ GeV, $\chi_e = 2$) for five different initial spin states (purple: $1$; red: $0.5$; green: $0$; orange: $-0.5$; blue: $-1$) in a uniform magnetic field $\bm{B} = (0, B_y, 0)$ versus the photon-to-electron energy ratio $\omega/\varepsilon$. Panels \textbf{a}, \textbf{b}: mean outgoing electron spin component $\langle \eta^{\prime 2}\rangle$ (along $\bm{B}$). Panels \textbf{c}, \textbf{d}: mean photon Stokes parameter $\langle \xi^{3}\rangle$. In panel \textbf{a} and \textbf{c} colored curves are obtained by averaging the spin and polarization inferred from Eqs.~\eqref{final_spin_angles_def} and \eqref{final_polarization_angles_def}, respectively, over $10^4$ sampled emission angles at each $\omega/\varepsilon$. In panels \textbf{b}, \textbf{d} colored curves show the same averages after discarding events with $|\bm{\eta}^{\,\prime}|>1$ or $|\bm{\xi}|>1$. Black dashed curves show the corresponding angle-integrated prediction from Eq.~\eqref{final_spin_angles_def_u} (panels~\textbf{a},~\textbf{b}) and Eq.~\eqref{final_polarization_angles_def_u} (panels~\textbf{c},~\textbf{d}).}
\label{fig:curves}
\end{figure}

In Sec.~\ref{subsec:sign} we have demonstrated that, even in a strictly uniform CCF, the local, instantaneous angle- and spin-resolved distribution can take both positive and negative values. Here we aim to visually identify the region of parameter space where this occurs. Figure~\ref{fig:scatter} displays the results obtained for an electron with energy $\varepsilon=1~\mathrm{GeV}$ and velocity perpendicular to the magnetic field $\bm{B}$ with quantum parameter $\chi_e=2$, and spin vector $\bm{\eta}$ anti-aligned with $\bm{B}$, as functions of the photon-to-electron energy ratio $\omega/\varepsilon$ and the polar angle $\theta$ or azimuthal angle $\varphi$, respectively. Each marker corresponds to one sample and is colored green when the physical bound is satisfied ($|\langle\bm{\eta}^{\,\prime}\rangle|\le 1$ in panels~\textbf{a}, \textbf{c} and $|\langle\bm{\xi}\rangle|\le 1$ in panels~\textbf{b}, \textbf{d}), and red otherwise. Note that, when translated into probabilities, a spin or polarization vector with magnitude exceeding unity corresponds to a negative value of the inferred probability. To obtain a clear picture across parameter space, we kept the electron state fixed and generated $10^7$ samples with the photon energy drawn uniformly from $0<\omega<\varepsilon$, rather than from the spectrum implied by Eq.~\eqref{spa_res_2}, which would strongly bias the sampling toward low-energy photons. The emission angles $(\theta,\varphi)$ were sampled from the spin- and polarization-summed angular distribution $dP^{(\eta)}_{\mathrm{NCS}}/(dt\,d\omega\,d\theta\,d\varphi)$ obtained from Eq.~\eqref{spa_res_2} after summing over the final electron spin and photon polarization states. For each sampled event, the inferred mean outgoing electron spin vector $\langle\bm{\eta}^{\,\prime}\rangle$ and mean photon Stokes vector $\langle\bm{\xi}\rangle$ were computed from Eqs.~\eqref{final_spin_angles_def} and \eqref{final_polarization_angles_def}, respectively.

Figure~\ref{fig:scatter} shows that the predicted larger-than-unity magnitudes of the spin and Stokes vectors occur predominantly in the forward direction for both electrons and photons, i.e., at small values of $\theta$ across essentially all photon energies (see panels~\textbf{a} and \textbf{b} of Fig.~\ref{fig:scatter}). As a function of the azimuthal angle, two well-defined regions centered at $\varphi=\pi/2$ and $\varphi=3\pi/2$ are clearly visible. These correspond to the axis orthogonal to the plane spanned by the incoming electron momentum and the transverse acceleration. The width of these regions is approximately independent of $\omega/\varepsilon$ for the electron spin (panel~\textbf{c}), while it increases toward low photon energies for the photon polarization (panel~\textbf{d}).

Figure~\ref{fig:curves} quantifies how events with physically admissible ($|\bm{\eta}^{\,\prime}|\le 1$, $|\bm{\xi}|\le 1$) and inadmissible ($|\bm{\eta}^{\,\prime}|>1$, $|\bm{\xi}|>1$) inferred spin/polarization contribute to the angle-integrated distribution in Eq.~\eqref{sp_res}. We considered an incoming electron with $\varepsilon=1~\mathrm{GeV}$ energy in a magnetic field $\bm{B}=(0,B_y,0)$ with quantum parameter $\chi_e=2$ and for several initial spin values $\eta^2$ (purple: $1$; red: $0.5$; green: $0$; orange: $-0.5$; blue:$-1$). Panels~\textbf{a} and \textbf{b} show the mean outgoing electron spin component $\langle \eta^{\prime 2}\rangle$ (along $\bm{B}$), while panels~\textbf{c} and \textbf{d} show the mean photon Stokes parameter $\langle \xi^{3}\rangle$, with all quantities plotted as functions of the photon-to-electron energy ratio $\omega/\varepsilon$. The parameter $\langle \xi^{3}\rangle$ quantifies the degree of linear polarization along $\bm{n} \times (\bm{s} \times \bm{n})$ (for $\langle \xi^{3}\rangle=+1$) or along $\bm{n} \times \bm{s}$ (for $\langle \xi^{3}\rangle=-1$), where the limiting values $\pm1$ correspond to pure states fully linearly polarized along the respective axis. In panels~\textbf{a} and \textbf{c}, the colored curves were obtained by sampling $10^4$ emission directions at each $\omega/\varepsilon$ and averaging the corresponding ensemble of spin and polarization inferred from Eqs.~\eqref{final_spin_angles_def} and \eqref{final_polarization_angles_def}. Panels~\textbf{b} and \textbf{d} show the same averages after discarding events with $|\bm{\eta}^{\,\prime}|>1$ or $|\bm{\xi}|>1$. In all panels of Fig.~\ref{fig:curves}, the black dashed curves show the corresponding angle-integrated (collinear-model) prediction from Eq.~\eqref{sp_res}, which always yields physically admissible results, i.e., spin and Stokes vectors with magnitude smaller than or equal to unity, and agrees with Ref.~\cite{SongPRL2023}. 

On the one hand, plots in panels~\textbf{a} and \textbf{c} of Fig.~\ref{fig:curves} confirm that sampling via Eqs.~\eqref{final_spin_angles_def} and \eqref{final_polarization_angles_def} reproduces the angle-integrated prediction of Eq.~\eqref{sp_res}. On the other hand, panels~\textbf{b} and \textbf{d} show that discarding inadmissible events affects the photon polarization much more strongly than the electron spin, particularly for photon energies $\omega \lesssim \varepsilon/2$. Therefore, simply ignoring inadmissible cases can be problematic, as it may lead to quantitatively incorrect results for the total spin and polarization vectors.


\subsection{From local to non-local differential distributions}

The sign-indefinite behavior demonstrated above implies that the \emph{local} differential distribution either starting from the Furry picture in a CCF [Eq.~\eqref{prob_tob_3sum_noint}] or from the ultrarelativistic limit obtained within the QM [Eq.~\eqref{spa_res_2}], cannot in general be interpreted as a probability density. While the lack of a \emph{local} probabilistic interpretation in space-time dependent backgrounds is well known (see, e.g., Refs.~\cite{ritusJSLR85, dipiazzaPRA18}) and is simply due to the breakdown of the LCFA, i.e., when the background-field variation length becomes comparable to the photon formation length ($a_0\lesssim 1$), the analysis above reveals a more fundamental limitation: even in a strictly uniform CCF, the angle-, spin-, and polarization-resolved NCS \emph{local} differential distribution does not admit a direct probabilistic interpretation. This result challenges a key assumption underlying MC event generators and particle-in-cell (PIC) implementations based on the LCFA, namely that local, instantaneous expressions can always be interpreted as probability densities and sampled to obtain spin-/polarization-resolved particle distributions.

A consistent treatment therefore requires a quantity with an unambiguous probabilistic meaning. Conceptually, the only object that can be rigorously identified as a probability is the fully phase-integrated expression in Eq.~\eqref{prob_tob_pre}, which yields the transition probability between asymptotic initial and final states. Equivalently, after changing variables to $(\phi_+,\phi_-)$, integrating over $\phi_-$, and summing over the photon polarizations (so as to focus on the lepton spin for simplicity), one must perform the integral over $\phi_+$ in Eq.~\eqref{matrix_element_squared_final} to obtain a distribution that is differential in the emitted-photon momentum and resolved in the lepton spin. By construction, the resulting phase-integrated distribution is non negative and yields physically admissible spin/polarization vector for any photon momentum $\bm{k}$.

Within the QM, this implies, following the same steps as in Sec.~\ref{subsec:QM}, that the mean outgoing spin vector is given by [cf.~Eq.~\eqref{final_spin_angles_def}]
\begin{equation}
\langle \bm{\eta}^{\,\prime} \rangle = \frac{\int_{-\infty}^{\infty}{dt\, C_{\mathrm{NCS}}(t) \bm{S}(t)}}{\int_{-\infty}^{\infty}{dt \, C_{\mathrm{NCS}}(t) w(t)}},
\label{final_spin_angles_int}
\end{equation}
and similarly that the mean photon Stokes vector is [cf.~Eq.~\eqref{final_polarization_angles_def}]
\begin{equation}
\langle \bm{\xi} \rangle =  \frac{\int_{-\infty}^{\infty}{dt\, C_{\mathrm{NCS}}(t) \bm{h}(t)}}{\int_{-\infty}^{\infty}{dt \, C_{\mathrm{NCS}}(t) w(t)}}.
\label{final_polarization_angles_int}
\end{equation}
Directly implementing the numerical calculation of these quantities in a code is unpractical, because it would require integration along the \emph{whole} trajectories of particles. However, as we will elaborate on below, although Eqs.~\eqref{final_spin_angles_int} and \eqref{final_polarization_angles_int} involve integrals over the entire particle's trajectory, for each specific photon momentum $\bm{k}$ and in a CCF the integrals receive appreciable contributions only from an extended but finite segment of the worldline (whose length depends also on the energy of the emitted photon, see Refs.~\cite{dipiazzaPRA18, dipiazzaPRA19}). For an ultrarelativistic emitter, the collimated nature of the radiation implies that this segment corresponds to a ``formation region'' over which the particle is deflected by a small angle ${\sim}1/\gamma$ \cite{Baier-book}.


\section{\label{sec:newMC}Modeling fully resolved NCS distributions}

We now identify the distributions required for MC simulations of NCS within the LCFA. Starting from Eq.~\eqref{spa_res_2_sum_pola}, summing over the outgoing electron spin states, one obtains
\begin{align}
\frac{dP^{(\eta)}_{\mathrm{NCS}}}{d\omega\, d\theta\, d\varphi} = & 4 \int dt\, C_{\mathrm{NCS}}(t)\, w(t) = \int dt\, \frac{\alpha\,\omega}{\pi^{2}} \sqrt{\frac{2\lambda}{3}} \frac{\gamma^{3} \sin\theta}{\chi_{e}\,\varepsilon^\prime} \notag \\
\times & \Bigg[ K_{\frac{1}{3}}(\xi_e) \Bigl( 2\lambda \frac{\varepsilon^2 + \varepsilon^{\prime 2}}{\varepsilon \varepsilon^\prime} - \frac{1}{\gamma^2} \Bigr) + K_{\frac{2}{3}}(\xi_e) \sqrt{2 \lambda} \notag \\
\times & \Bigl( \theta \sin\varphi \frac{\varepsilon^2 -\varepsilon^{\prime 2}}{ \varepsilon \varepsilon^\prime} \hat{\bm{v}} - \frac{\omega}{\gamma \varepsilon} \bm{b}  \Bigr) \cdot \bm{\eta} \Bigg], 
\label{eq:summed_ultrarel}
\end{align}
where
\begin{align}
\lambda=1-\bm{n}\cdot\bm{v},
\label{lambda_def} \\
\xi_e = \frac{4\sqrt{2}}{3} \frac{\omega \gamma^3}{\varepsilon^\prime \chi_e} \lambda^{\frac{3}{2}} \label{xi_e_def},
\end{align}
and $C_{\mathrm{NCS}}$ and $w$ are given in  Appendix~\ref{sec:yy-comp}. The time integral is evaluated along the classical lepton trajectory determined by the Lorentz force, while the spin evolves according to the BMT equation~\footnote{In the Furry-picture formulation, the momentum and spin evolution according to the Lorentz and BMT equations are encoded in the Volkov state (see Ref.~\cite{meurenBook14}).}. Integrating Eq.~\eqref{eq:summed_ultrarel} over the emission angles yields
\begin{align}
\frac{dP^{(\eta)}_{\mathrm{NCS}}}{d\omega} & = \frac{\alpha}{\sqrt{3}\pi\gamma^{2}} \int_{-\infty}^{+\infty} dt\, \Bigg[ K_{2/3}(z_q)\,\frac{\varepsilon^{2}+\varepsilon^{\prime 2}}{\varepsilon\,\varepsilon^\prime} \notag \\
-& \int_{z_q}^{\infty} dx\, K_{1/3}(x) - K_{1/3}(z_q)\,\frac{\omega}{\varepsilon}\,(\bm{\eta}\cdot\bm{b}) \Bigg],
\label{eq:summed_ultrarel_angleint}
\end{align}
where $\varepsilon^\prime=\varepsilon-\omega$ and $z_q=2\omega/[3(\varepsilon-\omega)\chi_e]$.

In a CCF, there is a one-to-one correspondence between a given photon momentum $\bm{k}=\omega\bm{n}$ and the time $t^*$ where the lepton velocity is parallel to $\bm{k}$~\cite{ritusJSLR85}. Accordingly, for fixed emission direction $\bm{n}$ the time integral in Eq.~\eqref{eq:summed_ultrarel_angleint} effectively receives contributions only over the corresponding formation region $t_f$, defined according to
\begin{align}
\frac{dP^{(\eta)}_{\mathrm{NCS}}(\bm{n})}{d\omega} & \approx \frac{\alpha}{\sqrt{3}\pi\gamma^{2}} \int_{t^* -t_f/2}^{t^* +t_f/2} dt\, \Bigg[ K_{2/3}(z_q)\,\frac{\varepsilon^{2}+\varepsilon^{\prime 2}}{\varepsilon\,\varepsilon^\prime}  \notag \\
- & \int_{z_q}^{\infty} dx\, K_{1/3}(x) - K_{1/3}(z_q)\,\frac{\omega}{\varepsilon}\,(\bm{\eta}\cdot\bm{b})
\Bigg].
\label{eq:summed_ultrarel_angleint_k}
\end{align}
More precisely, here $t_f$ is the direction-, and photon energy-dependent (see Refs.~\cite{dipiazzaPRA18, dipiazzaPRA19}) photon formation time, which on average corresponds to the time required for the lepton momentum $\bm{p}$ to turn by an angle ${\sim}1/\gamma$ with respect to $\bm{n}$~\cite{ritusJSLR85, Baier-book}. Moreover, in a CCF the integrand in Eq.~\eqref{eq:summed_ultrarel_angleint_k} depends only on $\varepsilon$, $\varepsilon^\prime=\varepsilon-\omega$, and the quantum parameter $\chi_e$ (constant in a CCF), as well as on the scalar product $\bm{\eta}\cdot\bm{b}$. Over the formation interval, $\varepsilon$ varies only at relative order $1/\gamma^{2}$ and $\bm{\eta}\cdot\bm{b}$ is approximately conserved in the ultrarelativistic regime because $\bm{\eta}$ and $\bm{b}$ co-rotate under the Lorentz and BMT dynamics~\footnote{The Lorentz and BMT precession frequency coincide except for the anomalous magnetic moment contribution, which is negligible over the photon formation region.}. As a result, the integrand can be taken as approximately constant over $t_f$, yielding
\begin{align}
\frac{dP^{(\eta)}_{\mathrm{NCS}}(\bm{n})}{d\omega} \approx & \frac{dP^{(\eta)}_{\mathrm{NCS}}(t^*)}{dt\, d\omega}\, t_f, \label{eq:spect} \\[4pt]
\frac{dP^{(\eta)}_{\mathrm{NCS}}(t^*)}{dt\, d\omega} \equiv & \frac{\alpha}{\sqrt{3}\pi\gamma^{2}} \Bigg[ K_{2/3}(z_q)\,\frac{\varepsilon^{2}+\varepsilon^{\prime 2}}{\varepsilon\,\varepsilon^\prime} - \int_{z_q}^{\infty} dx\, K_{1/3}(x) \notag \\
- & K_{1/3}(z_q)\,\frac{\omega}{\varepsilon}\,(\bm{\eta}\cdot\bm{b}) \Bigg]\Bigg\vert_{t=t^*}. \label{eq:summed_ultrarel_angleint_k2}
\end{align}
For an ultrarelativistic emitter radiation is strongly beamed and the photon formation time $t_f$ at a given direction is relatively short, so the photon spectrum observed along a fixed direction $\bm{n}$ is governed by the \emph{local} dynamics of the emitting lepton [see Eq.~\eqref{eq:spect}]. Accordingly, the photon spectrum at a given direction is determined by Eq.~\eqref{eq:summed_ultrarel_angleint_k2}. The dependence on the observation direction $\bm{n}$ enters only through the fact that, in a CCF, a given $\bm{k}=\omega\bm{n}$ selects the corresponding center of the formation region, i.e., a specific interval along the trajectory, and hence the local lepton state [in particular its energy $\varepsilon(t)$] within that interval. Equation~\eqref{eq:summed_ultrarel_angleint_k2} can therefore be interpreted as a local probability density per unit time and per unit photon energy. Finally, integrating Eq.~\eqref{eq:summed_ultrarel_angleint_k2} over the photon energy yields the local emission rate
\begin{equation}
\frac{dP^{(\eta)}_{\mathrm{NCS}}}{dt} = \int_{0}^{\varepsilon} d\omega\, \frac{dP^{(\eta)}_{\mathrm{NCS}}}{dt\, d\omega}.
\label{eq:rate}
\end{equation}

Under the assumptions stated above, each point \emph{within the formation region} gives essentially the same emission probability per unit time [Eq.~\eqref{eq:rate}] and the same spectral shape [Eq.~\eqref{eq:summed_ultrarel_angleint_k2}]. Similar conclusions apply to the angular distribution, provided the photon emission angles are defined with respect to the instantaneous $(\bm{s}, \bm{b}, \hat{\bm{v}})$ basis (see Fig.~\ref{angles_pic} and Sec.~\ref{subsec:QM}). As the emitting lepton propagates, the orthonormal triad $(\bm{s}, \bm{b}, \hat{\bm{v}})$ co-rotates with the lepton momentum $\bm{p}$, and the spin vector $\bm{\eta}$ precesses according to the BMT equation in a manner locked to the same rotation. Consequently, when angles are expressed with respect to the instantaneous $(\bm{s}, \bm{b}, \hat{\bm{v}})$ basis, the integrand of Eq.~\eqref{eq:summed_ultrarel} remains approximately unchanged across the formation region. Therefore, with sufficient statistics and for spin- and polarization-summed distributions, local MC sampling within the LCFA reproduces the time-integrated energy and angular distributions, namely, Eq.~\eqref{eq:summed_ultrarel}. This is consistent with established benchmarks of spin- and polarization-averaged spectra~\cite{dipiazzaPRA18,dipiazzaPRA19} and angular distributions~\cite{blackburnPRA20} of strong-field QED calculations against their MC implementation.

By contrast, for spin and polarization, one must instead work with formation-integrated quantities, as in Eqs.~\eqref{final_spin_angles_int} and~\eqref{final_polarization_angles_int}. As we have already mentioned, a direct numerical evaluation of these integrals along an emitter's trajectory is impractical in general, because it would require accurately integrating over the full formation region, whose length depends also on the emitted photon momentum. To overcome this difficulty, we compute the corresponding formation-integrated quantities analytically in a locally-matched CCF (see below). 

In a CCF, the one-to-one correspondence between a given photon momentum $\bm{k}$ and the segment of the trajectory contributing to its emission~\cite{ritusJSLR85} ensures that the relevant formation-region contribution is automatically selected once $\bm{k}$ is fixed. Concretely, we start from the phase-integrated transition amplitude in Eq.~\eqref{matrix_element}, specialized to a CCF with vector potential and gauge fixing as in Sec.~\ref{subsec:sign} [Eqs.~\eqref{gauge_fixing_0} and~\eqref{gauge_fixing}], but keeping arbitrary spin four-vectors $\zeta^\mu$ and $\zeta^{\prime\mu}$. Using the Airy-function representation in Eq.~\eqref{Ai}, the phase integral can be carried out analytically. After simplification, the amplitude and its modulus square read
\begin{widetext}
\begin{align}
\mkern-18mu \mkern-18mu \mathcal{M}_{fi} &= - e \frac{1}{\sqrt{8 V^3 \varepsilon \varepsilon^\prime \omega}} \frac{2\pi}{ ( 3 \kappa )^\frac{1}{3}} \exp\{ i \tilde{\alpha} \} \bar{u}_{p^\prime,\zeta^\prime} \Bigl\{ \Bigl[ \slashed{e}^* - \frac{e}{2}\Bigl( \frac{\slashed{a} \slashed{d} \slashed{e}^*}{p^\prime_-} + \frac{\slashed{e}^* \slashed{d} \slashed{a}}{p_-} \Bigr) \phi_0 \Bigr] \Ai(z) - \frac{i e}{2}\Bigl( \frac{\slashed{a} \slashed{d} \slashed{e}^*}{p^\prime_-} + \frac{\slashed{e}^* \slashed{d} \slashed{a}}{p_-} \Bigr) \frac{\Ai^{\prime}(z)}{(3  \kappa)^{\frac{1}{3}}}  \Bigr\} u_{p,\zeta},
\label{matrix_element_again_2}
\end{align}
and
\begin{align}
\mkern-18mu \mkern-18mu |\mathcal{M}_{fi}|^2 &= C_{\mathcal{M}} \, \frac{1}{4} \mathrm{Tr} \Big\langle \Bigl\{ \slashed{e}^* \Ai({z}) - \frac{e}{2}\Bigl( \frac{\slashed{a} \slashed{d} \slashed{e}^*}{p^\prime_-} + \frac{\slashed{e}^* \slashed{d} \slashed{a}}{p_-} \Bigr) \Bigl[ \phi_0 \Ai({z}) + i \frac{\Ai^\prime({z})}{(3  \kappa)^{\frac{1}{3}}} \Bigr] \Bigr\} (\slashed{p} + m)(1+ \gamma^5 \slashed{\zeta}) \times \notag \\
& \qquad \qquad \qquad \times \Bigl\{ \slashed{e} \Ai({z}) - \frac{e}{2}\Bigl( \frac{\slashed{e} \slashed{d} \slashed{a}}{p^\prime_-} + \frac{\slashed{a} \slashed{d} \slashed{e}}{p_-} \Bigr) \Bigl[ \phi_0 \Ai({z}) - i \frac{\Ai^\prime({z})}{(3  \kappa)^{\frac{1}{3}}} \Bigr] \Bigr\} (\slashed{p}^\prime + m)(1+ \gamma^5 \slashed{\zeta}^\prime) \Big\rangle,
\label{matrix_element_squared_simpl_2}
\end{align}
\end{widetext}
with 
\begin{align}
\phi_0 &= \frac{\beta}{3 \kappa}, \\
a^\nu &= -|\bm{E}| a_1^\nu, \\
\mu &= p^\prime_+ + k_+ - p_+, \label{mu_def} \\  
\beta &= \frac{e}{2} \Bigl( \frac{p^\prime_\mu}{p^\prime_-} - \frac{p_\mu}{p_-} \Bigr) a^\mu, \label{beta_def} \\  
\kappa &= - \frac{e^2}{6} a^2 \Bigl( \frac{1}{p^\prime_-} - \frac{1}{p_-} \Bigr), \label{kappa_def} \\
\tilde{\alpha} &= \frac{2 \beta^3}{27 \kappa^2} - \frac{\mu \beta}{3 \kappa}, \\
{z} &= \left(\mu - \frac{\beta^2}{3 \kappa} \right) \left( \frac{1}{3\kappa} \right)^\frac{1}{3}, \\
C_{\mathcal{M}} & = \frac{e^2 \pi^2}{2 V^3 \varepsilon \varepsilon^\prime \omega \,(3\kappa)^{\frac{2}{3}}} .
\label{defintions_integral}  
\end{align}
The modulus square of the amplitude summed over the photon polarization quantum numbers  can be written as
\begin{equation}
\sum_{\xi} |\mathcal{M}_{fi}|^2 = C_{\mathcal{M}} ( \underline I + (\zeta^\prime \, \underline{l})),
\label{matrix_element_squared_simpl_3}
\end{equation}
where explicit expressions for $\underline{I}$ and $\underline{l}^\mu$ are provided in Appendix~\ref{app-calc}. 

For practical use in kinetic simulations, it is convenient to express the spin four-vector $\zeta^{\prime\mu}$ characterizing the detector in terms of its rest-frame spin vector $\bm{\eta}^\prime$. Decomposing with respect to the final-momentum direction $\bm{p}^\prime$, one has
\begin{align}
\zeta^{\prime 0} = \frac{|\bm{p}^{\,\prime}|}{m}\,\eta^{\prime}_{\parallel},
\qquad
\bm{\zeta}^{\,\prime}_{\perp}=\bm{\eta}^{\,\prime}_{\perp},
\qquad
\zeta^{\prime}_{\parallel}=\frac{\varepsilon^\prime}{m}\,\eta^{\prime}_{\parallel},
\label{4dto3drest}
\end{align}
where the suffixes $\parallel$ and $\perp$ denote the components of the vectors $\bm{\zeta}'$ and $\bm{\eta}'$ parallel and perpendicular to $\bm{p}'$, respectively \cite{Berestetskii-book}. With this, Eq.~\eqref{matrix_element_squared_simpl_3} becomes
\begin{align}
\sum_{\xi} |\mathcal{M}_{fi}|^2
&=
C_{\mathcal{M}}\left( \underline{I} + \bm{\eta}^{\,\prime}\cdot\bm{\delta} \right),
\label{matrix_element_squared_simpl_4}
\\
\bm{\delta}
&\equiv
\left(\frac{|\bm{p}^{\,\prime}|}{m} \underline{l}^{0}-\frac{\varepsilon^\prime}{m} \underline{l}_{\parallel}\right)\hat{\bm{v}}^\prime
-\underline{\bm{l}}_{\perp},
\label{delta def}
\end{align}
and, by direct analogy with the derivation of Eq.~\eqref{final_spin_angles_int}, the mean outgoing spin vector is
\begin{equation}
\langle \bm{\eta}^{\,\prime} \rangle = \frac{\bm{\delta}}{\underline{I}}.
\label{legitimate_SPA}
\end{equation}
An analogous procedure yields the photon polarization, with only minor modifications. For each emission, we define a polarization basis
\begin{equation}
e_{(i)}^{\mu} = f_{(i)}^{\mu} -
\frac{(f_{(i)} d)}{(k d)}\,k^{\mu},
\qquad i=1,2,
\label{double_gauge_fixed_pol}
\end{equation}
where the auxiliary vectors $f_{(i)}^\mu$ have vanishing time component and are constructed by orthogonalizing a reference direction $\bm{e}_{(x)}=(1,0,0)$ with respect to $\bm{k}$:
\begin{align}
\bm{e}_{(x)\perp} &\equiv \bm{e}_{(x)}-\frac{\bm{e}_{(x)}\cdot\bm{k}}{\bm{k}\cdot\bm{k}}\,\bm{k} \label{pre_fix}\\
f^\mu_{(1)} &=
\left(0,\, \frac{\bm{e}_{(x)\perp}}{\left|\bm{e}_{(x)\perp} \right|}\right),
\label{f_1}
\\
f^\mu_{(2)} &= \left(0,\,\frac{\bm{k}\times \bm{f}_{(1)}}{|\bm{k}\times \bm{f}_{(1)}|}\right).
\label{f_2}
\end{align}
This construction enforces mutual orthogonality and transversality to both $k^\mu$ and $d^\mu$ [see Eqs.~\eqref{gauge_fixing_0} and~\eqref{gauge_fixing}]. Writing the polarization vector as
\begin{equation}
e^\mu=g_1 e_{(1)}^\mu + g_2 e_{(2)}^\mu,
\label{pola_def}
\end{equation}
and defining the Stokes parameters via
\begin{align}
\xi_1 & = g^*_2 g_1 + g_2 g^*_1,
\notag\\
\xi_2 & = i(g^*_2 g_1 - g_2 g^*_1),
\notag\\
\xi_3 & = g^*_1 g_1 - g^*_2 g_2,
\label{stokes_param_def}
\end{align}
with the normalization condition $g^*_1 g_1 + g^*_2 g_2=1$, the photon density matrix $e^*_\mu e_\nu$ can be written in the standard form \footnote{For an outgoing photon, $e^*_\mu e_\nu$ corresponds to the Hermitian of the polarization density matrix defined in Eq.~\eqref{phtn_dens_mtrx}.}
\begin{align}
e^*_\mu e_\nu & = \frac{1}{2}( e^{(1)}_\mu e^{(1)}_\nu + e^{(2)}_\mu e^{(2)}_\nu ) + \frac{\xi_1}{2}( e^{(2)}_\mu e^{(1)}_\nu + e^{(1)}_\mu e^{(2)}_\nu) \notag \\
+ & i \frac{\xi_2}{2}(- e^{(2)}_\mu e^{(1)}_\nu + e^{(1)}_\mu e^{(2)}_\nu ) + \frac{\xi_3}{2} ( e^{(1)}_\mu e^{(1)}_\nu - e^{(2)}_\mu e^{(2)}_\nu ).
\end{align}
Summing Eq.~\eqref{matrix_element_squared_simpl_2} over the final electron spins yields
\begin{equation}
\sum_{\zeta^\prime}|\mathcal{M}_{fi}|^2 =
C_{\mathcal{M}}\left(\underline{I} + \bm{\xi}\cdot\bm{y}\right),
\label{matrix_element_squared_simpl_5}
\end{equation}
where the components of $\bm{y}=(y^1, y^2, y^3)$ are given in Appendix~\ref{app-calc}. By analogy with the derivation of Eq.~\eqref{final_polarization_angles_int}, the mean photon Stokes vector is
\begin{equation}
\langle \bm{\xi} \rangle =\frac{\bm{y}}{\underline{I}}.
\label{legitimate_PPA}
\end{equation}
For any given $\bm{k}$, Eqs.~\eqref{legitimate_SPA} and~\eqref{legitimate_PPA} provide formation-region-integrated outgoing lepton spin and photon polarization with direct physical interpretation.

Our analytical building blocks are derived in a CCF, consistently with the LCFA employed in simulations, which is formulated locally in terms of particle and field invariants associated to the instantaneous rest frame of the emitting particle. In particular, emission probabilities depend on the local quantum parameter $\chi_e=\sqrt{|(F^{\mu\nu}p_\nu)^2|}/(mF_{\mathrm{cr}})$ and on the instantaneous transverse acceleration direction (see Sec.~\ref{sec:new_method} for details). 


\subsection{\label{stokes_param}Selection of the polarization basis}

The photon polarization basis in Eq.~\eqref{double_gauge_fixed_pol} depends on the propagation direction $\bm{d}$ of the background and on the emitted-photon momentum $k^\mu=(\omega,\bm{k})$. As a consequence, the Stokes vector obtained from Eq.~\eqref{legitimate_PPA} is defined \emph{event by event} with respect to a basis that generally changes with $\bm{k}$. For this reason, in general, when polarization observables are combined across events each Stokes vector must first be rotated into a common reference basis (e.g., one fixed in the laboratory frame or tied to a detector plane). 

Finally, the construction in Eqs.~\eqref{double_gauge_fixed_pol}--\eqref{f_2} becomes ill-defined when the reference direction $\bm{e}_{(x)}$ is collinear with $\bm{k}$, because the orthogonalization step yields a vanishing vector. In this case, we switch to an alternative reference axis (e.g., $\bm{e}_{(y)}$) and repeat the construction, thereby ensuring a well-defined transverse polarization basis for all emission directions.


\subsection{\label{sec:new_method}Implementing spin- and polarization-resolved distributions}

We now introduce our emission algorithm for NCS that enables the consistent sampling of \emph{both} kinematics (energy and angles) and internal degrees of freedom (electron spin and photon polarization) in PIC and MC codes within the LCFA. 

The algorithm preserves the established LCFA workflow for the \emph{momentum}, for which local sampling has been shown to be accurate under the assumptions discussed in Sec.~\ref{sec:newMC}, while replacing the ill-defined local treatment of spin and polarization by \emph{phase-integrated} distributions evaluated in an auxiliary, event-by-event matched constant crossed field. The resulting algorithm is readily integrable in existing SFQED-PIC/MC frameworks. The required routines for angle-resolved emission and formation-integrated spin/polarization calculation have been implemented in \texttt{SFQEDtoolkit}~\cite{montefioriCPC2023}.

\textit{Algorithm 1}: For an incoming lepton with momentum $\bm{p}$ and spin $\bm{\eta}$, each computational cycle comprises the following sequence of steps:
\begin{enumerate}[start=0]
\item[(0)] \textit{Particle push:} advance the particle position
$\bm{x}$, momentum $\bm{p}$, and spin $\bm{\eta}$ by integrating the Lorentz force and the BMT equation, respectively.
\item[(1)] \textit{Event triggering:} sample whether an emission occurs in the time step $\Delta t$ using the local (total) emission rate $dP^{(\eta)}_{\mathrm{NCS}}/dt$ in Eq.~\eqref{eq:rate}.
\end{enumerate}
If an event occurs
\begin{enumerate}[start=2]
\item[(2)] \textit{Energy sampling:} sample the photon energy $\omega$ from the local spectrum $dP^{(\eta)}_{\mathrm{NCS}}/(dt\,d\omega)$ in Eq.~\eqref{eq:summed_ultrarel_angleint_k2}.
\item[(3)] \textit{Angle sampling:} using the photon energy $\omega$ determined at step~(2), sample the emission angles $\theta,\,\varphi$ with respect to the instantaneous QM basis of Sec.~\ref{subsec:QM} from the local angular distribution $dP^{(\eta)}_{\mathrm{NCS}}/(dt\,d\omega\,d\theta\,d\varphi)$, i.e., the \emph{integrand} in Eq.~\eqref{eq:summed_ultrarel}. Once the photon momentum is fully sampled, update the electron momentum by enforcing momentum conservation (photon emission recoil).
\item[(4)] \textit{Spin/polarization vector:} given the photon momentum $\bm{k}$ obtained at steps (2)--(3), determine the outgoing lepton mean spin $\langle \bm{\eta}^{\,\prime} \rangle$ and photon mean Stokes vector $\langle \bm{\xi} \rangle$ from Eq.~\eqref{legitimate_SPA} and Eq.~\eqref{legitimate_PPA}, respectively (see also below).
\end{enumerate}
Otherwise, if no emission is deemed
\begin{enumerate}[start=5]
\item[(5)] \textit{No-emission spin evolution}: using the no-emission probability $1 - (dP^{(\eta)}_{\mathrm{NCS}}/dt) \Delta t$, update the lepton mean spin according to the non-radiative mixed-states transition prescribed in Refs.~\cite{CAIN, liPRD23, montefioriPhD25}. This ``no-emission'' spin evolution physically arises from interference between the one‑loop electron self‑energy and the forward‑scattering amplitude, acting even in the absence of emission; see Refs.~\cite{torgrimssonNJP21, torgrimssonPRL21, CAIN, liPRD23, montefioriPhD25, baierSPU1972} for details.
\end{enumerate}
Steps (1)--(3) of \textit{Algorithm 1} are performed by evaluating time-local expressions on the instantaneous particle state and fields. This procedure is well benchmarked for spin- and polarization-averaged observables. In step (4), the \emph{phase-integrated} expressions in Eqs.~\eqref{legitimate_SPA} and \eqref{legitimate_PPA} are evaluated in an auxiliary constant crossed field (ACCF) matched to the instantaneous local invariants and lepton dynamics. The ACCF is employed \emph{only} to compute the phase-integrated $\langle \bm{\eta}^{\,\prime} \rangle$ and $\langle \bm{\xi} \rangle$ for the sampled photon energy $\omega$ and momentum $\bm{k}$.

The ACCF is constructed as follows: at emission, let the emitting lepton have energy $\varepsilon$, momentum $\bm{p}$, and quantum parameter $\chi_e$ (computed from the local electromagnetic field and the instantaneous momentum). The ACCF is fixed by:
\begin{enumerate}
\item[(1)] {Field magnitude (matching $\chi_e$).} As discussed at next step, the ACCF is oriented opposite to the emitting lepton momentum. In this case, to leading order in $1/\gamma$, $\chi_e \approx 2\varepsilon |e|E/m^3$ for a counterpropagating ultrarelativistic
lepton. We therefore set the ACCF field strength to reproduce the same instantaneous $\chi_e$,
\begin{equation}
E_{\mathrm{ACCF}}
=
\frac{\chi_e}{2\varepsilon |e|}\,m^3 .
\label{eq:ACCF_strength}
\end{equation}
\item[(2)] {Field orientation (matching the local transverse acceleration).} We define an ACCF with normalized Poynting vector $\bm{d}\equiv(\bm{E}\times \bm{B})/|\bm{E}|^2$ such that $\bm{d}=-\bm{p}/|\bm{p}|$. We align the ACCF electric-field direction with the instantaneous unit vector $\bm{s}$ along the lepton's
transverse acceleration [$\bm{s} \cdot \bm{p} = \bm{s} \cdot \bm{d} = 0$] induced by the local background field,
\begin{align}
\bm{E}_{\mathrm{ACCF}}
= &
\mathrm{sgn}(e)\,E_{\mathrm{ACCF}}\,\bm{s},
\notag \\
\bm{B}_{\mathrm{ACCF}}=& \bm{E}_{\mathrm{ACCF}} \times \frac{\bm{p}}{|\bm{p}|},
\label{eq:ACCF_orientation}
\end{align}
$\mathrm{sgn}(e)=+1$ for positrons and $-1$ for electrons.
\end{enumerate}
Matching $\chi_e$ ensures that the ACCF reproduces the relevant strong-field QED photon-energy spectrum, while the field orientation preserves the local geometric structure. Indeed, for an ultrarelativistic particle moving in an \emph{arbitrary} slowly varying background field $(a_0 \gg 1)$, e.g., a constant and uniform magnetic field, the field in the instantaneous rest frame contains orthogonal electric and magnetic components of approximately equal magnitude, so that the local emission process is equivalent (in the LCFA/QM sense) to emission in a crossed-field configuration. This is precisely the rationale behind the auxiliary constant crossed field (ACCF) used here: at each emission we construct a locally matched CCF that reproduces the same $\chi_e$ (instantaneous rest frame field strength) and the same local geometry (instantaneous rest frame field geometry via the transverse-acceleration direction) as the background electromagnetic field, and we evaluate formation-integrated spin and polarization in that matched CCF.

Importantly, the phase controlling the formation region is independent of spin and polarization [see Eq.~\eqref{exponential}], so the accuracy of neglecting any local space and time dependence of the background field within the formation length in the probability of single-photon emission $P_{\text{NCS}}$ is that of the LCFA itself, i.e., of order $1/a_0^2$ \cite{baierNPB1989, dipiazzaPRA18, dipiazzaPRA19, ildertonPRA2019}. We note that MC simulations based on single-photon NCS, as those considered here, neglect coherent multi-photon contributions. By considering the exemplary case of  nonlinear double Compton scattering \cite{seiptPRD2012, mackenrothPRL2013, kingJoP2015, dinuPRD2019}, the ratio of the one-step (coherent) to the two-step (incoherent cascade) probability scales as $1/(a_0 \Delta \psi)$, where $\Delta \psi$ is the total laser phase accumulated over the interaction ($\Delta \psi \approx a_0$ for the simulations in Sec.~\ref{e_laser}). Analogous considerations can be made for the emission of more than two photons, such that the cascade approximation, in which the coherent contributions are ignored, is well justified in the considered regime.

In summary, the new algorithm: (i) retains the local LCFA treatment for event triggering, photon-energy sampling, and photon-emission-angle sampling; and (ii) replaces the problematic time-local spin/polarization expressions by a non-negative, formation-region-integrated expression evaluated in a matched ACCF. This preserves compatibility with existing PIC/MC workflows while restoring a rigorous probabilistic interpretation for spin and polarization at the event level. An analogous strategy can be extended to, e.g., nonlinear Breit-Wheeler.

\begin{figure}[tb]
\centering
\includegraphics[width=1\linewidth]{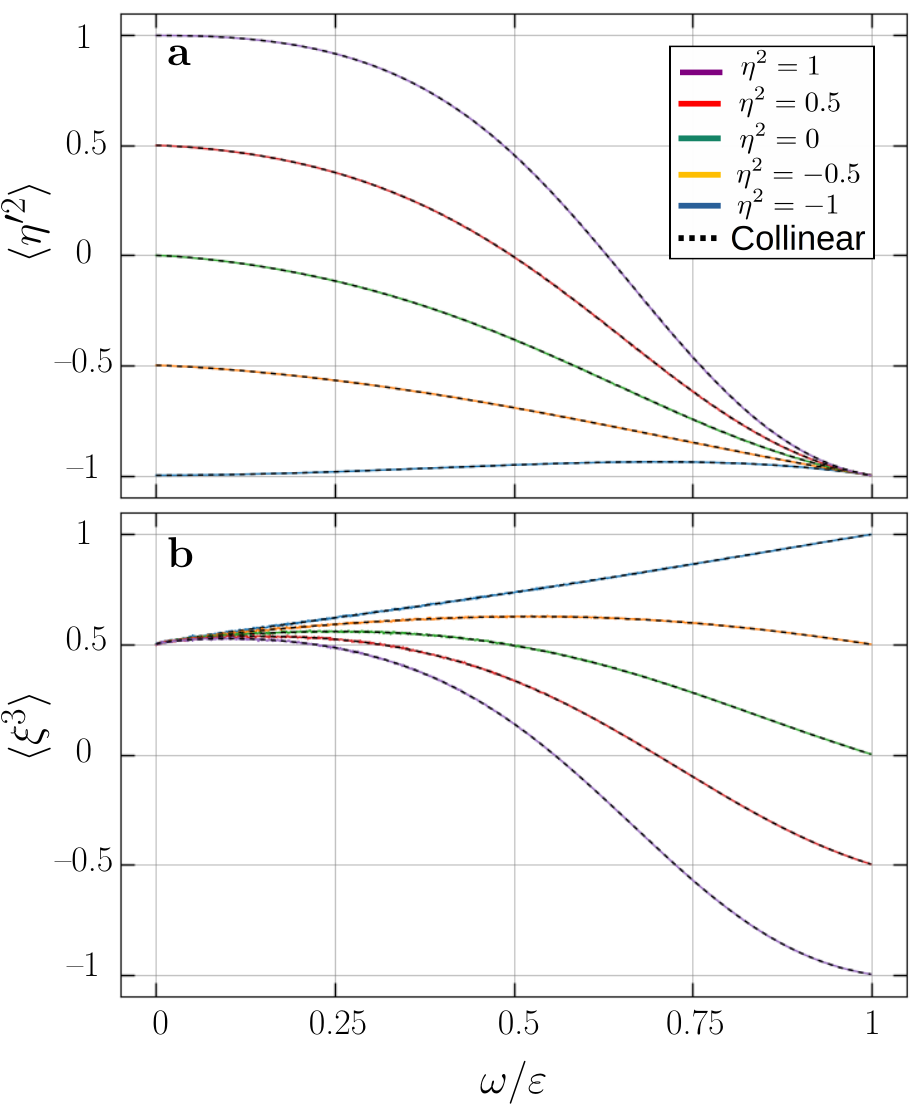}
\caption{Spin and polarization curves for NCS of an ultrarelativistic electron ($\varepsilon = 1$ GeV, $\chi_e = 2$) for five different initial spin states (purple: $1$; red: $0.5$; green: $0$; orange: $-0.5$; blue: $-1$) in a uniform magnetic field $\bm{B} = (0, B_y, 0)$ versus the photon-to-electron energy ratio $\omega/\varepsilon$. Panel \textbf{a}: mean outgoing electron spin component $\langle \eta^{\prime 2}\rangle$ (along $\bm{B}$) according to Eq.~\eqref{legitimate_SPA}. Panel \textbf{b}: mean photon Stokes parameter $\langle \xi^{3}\rangle$ according to Eq.~\eqref{legitimate_PPA}. Black dashed curves show the corresponding local- angle-integrated prediction according to Eq.~\eqref{final_spin_angles_def_u} (panel~\textbf{a}) and Eq.~\eqref{final_polarization_angles_def_u} (panel~\textbf{b}).}
\label{fig:curves_int}
\end{figure}

As a consistency check, we repeat the calculation of the angle-averaged spin and polarization curves as functions of the photon-to-electron energy ratio $\omega/\varepsilon$, using the same parameters as in Fig.~\ref{fig:curves} but using the phase-integrated expressions in Eqs.~\eqref{legitimate_SPA} and~\eqref{legitimate_PPA}. Specifically, for each value of $\omega/\varepsilon$ we generate $10^4$ emission events with angles sampled from the \emph{integrand} in Eq.~\eqref{eq:summed_ultrarel}, and for each event we compute the final electron spin and photon Stokes vector using the phase-integrated formulas in Eqs.~\eqref{legitimate_SPA} and~\eqref{legitimate_PPA}. All sampled states are physically admissible ($|\bm{\eta}^{\,\prime}|\le 1$ and $|\bm{\xi}|\le 1$). We then compare these results with the local (instantaneous) angle-integrated predictions from Eqs.~\eqref{final_spin_angles_def_u} and~\eqref{final_polarization_angles_def_u}, and find that the corresponding curves agree within numerical accuracy, as shown in Fig.~\ref{fig:curves_int}. This agreement is consistent with the fact that, within the LCFA, each point inside the formation region contributes essentially the same probability per unit time and exhibits the same spectral and angular dependence, as discussed in Sec.~\ref{sec:newMC}.
%


\section{\label{sec:sims}Simulations}

The angle-resolved, phase-integrated model developed here is conceptually distinct from the collinear, local model commonly employed to date and it uses only non-negative probability distributions by construction.
In fact, although the angle-integrated spin- and polarization-resolved distributions are non-negative, they receive essential contributions from regions where the corresponding fully resolved local differential is negative (see Figs.~\ref{fig:scatter} and~\ref{fig:curves}). 
To quantify the impact on ensemble-averaged spin and polarization observables, we consider two representative simulation configurations, presented in the following subsections.  In the first case, an ultrarelativistic electron bunch propagates through an intense electromagnetic laser pulse (see Sec.~\ref{e_laser}) and in the second case through a constant magnetic field (see Sec.~\ref{subsec:B}).  In both cases, we perform MC simulations using both (i) the standard collinear, local (CL) model based on time-local angle-integrated distributions [Eqs.~\eqref{final_spin_angles_def_u} and~\eqref{final_polarization_angles_def_u}], and (ii) the angle-resolved, nonlocal (AN) model based on phase-integrated distributions [Eqs.~\eqref{legitimate_SPA} and~\eqref{legitimate_PPA}]. The field profile and bunch parameters are specified in each subsection.

As discussed in the introduction, the polarization of the emitted photon and the spin of the outgoing electron are tied to each particle's momentum.  In individual events, for a photon, $\langle \xi^{2}\rangle$ and $|\langle\bm{\xi}_\perp\rangle| \equiv \sqrt{\langle\xi^{1}\rangle^{2}+\langle\xi^{3}\rangle^{2}}$ quantify the degree of circular and linear polarization, respectively, in the plane perpendicular to the photon momentum.  Additionally, these combinations of Stokes parameters are Lorentz invariant~\cite{Berestetskii-book}.  For an electron, the analogous quantities are the helicity $\langle h \rangle$ and the magnitude of the spin component transverse to the electron momentum $|\langle\bm{\eta}^{\prime}\rangle_{\perp}|$.  The electron helicity operator $\hat{h}$ is defined as the scalar product of the Dirac spin operator $\hat{\bm{\Sigma}}$ and of the momentum operator $\hat{h} \equiv \hat{\bm{\Sigma}}\cdot\hat{\bm{P}}/|\hat{\bm{P}}|$.  For an electron in a state of definite momentum and definite spin (as is the outgoing electron in the laboratory frame considered here), the expectation value of the helicity is $\langle h \rangle = (m/\varepsilon') \langle \bm{\zeta}^{\,\prime} \rangle \cdot \bm{p}^{\,\prime} / |\bm{p}^{\,\prime}| = \langle \bm{\eta}^{\,\prime} \rangle \cdot \bm{p}^{\,\prime} / |\bm{p}^{\,\prime}|$, where $\bm{\zeta}^{\,\prime}$ is the spatial part of the outgoing electron's spin four-vector.

In practice, since we are interested in beam-level observables, we group particles into angular bins such that all particles in a given bin have nearly the same propagation direction.  For photons, we then compute the binned averages
\begin{align}
\bar{\xi}^{2} &\equiv \frac{1}{N}\sum_{\ell=1}^{N}
    \langle \xi^{2} \rangle_\ell, \\
|\bar{\bm{\xi}}_\perp| &\equiv
    \sqrt{\left(\frac{1}{N}\sum_{\ell=1}^{N}
               \langle \xi^{1} \rangle_\ell \right)^{2}
        + \left(\frac{1}{N}\sum_{\ell=1}^{N}
               \langle \xi^{3} \rangle_\ell \right)^{2}} \label{xi_bar_perp_def},
\end{align}
where the index $\ell$ runs over the $N$ photons in the selected bin. 
This procedure is meaningful because, within a sufficiently narrow angular bin, photons share essentially the same emission direction $\bm{k}/|\bm{k}|$ and, in the ultrarelativistic regime considered here, are produced by electrons with nearly the same momentum direction $\bm{p}/|\bm{p}|$. This results in a common polarization basis [Eqs.~\eqref{double_gauge_fixed_pol}–\eqref{stokes_param_def}], such that the sums in Eq.~\eqref{xi_bar_perp_def} are well defined. For electrons, we analogously report the binned average helicity,
\begin{align}
\bar{h} &\equiv \frac{1}{N}\sum_{\ell=1}^{N} \langle h \rangle_\ell,
\label{bar_h}
\end{align}
and the magnitude of the averaged transverse spin-polarization vector,
\begin{align}
|\bar{\bm{\eta}}^\prime_\perp| &\equiv
    \left|\frac{1}{N}\sum_{\ell=1}^{N}
        \langle\bm{\eta}^{\prime}\rangle_{\perp \, \ell} \right|\,.
\end{align}
Again, for a sufficiently collimated beam, the ensemble average in Eq.~\eqref{bar_h} directly represents the beam polarization along the common propagation direction.  Indeed, angular collimation provides a shared reference direction, so the per-particle helicities $\langle h \rangle_\ell = \langle u^{\dagger}_{p',\zeta'}\,\bm{\Sigma}\,u_{p',\zeta'}\rangle_\ell \cdot \bm{p}^{\,\prime}_\ell / (2\varepsilon'_\ell|\bm{p}'_\ell|)$ can be consistently averaged.  For ultrarelativistic and almost collinear electrons the situation is similar to photons in the sense that, whereas for a photon the helicity is a strictly Lorentz-invariant quantity, for an ultrarelativistic electron the helicity is Lorentz invariant except for the restricted class of Lorentz boosts that do invert the direction of electron momentum.


\subsection{\label{e_laser}Electron beam--laser interaction}
\begin{figure*}[t]
\centering
\includegraphics[width=0.95\linewidth]{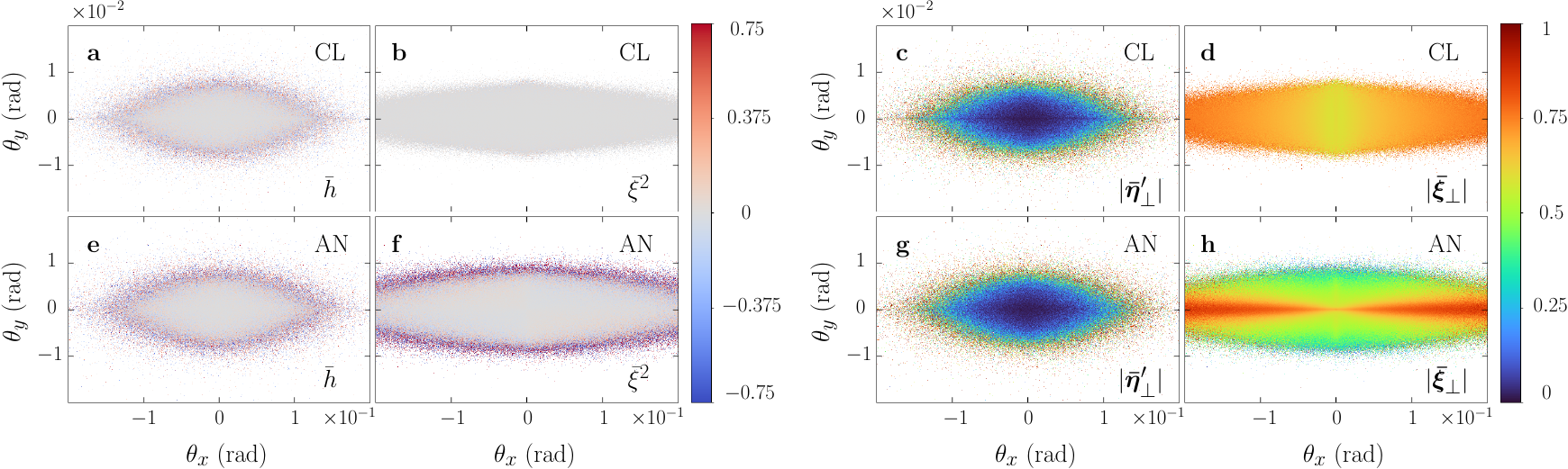}
\caption{Angle-resolved distribution of electron helicity $\bar{h}$ (panels~\textbf{a},~\textbf{e}), degree of photon circular polarization $\bar{\xi}^{2}$(panels~\textbf{b},~\textbf{f}), transverse spin magnitude $|\bar{\bm{\eta}}^{\prime}_{\perp}|$ (panels~\textbf{c},~\textbf{g}), and degree of photon linear polarization $|\bar{\bm{\xi}}_\perp|$ (panels~\textbf{d},~\textbf{h}) in the head-on collision of an ultrarelativistic electron bunch ($2.0~\mathrm{pC}$, $2~\mathrm{GeV}$ mean energy, 4\% energy spread, $\mathrm{FWHM}_{\parallel}=3~\mu\mathrm{m}$, $\sigma_{\perp}=4~\mu\mathrm{m}$, $\sigma_{\theta}=2~\mathrm{mrad}$) with a tightly focused, linearly polarized laser pulse propagating along $+z$ ($\lambda_0=1~\mu\mathrm{m}$, $a_0=80$, $\tau_{\mathrm{FWHM}}\approx 27~\mathrm{fs}$, $w_0=2~\mu\mathrm{m}$, $I_0\approx 8.8\times10^{21}\,\mathrm{W/cm^2}$). Results are shown for both the collinear local (CL, panels~\textbf{a}--\textbf{d}) and angle-resolved nonlocal (AN, panels~\textbf{e}--\textbf{h}) models; only photons with energy greater than $10~\mathrm{MeV}$ are plotted.}
\label{fig:h_sperp_stokes_lin}
\end{figure*}
\begin{figure*}[t]
\centering
\includegraphics[width=0.95\linewidth]{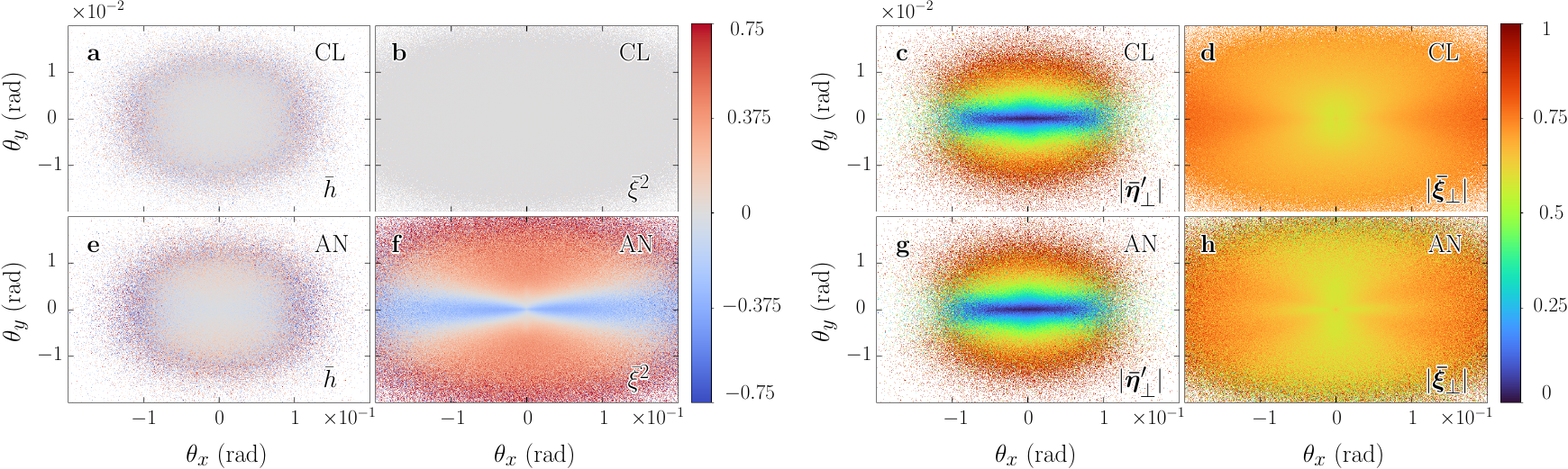}
\caption{Same as Fig.~\ref{fig:h_sperp_stokes_lin}, but with the laser pulse having a modest ellipticity ($\epsilon=|E_y|/|E_x|=0.05$). Only photons with energy larger than $10~\mathrm{MeV}$ are plotted.}
\label{fig:h_sperp_stokes_ell}
\end{figure*}

As a first test case, we consider the head-on collision of an ultrarelativistic electron bunch and a tightly focused laser pulse propagating along $+z$. 
{The electron bunch carries a total charge $Q=2.0$ pC, with mean energy $\varepsilon=2$ GeV and relative energy spread $\Delta\varepsilon/\varepsilon=0.04$.
Its spatial charge density is taken to be Gaussian
\begin{equation} q(x,y,z)=\frac{Q_0}{(2\pi)^{3/2}\sigma_\perp^2\sigma_\parallel} \exp\left(-\frac{x^2+y^2}{2\sigma_\perp^2}-\frac{z^2}{2\sigma_\parallel^2}\right),
\label{eq:beam_density}
\end{equation}
with transverse rms size $\sigma_\perp = 4$ $\mu$m and longitudinal full width at half maximum $FWHM_\parallel = 2 \sqrt{2 \ln{(2)}} \, \sigma_\parallel = 3$ $\mu$m.
The normalization constant $Q_0$ is chosen such that $\int q(x,y,z) dxdydz = Q$.
The beam angular profile is also taken to be Gaussian, with rms divergence $\sigma_\theta = 2$ mrad.
The distribution of the $\theta_x$, $\theta_y$ angles about the mean propagation direction is
\begin{equation}
f(\theta_x,\theta_y)=\frac{1}{2\pi\sigma_\theta^2} \exp\left(-\frac{\theta_x^2+\theta_y^2}{2\sigma_\theta^2}\right). \label{eq:beam_angular}
\end{equation}}
Note that electron beams with GeV-scale energies, pC-level charge, mrad-scale divergence, femtosecond durations, and quasi-monoenergetic spectra at the few-percent level are routinely produced in laser-wakefield accelerators in optimized operating regimes~\cite{lundhNP11, wangNC13, leemansPRL14, gonsalvesPRL19}.

{The laser pulse used in the simulations is modeled as a focused Gaussian beam following Ref. \cite{salaminPRSTAB02}, using the full field expression given there.
Its parameters are chosen as follows: central wavelength $\lambda_0 = 2 \pi c /\omega_0 = 1$ $\mu$m, normalized amplitude $a_0 = 80$, focal waist is $w_0 = 2$ $\mu$m, and FWHM pulse duration $\tau_{\text{FWHM}} \approx 27$ fs. 
These values correspond to a cycle-averaged peak intensity and power $I_0 \approx 8.8 \times 10^{21}$ W/cm$^2$ and $P_0 \approx 0.55$ PW, respectively, with total pulse energy $E \approx 16$~J.

For reference, to lowest order in the diffraction angle, the corresponding electric field can be written as
\begin{align}
\bm{E}(x,y,z,t) &= E_0 \, \exp\left( - \frac{\psi^2}{\omega_0^2\tau^2} - \frac{x^2+y^2}{w_z^2} \right) \notag \\
&\qquad \times \left[\sin(\psi)\,{\bm{x}} + \epsilon\sin(\psi+\pi/2)\,{\bm{y}} \right],
\label{electric field equation}
\end{align}
where $E_0 = m\omega_0 a_0/|e|$ is the peak field amplitude, $\psi$ the plane-wave phase, $\tau = \tau_{\text{FWHM}}/\sqrt{2 \ln(2)}$ the characteristic pulse duration, $w_z$ the beam waist at axial position $z$, and $\epsilon \geq 0$ is the laser ellipticity parameter ($\epsilon = 0$ corresponds to linear polarization along $\bm{x}$).
The unit vectors ${\bm{x}}$ and ${\bm{y}}$ span the plane transverse to the propagation direction.}

Petawatt-class laser systems routinely deliver pulses with energies in the 10--30~$\mathrm{J}$ range and durations of a few tens of femtoseconds, corresponding to powers in the 0.1--1~$\mathrm{PW}$ range~\cite{dansonHPLSE19}. Tight focusing of such pulses to few-micron waists, reaching on-target intensities in the $10^{21}\,\mathrm{W/cm^2}$ class required for strong-field QED experiments, has been demonstrated in recent high-intensity laser--electron collision measurements~\cite{colePRX18, poderPRX18, mirzaieNP24, losNC26}. Moreover, the experimental realization of laser intensities exceeding $10^{23}\,\mathrm{W/cm^2}$ has been reported~\cite{yoonO2021}, and several laser facilities are under construction or design that are expected to surpass this level by a substantial margin~\cite{NSF_OPAL, Vulcan_20-20, SEL}.

We perform two simulations: one with a laser pulse linearly polarized along the $x$~axis, and one with the addition of a small degree of ellipticity $\epsilon=|E_y|/|E_x|=0.05$. Such a modest ellipticity at the considered power and intensity can be implemented straightforwardly at petawatt-class facilities, e.g., by inserting large-aperture reflective quarter-wave retarders after the compressor and before the focusing optics~\cite{kepplerOE12}.
Although the simulations include photon emissions over the full energy range (with no lower energy cutoff), in both the linear and elliptical laser-polarization cases we plot only photons with energies above $10~\mathrm{MeV}$, since this range is more accessible experimentally for photon detection. Plotting the full photon sample does not qualitatively
alter the displayed polarization pattern.

Figure~\ref{fig:h_sperp_stokes_lin} shows angle-resolved distributions of electron helicity $\bar{h}$ (panels~\textbf{a},~\textbf{e}), degree of photon circular polarization $\bar{\xi}^{2}$ (panels~\textbf{b},~\textbf{f}), transverse spin magnitude $|\bar{\bm{\eta}}^{\prime}_{\perp}|$ (panels~\textbf{c},~\textbf{g}), and linear polarization $|\bar{\bm{\xi}}_\perp|$ (panels~\textbf{d},~\textbf{h}) in the head-on collision of an ultrarelativistic electron bunch with a linearly polarized laser pulse. Distributions are shown as functions of the deflection angles ($\theta_x, \theta_y$), defined for electrons (photons) as $\theta_x = \atan(p^\prime_x,- p^\prime_z)$ ($\theta_x = \atan(k_x,- k_z)$)
and $\theta_y = \atan(p^\prime_y,- p^\prime_z)$ ($\theta_y = \atan(k_y,-k_z)$), where the 2-argument arctangent function is used, and the minus sign for the $z$-component of the momentum accounts for the incoming electron bunch propagation along the negative $z$ direction. In these simulations, the bin size is $\Delta\theta_x = 4 \times 10^{-4}$ rad and $\Delta\theta_y = 4 \times 10^{-5}$ rad. Both the CL model (panels~\textbf{a}-\textbf{d}) and the AN model (panels~\textbf{e}-\textbf{h}) results are displayed. As expected, the symmetric, oscillatory structure of the field of a many-cycle linearly-polarized laser pulse yields no net electron spin polarization in either model. Similarly, both approaches predict negligible circular photon polarization ($\bar{\xi}^{2}\approx 0$) and a high degree of linear polarization. Yet, notably, the AN model yields a stronger (weaker) linear polarization along (orthogonal to) the laser polarization axis (see panel~\textbf{d} for CL and panel~\textbf{h} for AN). Note that the initial angular spread of the electron bunch, together with the fact that the laser focal waist is half the bunch transverse rms size ($w_0=2~\mu\mathrm{m}$, $\sigma_{\perp}=4~\mu\mathrm{m}$), implies that a large fraction of electrons of the bunch experience field amplitudes well below the peak value. This reduces the sensitivity of the electron spin and angular distributions to the emission model and contributes to the limited differences observed between CL and AN in this setup.

Figure~\ref{fig:h_sperp_stokes_ell} shows the same observables as Fig.~\ref{fig:h_sperp_stokes_lin} with the same laser and bunch parameters, but for a colliding laser pulse with a modest ellipticity $\epsilon=|E_y|/|E_x|=0.05$. For the outgoing electrons, the CL and AN models yield similar results for both the helicity $\bar{h}$  (panels~\textbf{a} and \textbf{e}) and the transverse spin magnitude $|\bar{\bm{\eta}}^{\prime}_{\perp}|$ (panels~\textbf{c} and \textbf{g}). A small helicity bias is nevertheless present in the AN result (panel~\textbf{e}), indicating a distinctive preference for the final spin to align or anti-align with the propagation direction in the AN model. We discuss this effect further in the pulsar-like magnetic-field setup (see Sec.~\ref{subsec:B}). Photon polarization, by contrast, is strongly dependent on the emission model. Both models predict a comparably high degree of linear polarization across the explored angular range (panels~\textbf{d} and \textbf{h}). However, while the CL model yields a negligible circular component (panel~\textbf{b}), the AN model predicts a qualitatively different pattern with a significant circular polarization that changes with the emission direction (panel~\textbf{f}). In particular, within the AN description, photons emitted along the major laser polarization axis $x$ are predominantly left-handed ($\bar{\xi}^{2}<0$), whereas those emitted along the minor laser polarization axis $y$ are predominantly right-handed ($\bar{\xi}^{2}>0$).

The outer regions of the distributions shown in Figs.~\ref{fig:h_sperp_stokes_lin} and \ref{fig:h_sperp_stokes_ell} correspond to lower particle statistics and may be affected by larger fluctuations; we ensured the spin and polarization patterns remain clearly identifiable and are governed by the underlying angularly/spin/polarization-resolved dynamics.
The particle-density maps provide complementary information, without altering the physical conclusions, and are therefore not included.


\subsection{\label{subsec:B} Emission in a pulsar-like magnetic field}

In the polar-cap region of ordinary pulsars, freshly produced secondary pairs can reach energies of several hundred MeV~\cite{daughertyAJ82}. For a short time after creation, these particles may also possess appreciable pitch angles $\theta_{\mathrm{pitch}}\gtrsim 0.1$, and, when immersed in a magnetic field of typical strength $B\sim F_{\mathrm{cr}}/100$, radiate copiously.  To probe electron and photon polarization under comparable conditions, we consider a constant and uniform magnetic field with $B=F_{\mathrm{cr}}/100$, and electrons with momentum components perpendicular $|p_\perp|=100\,m$ and parallel $|p_\parallel|=400\,m$ to the magnetic field, corresponding to an energy of $210~\mathrm{MeV}$, a pitch angle $\theta_{\mathrm{pitch}}\approx 0.25~\mathrm{rad}$, and an initial quantum parameter $\chi_e=1$. All remaining bunch parameters are the same as in the electron--laser setup. The simulation is evolved until the bunch center completes approximately one full gyration in the $x$--$y$ plane (in absence of recoil due to emission). 

\begin{figure*}[t!]
\centering
\includegraphics[width=0.95\linewidth]{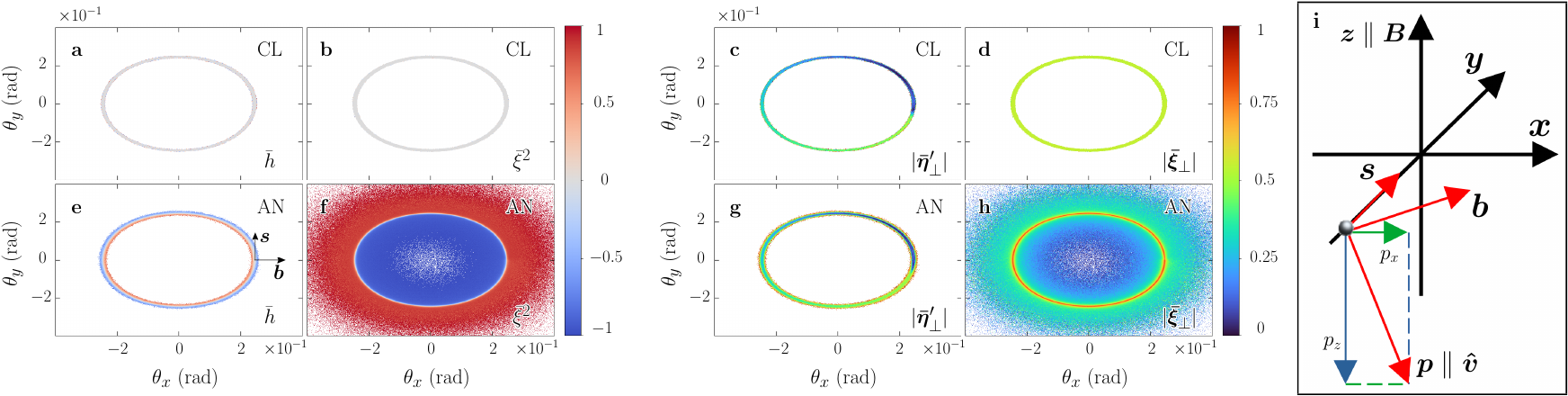}
\caption{Angle-resolved distribution of electron helicity $\bar{h}$ (panels~\textbf{a},~\textbf{e}), degree of photon circular polarization $\bar{\xi}^{2}$ (panels~\textbf{b},~\textbf{f}), transverse spin magnitude $|\bar{\bm{\eta}}^{\prime}_{\perp}|$ (panels~\textbf{c},~\textbf{g}), and degree of photon linear polarization $|\bar{\bm{\xi}}_\perp|$ (panels~\textbf{d},~\textbf{h}) for the constant uniform magnetic-field setup ($\bm{B}=B\,\bm{z}$, $B=F_{\mathrm{cr}}/100$), representative of secondary leptons in pulsar polar-cap conditions. The electron bunch is initialized with average $|p_{\perp}|=100\,m$ and $|p_{\parallel}|=400\,m$ ($210~\mathrm{MeV}$, $\theta_{\mathrm{pitch}}\approx 0.25~\mathrm{rad}$, $\chi_e = 1$) at $\atan(p_x,-p_z) =\theta_{\mathrm{pitch}}$ and $\atan(p_y,- p_z) = 0$ and evolved for one full gyration in the $x$--$y$ plane. Panel~\textbf{i} shows the three-dimensional orientation of the triad $(\bm{s}, \bm{b}, \hat{\bm{v}})$ for electrons at the initial position; panel~\textbf{e} displays the projections of $\bm{s}$ and $\bm{b}$ onto the $\theta_x$--$\theta_y$ plane. At every other point on the momentum ring, $\bm{s}$ and $\bm{b}$ preserve the same orientation relative to the local electron momentum: $\bm{s}$ is tangential to the ring in the counterclockwise direction and $\bm{b}$ points radially outward.}
\label{fig:h_bconst}
\end{figure*}

Figure~\ref{fig:h_bconst} shows angle-resolved distributions of electron helicity $\bar{h}$ (panels~\textbf{a},~\textbf{e}), degree of photon circular polarization $\bar{\xi}^{2}$ (panels~\textbf{b},~\textbf{f}), transverse spin magnitude $|\bar{\bm{\eta}}^{\prime}_{\perp}|$ (panels~\textbf{c},~\textbf{g}), and linear polarization $|\bar{\bm{\xi}}_\perp|$ (panels~\textbf{d},~\textbf{h}). The electron bunch has an initial angular position of $\atan(p_x,-p_z)=\theta_{\mathrm{pitch}}$ and $\atan(p_y,-p_z)=0$ and, under the action of the constant magnetic field, its center rotates anticlockwise in both the $\theta_x$--$\theta_y$ and $x$--$y$ planes while moving with constant velocity along the $z$ direction.  Electrons that radiate lose energy, thus increasing the cyclotron frequency $\omega_c=|e|B/(\gamma m)$. These particles advance more rapidly along the angular trajectory in the $(\theta_x,\theta_y)$ deflection map and appear ahead of nonradiating electrons.  Consequently, by the end of the simulation the electron momenta form a ring of radius $\theta_{\mathrm{pitch}}$ in the $\theta_x$--$\theta_y$ plane (see panels~\textbf{a}, \textbf{c}, \textbf{e}, and~\textbf{g}; note the different horizontal and vertical axis scales).  In this simulation, the bin size is $\Delta\theta_x = \Delta\theta_y = 8 \times 10^{-4}$~rad.

\begin{figure*}[ht!]
\centering
\includegraphics[width=0.95\linewidth]{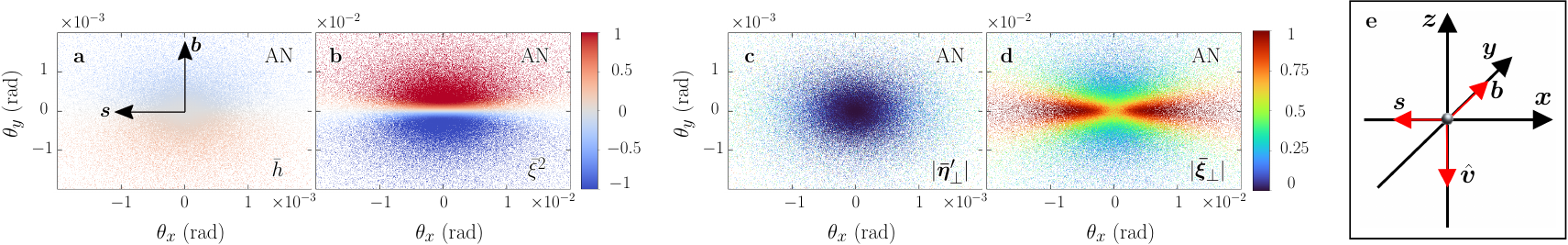}
\caption{Angle-resolved distribution of electron helicity $\bar{h}$ (panel~\textbf{a}), degree of photon circular polarization $\bar{\xi}^{2}$ (panel~\textbf{b}), transverse spin magnitude $|\bar{\bm{\eta}}^{\prime}_{\perp}|$ (panel~\textbf{c}), and degree of photon linear polarization $|\bar{\bm{\xi}}_\perp|$ (panel~\textbf{d}), computed with the AN model and averaged over $10^6$ independent NCS events all initiated from the same unpolarized ($\langle \bm{\eta} \rangle = \bm{0}$) electron with energy $210$~MeV and $\chi_e=1$. The background field is arbitrary: without loss of generality, the instantaneous triad $(\bm{s},\bm{b},\bm{\hat{v}})$ associated with the emitting electron is set to $(-\bm{x},\bm{y},-\bm{z})$. The AN model predicts $\bar{h}$ and $\bar{\xi}^{2}$ to track the sign of $-(\bm{p}^\prime \cdot \bm{b})$ and $\bm{k} \cdot \bm{b}$, respectively; $|\bar{\bm{\eta}}^{\prime}_{\perp}|$ is negligible, while $|\bar{\bm{\xi}}_\perp|$ approaches unity along $\bm{s}$. Panel~\textbf{a} also shows the projections of $\bm{s}$ and $\bm{b}$ onto the $\theta_x$--$\theta_y$ plane; panel~\textbf{e} shows their three-dimensional orientation.}
\label{fig:h_single_test}
\end{figure*}

To interpret the results in Fig.~\ref{fig:h_bconst}, we first summarize the main predictions of the AN model (see Sec.~\ref{sec:newMC}) by simulating a large number of independent NCS events for a \emph{fixed} incoming electron. Since the electron does not evolve, it is not necessary to specify the spacetime structure of the background field. The only inputs required by the AN model are the electron energy $\varepsilon=210~\mathrm{MeV}$, the quantum parameter $\chi_e=1$, the initial spin state (unpolarized, $\langle\bm{\eta}\rangle=\bm{0}$), and the instantaneous triad $(\bm{s},\bm{b},\bm{\hat{v}})$, which we set to $(-\bm{x},\bm{y},-\bm{z})$ without loss of generality (see panel~\textbf{e} of Fig.~\ref{fig:h_single_test}).  We sampled $10^6$ events and for each event used the method introduced in Sec.~\ref{sec:new_method} to sample the photon energy, emission direction, final electron spin $\bm{\eta}^\prime$, and photon Stokes vector $\bm{\xi}$.

The results are summarized in Fig.~\ref{fig:h_single_test}, where the angles $\theta_x$ and $\theta_y$ have the same meaning as in the previous section.  Thus, a point along the horizontal axis $\theta_y=0$ corresponds to an electron or a photon emitted along $+\bm{s}$ for $\theta_x<0$ or $-\bm{s}$ for $\theta_x>0$.  Analogously, a point along the vertical axis $\theta_x=0$ corresponds to an electron or a photon emitted along $+\bm{b}$ for $\theta_y>0$ or $-\bm{b}$ for $\theta_y<0$ [see the orientation of $\bm{s}$ and $\bm{b}$ in panel~\textbf{a} of Fig.~\ref{fig:h_single_test}]. We observe that:
\begin{itemize}
\item \textbf{Electron helicity}: the mean helicity $\bar{h}$ is negative (positive) for $\theta_y>0$ ($\theta_y<0$), i.e., for the momentum $\bm{p}'$ having a component along $+\bm{b}$ ($-\bm{b}$) (panel~\textbf{a}).  In other words, the larger the projection of $\bm{p}^\prime$ along $\bm{b}$, the more the spin tends to anti\mbox{-}align with $\bm{p}^\prime$.

\item \textbf{Photon circular polarization}: an analogous pattern appears for the circular Stokes parameter $\bar{\xi}^2$ (panel~\textbf{b}). The right-handedness (left-handedness) increases as the emission becomes more parallel (anti-parallel) to $\bm{b}$, with $\bar{\xi}^2$ growing with $\bm{k} \cdot \bm{b}$.

\item \textbf{Transverse spin and linear polarization}: no net transverse spin is expected ($|\bm{\bar{\eta}}^\prime_\perp| \approx 0$, panel~\textbf{c}).  By contrast, the degree of linear polarization approaches unity along $\pm\bm{s}$, with $|\bm{\bar{\xi}}_\perp| \rightarrow 1$ (panel~\textbf{d}).
\end{itemize}

Now, with the help of the above general analysis, corresponding to Fig.~\ref{fig:h_single_test}, we note that, since the electrons gyrate around the $z$ axis (and move with constant velocity along $z$), on each point of the momentum ring of radius $\theta_{\mathrm{pitch}}$ in Fig.~\ref{fig:h_bconst}:
\begin{enumerate}
\item The projection of the velocity of either the final electron or of the photon onto the $\theta_x$-$\theta_y$ plane is normal to the momentum ring and points outwards.
\item The acceleration unit vector $\bm{s}$ is tangent to the momentum ring and points counterclockwise.
\item The projection of the unit vector $\bm{b}$ onto the $\theta_x$-$\theta_y$ plane points radially outwards.
\end{enumerate}
The three-dimensional orientation of $\bm{s}$ and $\bm{b}$ for electrons emitting at $(\theta_x,\theta_y)=(\theta_{\mathrm{pitch}},0)$ is shown in panel~\textbf{i} of Fig.~\ref{fig:h_bconst}, while panel~\textbf{e} shows their projections onto the $\theta_x$--$\theta_y$ plane. Consequently, if the $\bm{b}$ ($y$) axis in panels~\textbf{a}--\textbf{d} of Fig.~\ref{fig:h_single_test} is reoriented to align with any chosen radial direction of the ring in Fig.~\ref{fig:h_bconst}, the spin and polarization patterns in panels~\textbf{e}--\textbf{h} of Fig.~\ref{fig:h_bconst} are obtained by rotating the corresponding pattern from Fig.~\ref{fig:h_single_test} about the origin $(\theta_x, \theta_y) = (0,0)$ (see panel~\textbf{i} in Fig.~\ref{fig:h_bconst}).

In the CL model, photons are emitted parallel to the electron momenta and therefore inherit the same ring-like distribution (panels~\textbf{b} and~\textbf{d} of Fig.~\ref{fig:h_bconst}). In the AN description (panels~\textbf{f} and~\textbf{h}), the photon ring is broadened because the AN model assigns a nontrivial emission cone of half-width ${\sim}1/\gamma$ around the instantaneous electron velocity, in contrast to the strictly collinear CL emission.

Figure~\ref{fig:h_bconst} shows that the CL model predicts no spin--momentum alignment (panel~\textbf{a}), whereas the AN model (panel~\textbf{e}) exhibits a pronounced helicity bias. In particular, and according to the considerations in the previous paragraph, when the recoiling electron momentum has a positive component along the instantaneous unit vector $\bm{b}$ (defined in Fig.~\ref{angles_pic}), with $\bm{p}^\prime \cdot \bm{b} > 0$, the spin tends to anti-align with the final momentum (blue outer circle); when that momentum component points in the opposite direction ($\bm{p}^\prime \cdot \bm{b} < 0$), the spin preferentially aligns instead (red inner circle).  The red and blue circles bracket a gray, spin-neutral ring corresponding to those electrons that acquire no momentum component along $\bm{b}$ at emission (i.e., $\bm{p}^\prime \cdot \bm{b} \approx 0$).

Photon polarization provides an even clearer signature of this dynamics within the AN model. Indeed, the angular distribution of the Stokes parameter $\bar{\xi}^{2}$ predicted by the AN model (panel~\textbf{f} of Fig.~\ref{fig:h_bconst}) closely mirrors the electron-helicity pattern: when the projection of the emitted photon momentum along the instantaneous electron $\bm{b}$ versor is positive ($\bm{k} \cdot \bm{b} > 0$), photons are strongly right-handed ($\bar{\xi}^{2}>0$), whereas the opposite orientation ($\bm{k} \cdot \bm{b} < 0$) corresponds to left-handed polarization ($\bar{\xi}^{2}<0$).  In the AN model, for photons emitted nearly along the bunch direction, the polarization becomes predominantly linear, with $|\bar{\bm{\xi}}_\perp| \gtrsim 0.7$ (panel~\textbf{h}) and $\bar{\xi}^{2}\approx 0$.  In sharp contrast, the CL model yields an almost uniform degree of linear polarization, $|\bar{\bm{\xi}}_\perp| \approx 0.5$, and no appreciable circular component, $\bar{\xi}^{2} \approx 0$, consistent with its use of angle-integrated (collinear) emission distributions.


\section{\label{sec:conclusion}Conclusion and Outlook}

In this work, focusing on NCS as the paradigmatic process of SFQED, we have shown that the instantaneous, angle-resolved spin- and polarization-resolved differential distribution, obtained by treating the LCFA integrand as a probability density per unit time/phase, is \emph{sign-indefinite} even in a strictly uniform constant crossed field. We stress that this occurs already in a uniform constant field, indicating that it is not related to a failure of the LCFA. As a consequence, the commonly used inference of outgoing electron spin and photon Stokes vectors from local fully differential expressions can yield unphysical states with $|\langle\bm{\eta}'\rangle|>1$ and/or $|\langle\bm{\xi}\rangle|>1$, i.e., negative inferred probabilities. This establishes an \emph{intrinsic nonlocality} of spin and polarization in strong-field QED. We have then demonstrated that a consistent probabilistic interpretation is restored only at the level of \emph{phase-integrated} quantities: the spin- and polarization-resolved distributions become non-negative and yield admissible, normalizable outgoing spin and Stokes vectors once the full photon formation region is integrated over. Building on this observation, we derived closed analytical expressions for phase-integrated spin and polarization by evaluating phase-integrated amplitudes in a constant crossed field, leading to analytic formulae for $\langle\bm{\eta}'\rangle$ and $\langle\bm{\xi}\rangle$ at fixed emitted photon momentum. We have put forward and implemented an algorithm which preserves the established and well-benchmarked \emph{local} LCFA workflow for event triggering and photon momentum (rates, photon energy, and emission angles), but replaces the ill-defined time-local treatment of spin and polarization by \emph{phase-integrated} spin/polarization vectors computed in an event-by-event matched auxiliary constant crossed field. By construction, the resulting emission events carry physically admissible spin and Stokes vectors while remaining compatible with LCFA-based SFQED MC and QED--PIC methods.

Benchmark simulations in two representative configurations, (i) a GeV-class electron--laser collision at current experimental parameters, and (ii) emission in a pulsar-like magnetic field, demonstrate that this nonlocal (phase-integrated) treatment can produce qualitatively different results in electron spin and photon polarization compared with the standard local collinear model. These findings provide a new framework for predictive modeling of spin-, and polarization-resolved observables in strong-field QED experiments and for interpreting energetic polarized radiation from astrophysical environments.

Closely analogous issues arise for NBW pair production. Preliminary results reported in Ref.~\cite{montefioriPhD25} confirm that the angle- and spin/polarization-resolved local differential rate in NBW is likewise sign-indefinite in a uniform constant field, indicating that the structural nonlocality identified here is not specific to NCS but is a generic feature of strong-field QED processes whenever spin and polarization are resolved. Extending the phase-integrated, LCFA-compatible sampling strategy introduced here to NBW therefore represents a natural and  pressing direction for future work.


\begin{acknowledgments}

This paper is part of the Ph.D. work of S.M. at Heidelberg University (see Ref.~\cite{montefioriPhD25}).

We gratefully acknowledge helpful discussions with Yue-Yue Chen and Karen Z. Hatsagortsyan. A.D.P. would like to thank the Max Planck Institute for Nuclear Physics for the generous hospitality during several research visits that were important for the completion of this project.

This material is based upon work supported by the U.S. Department of Energy [National Nuclear Security Administration] University of Rochester ‘National Inertial Confinement Fusion Program’ under Award Number DE-NA0004144. 

This report was prepared as an account of work sponsored by an agency of the United States Government. Neither the United States Government nor any agency thereof, nor any of their employees, makes any warranty, express or implied, or assumes any legal liability or responsibility for the accuracy, completeness, or usefulness of any information, apparatus, product, or process disclosed, or represents that its use would not infringe privately owned rights. Reference herein to any specific commercial product, process, or service by trade name, trademark, manufacturer, or otherwise does not necessarily constitute or imply its endorsement, recommendation, or favoring by the United States Government or any agency thereof. The views and opinions of authors expressed herein do not necessarily state or reflect those of the United States Government or any agency thereof.
\end{acknowledgments}


\appendix


\begin{widetext}

\section{\label{sec:airy} Airy functions}

For reference, we collect the definitions and identities for the Airy function $\Ai(z)$ used in this work. Its Fourier integral representation and derivatives are
\begin{align}
\Ai(z) &= \int_{-\infty}^\infty \frac{d\phi}{2 \pi} e^{iz\phi + i\frac{\phi^3}{3}}, \notag \\
\Ai^\prime(z) &= i\int_{-\infty}^\infty \frac{d\phi}{2 \pi} \phi e^{iz\phi + i\frac{\phi^3}{3}}, \notag \\
\Ai^{\prime\prime}(z) &= i^2\int_{-\infty}^\infty \frac{d\phi}{2 \pi} \phi^2 e^{iz\phi + i\frac{\phi^3}{3}} = z \Ai(z).
\label{Ai}
\end{align}
Their relations to the modified Bessel functions of the second kind $K_{\nu}(z)$ (for $z>0$) are
\begin{align}
\Ai(z) &= \frac{\sqrt{z}}{\pi \sqrt{3}} K_\frac{1}{3} \Bigl(\frac{2}{3} z^\frac{3}{2}\Bigr), \label{Airy to Bessel 1} \\
\Ai^\prime(z) &= - \frac{z}{\pi \sqrt{3}} K_\frac{2}{3} \Bigl(\frac{2}{3} z^\frac{3}{2}\Bigr).
\label{Airy to Bessel 2}
\end{align}
Finally, the following integral representations may prove to be useful
\begin{align}
\Ai^2(z) &= \frac{4^\frac{1}{3}}{2 \pi} \int_{- \infty}^{+ \infty} \Ai(4^\frac{1}{3}(z+x^2)) dx, \notag \\
\Ai^{\prime 2}(z) &= \frac{4^\frac{1}{3}}{\pi} \int_{- \infty}^{+ \infty} x^2 \Ai(4^\frac{1}{3}(z+x^2)) dx + z \frac{4^\frac{1}{3}}{2 \pi} \int_{- \infty}^{+ \infty} \Ai(4^\frac{1}{3}(z+x^2)) dx.
\label{airy-squared}
\end{align}
%


\section{\label{sec:yy-comp}Explicit expressions for the local angle- and spin/polarization-resolved distribution}

Within the quasiclassical method (QM) and the locally constant field approximation (LCFA), the nonlinear Compton scattering (NCS) probability in Eq.~\eqref{prob_tob_pre} leads to the time-local, fully differential distribution reported in Eq.~\eqref{spa_res_2}, which we reproduce here for convenience:
\begin{equation}
\frac{dP^{(\eta\eta^\prime\xi)}_{\mathrm{NCS}}}{dt\, d\omega\, d\theta\, d\varphi} = C_{\mathrm{NCS}} \left(w + \xi^{i} h^{i} + \eta^{\prime j} S^{j} + \xi^{i}\eta^{\prime j} P^{ij}\right). \notag
\end{equation}
Here repeated indices $i,j=1,2,3$ are summed over (Einstein convention). The vectors $h^{i}$ and $S^{j}$ determine the polarization- and spin-dependent parts of the distribution, respectively, while $P^{ij}$ is the spin-polarization correlation matrix. Derivation details can be found in Refs.~\cite{daiPRD23,montefioriPhD25}; below we provide the explicit expressions for $w$, $h^{i}$, $S^{j}$, $P^{ij}$, and for the prefactor $C_{\mathrm{NCS}}$.

The prefactor reads
\begin{equation}
C_{\mathrm{NCS}} = \frac{\alpha\,\omega}{4\pi^{2}} \sqrt{\frac{2}{3}}\, \lambda^{1/2} \frac{\gamma^{3}}{\chi_{e}\,\varepsilon^\prime}\, \sin\theta ,
\label{eq:CNCS_def}
\end{equation}
where
\begin{equation}
\lambda(t)=1-\bm{n}\cdot\bm{v}(t) \notag
\end{equation}
encodes the instantaneous angle between the electron velocity $\bm{v}(t)$ and the photon emission direction $\bm{n}$. The explicit expressions for $w$, $h^{i}$, $S^{j}$, $P^{ij}$ are:
\begin{align}
w & = K_{\frac{1}{3}}(\xi_e) \Bigl[ 2\lambda \frac{\varepsilon^2 + \varepsilon^{\prime 2}}{\varepsilon \varepsilon^\prime} - \frac{1}{\gamma^2} \Bigr] + K_{\frac{2}{3}}(\xi_e) \sqrt{2 \lambda} \Bigl[ \theta \sin\varphi \frac{\varepsilon^2 -\varepsilon^{\prime 2}}{ \varepsilon \varepsilon^\prime} \hat{\bm{v}} - \frac{\omega}{\gamma \varepsilon} \bm{b}  \Bigr] \cdot \bm{\eta}, \label{w_def}\\
h^1 &= K_{\frac{1}{3}}(\xi_e) \theta^2\sin2\varphi + K_{\frac{2}{3}}(\xi_e) \sqrt{2 \lambda} \frac{\omega}{\gamma \varepsilon^\prime} (\bm{\eta} \cdot \bm{s} ), \label{pe_ang_pol_spin_res_1_1_spin_summed} \\
h^2 &= K_{\frac{1}{3}}(\xi_e) \Bigl[ \Bigl( 2\lambda \frac{\varepsilon^2 -\varepsilon^{\prime 2}}{ \varepsilon \varepsilon^\prime} - \frac{\omega}{\gamma^2 \varepsilon} \Bigr) \hat{\bm{v}} - \frac{\theta \omega}{\gamma \varepsilon} (\sin\varphi \bm{b} + \cos\varphi \bm{s}) \Bigr] \cdot \bm{\eta} + K_{\frac{2}{3}}(\xi_e) \sqrt{2 \lambda} \theta \sin\varphi \frac{\varepsilon^2 + \varepsilon^{\prime 2}}{ \varepsilon \varepsilon^\prime}, \label{pe_ang_pol_spin_res_2_1_spin_summed} \\
h^3 &= K_{\frac{1}{3}}(\xi_e) (\theta^2\cos2\varphi + 2\lambda) - K_{\frac{2}{3}}(\xi_e) \sqrt{2 \lambda} \frac{\omega}{\gamma \varepsilon^\prime} (\bm{\eta} \cdot \bm{b}), \label{pe_ang_pol_spin_res_3_1_spin_summed} \\
\bm{S} & = K_{\frac{1}{3}}(\xi_e) \Bigl[ \Bigl(4\lambda - \frac{1}{\gamma^2}\Bigr) \bm{\eta} + \frac{\varepsilon^2 -\varepsilon^{\prime 2}}{2 \varepsilon \varepsilon^\prime}\frac{\theta}{\gamma} \Bigl( \sin\varphi (\bm{s} \times \bm{\eta}) - \cos\varphi (\bm{b} \times \bm{\eta}) \Bigr) \Bigr] \notag\\
& \quad + K_{\frac{1}{3}}(\xi_e) \frac{\omega^2}{\varepsilon \varepsilon^\prime} \Bigl[ \theta^2 (\bm{\eta}\cdot \hat{\bm{v}}) - \frac{\theta}{2\gamma} \Bigl( \sin\varphi(\bm{\eta} \cdot \bm{b}) + \cos\varphi(\bm{\eta} \cdot \bm{s}) \Bigr) \Bigr] \hat{\bm{v}} + K_{\frac{2}{3}}(\xi_e) \sqrt{2 \lambda} \theta \sin\varphi \frac{\varepsilon^2 -\varepsilon^{\prime 2}}{ \varepsilon \varepsilon^\prime} \hat{\bm{v}} \notag \\
& \quad - K_{\frac{1}{3}}(\xi_e)  \frac{\omega^2}{2 \varepsilon \varepsilon^\prime} \frac{\theta \cos\varphi}{\gamma} (\bm{\eta}\cdot \hat{\bm{v}}) \bm{s} - K_{\frac{1}{3}}(\xi_e)\frac{\omega^2}{2 \varepsilon \varepsilon^\prime} \frac{\theta \sin\varphi}{\gamma} (\bm{\eta}\cdot \hat{\bm{v}})\bm{b} - K_{\frac{2}{3}}(\xi_e) \sqrt{2 \lambda} \frac{\omega}{\gamma\varepsilon^\prime} \bm{b}, \label{pe_ang_pol_spin_res_0_1_for_pola} \\
P^{1j} &= K_{\frac{1}{3}}(\xi_e) \Bigl\{ \theta^2\sin2\varphi \frac{\varepsilon^2 + \varepsilon^{\prime 2}}{2\varepsilon\varepsilon^\prime}\bm{\eta} \notag \\
& \qquad \qquad + \frac{\varepsilon^2 -\varepsilon^{\prime 2}}{2 \varepsilon \varepsilon^\prime}\Bigl[ (- \theta^2\cos2\varphi - 2\lambda) (\hat{\bm{v}} \times \bm{\eta}) + \frac{\theta \cos\varphi}{\gamma} (\bm{s} \times \bm{\eta}) - \frac{\theta \sin\varphi}{\gamma} (\bm{b} \times \bm{\eta}) \Bigr] \notag \\
& \qquad \qquad - \theta^2\sin2\varphi\frac{\omega^2}{2 \varepsilon \varepsilon^\prime}(\bm{\eta}\cdot \hat{\bm{v}}) \hat{\bm{v}} + \frac{\omega^2}{2 \varepsilon \varepsilon^\prime} \frac{\theta \sin\varphi}{\gamma} \bigl[(\bm{\eta}\cdot \hat{\bm{v}}) \bm{s}) + (\bm{\eta} \cdot \bm{s}) \hat{\bm{v}} \bigr] \notag \\
& \qquad \qquad + \frac{\omega^2}{2\varepsilon \varepsilon^\prime} \frac{\theta \cos\varphi}{\gamma} \bigl[(\bm{\eta}\cdot \hat{\bm{v}}) \bm{b} + (\bm{\eta} \cdot \bm{b}) \hat{\bm{v}} \bigr] - \frac{\omega^2}{2 \varepsilon \varepsilon^\prime \gamma^2} \bigl[(\bm{\eta}\cdot \bm{s})\bm{b} + (\bm{\eta} \cdot \bm{b})\bm{s} \bigr] \Bigr\} + K_{\frac{2}{3}}(\xi_e) \sqrt{2 \lambda} \frac{\omega}{\gamma\varepsilon} \bm{s}, \label{pe_ang_pol_spin_res_1_1_res} \\
P^{2j} &=  K_{\frac{1}{3}}(\xi_e) \Bigl[ \Bigl( 2\lambda \frac{\varepsilon^2 -\varepsilon^{\prime 2}}{ \varepsilon \varepsilon^\prime} - \frac{\omega}{\gamma^2 \varepsilon^\prime} \Bigr) \hat{\bm{v}} - \frac{\theta\sin\varphi}{\gamma} \frac{\omega}{\varepsilon^\prime} \bm{b} - \frac{\theta\cos\varphi}{\gamma} \frac{\omega}{\varepsilon^\prime} \bm{s} \Bigr] \notag \\
& \quad + K_{\frac{2}{3}}(\xi_e) \sqrt{2 \lambda} \Bigl\{ 2 \theta \sin\varphi \bm{\eta} + \frac{\varepsilon^2 -\varepsilon^{\prime 2}}{2 \varepsilon \varepsilon^\prime} \frac{1}{\gamma} (\bm{s} \times \bm{\eta} ) + \theta \sin\varphi \frac{\omega^2}{\varepsilon\varepsilon^\prime} (\bm{\eta}\cdot \hat{\bm{v}}) \hat{\bm{v}} - \frac{\omega^2}{2\varepsilon\varepsilon^\prime\gamma}\bigl[(\bm{\eta}\cdot \hat{\bm{v}})\bm{b} + (\bm{\eta} \cdot \bm{b}) \hat{\bm{v}} \bigr] \Bigr\}, \label{pe_ang_pol_spin_res_2_1_res} \\
P^{3j} &= K_{\frac{1}{3}}(\xi_e) \Bigl\{ \Bigl[\frac{\varepsilon^2 + \varepsilon^{\prime 2}}{2\varepsilon\varepsilon^\prime} (\theta^2\cos2\varphi + 2\lambda) \bm{\eta} \notag \\
& \qquad \qquad + \frac{\varepsilon^2 -\varepsilon^{\prime 2}}{2 \varepsilon \varepsilon^\prime} \Bigl[ \theta^2 \sin2\varphi (\hat{\bm{v}} \times \bm{\eta}) - \frac{\theta\sin\varphi}{\gamma} (\bm{s} \times \bm{\eta}) - \frac{\theta\cos\varphi}{\gamma} (\bm{b} \times \bm{\eta}) \Bigr] \notag \\
& \qquad \qquad - \frac{\omega^2}{2\varepsilon \varepsilon^\prime}(\theta^2\cos2\varphi + 2\lambda)(\bm{\eta}\cdot \hat{\bm{v}}) \hat{\bm{v}} + \frac{\omega^2}{2 \varepsilon \varepsilon^\prime} \frac{\theta\cos\varphi}{\gamma} \bigl[(\bm{\eta} \cdot \hat{\bm{v}}) \bm{s} + (\bm{\eta} \cdot \bm{s}) \hat{\bm{v}} \bigr] \notag \\
& \qquad \qquad - \frac{\omega^2}{2\varepsilon \varepsilon^\prime} \frac{\theta\sin\varphi}{\gamma} \bigl[(\bm{\eta} \cdot \hat{\bm{v}}) \bm{b} + (\bm{\eta} \cdot \bm{b}) \hat{\bm{v}} \bigr] - \frac{\omega^2}{2\varepsilon\varepsilon^\prime\gamma^2} \bigl[(\bm{\eta} \cdot \bm{s}) \bm{s} - (\bm{\eta} \cdot \bm{b}) \bm{b} \bigr] \Bigr\} - K_{\frac{2}{3}}(\xi_e) \sqrt{2 \lambda} \frac{\omega}{\gamma \varepsilon} \bm{b},
\label{pe_ang_pol_spin_res_3_1_res} 
\end{align}
where $K_\nu$ denotes the modified Bessel function of the second kind, and $\xi_e$ has been defined in Eq.~\eqref{xi_e_def} and is reported below for convenience
\begin{equation}
\xi_e = \frac{4\sqrt{2}}{3} \frac{\omega \gamma^3}{\varepsilon^\prime \chi_e} \lambda^{\frac{3}{2}} \notag
\end{equation}
is their argument. All other quantities are defined in the main text (see Secs. \ref{sec:intro} and \ref{subsec:QM}).


\section{\label{sec:and_integrate}Explicit expressions for the local angle-integrated spin/polarization-resolved distribution}

The local, angle-integrated distribution is obtained by integrating the fully differential expression in Eq.~\eqref{spa_res_2} over the photon emission solid angle. Using the angular-integration identities collected in the Appendix of Ref.~\cite{chenPRD22} (see also Ref.~\cite{montefioriPhD25}), one finds ~Eq.~\eqref{sp_res}, which we report below for convenience
\begin{align}
\frac{dP^{(\eta\eta^\prime\xi)}_{\mathrm{NCS}}}{dt\, d\omega}
&=
\int_{0}^{\pi} d\theta \int_{0}^{2\pi} d\varphi\, \frac{dP^{(\eta\eta^\prime\xi)}_{\mathrm{NCS}}}{dt\, d\omega\, d\theta\, d\varphi} \notag \\
&= \underline{C}_{\mathrm{NCS}} \left( \underline{w} + \xi^{i}\underline{h}^{\,i} + \eta^{\prime j}\underline{S}^{\,j} + \xi^{i}\eta^{\prime j}\underline{P}^{\,ij} \right). \notag
\end{align}
The explicit expressions for $\underline{C}_{\mathrm{NCS}}$, 
$\underline{w}$, $\underline{h}^{i}$, $\underline{S}^{j}$, $\underline{P}^{ij}$ are:
\begin{align}
\underline{C}_{\mathrm{NCS}} &= \frac{\alpha}{4\sqrt{3}\pi\gamma^2}, \\
\underline w &= K_{\frac{2}{3}}(z_q) \frac{\varepsilon^2 + \varepsilon^{\prime 2}}{\varepsilon \varepsilon^\prime} - \int^\infty_{z_q} dx K_\frac{1}{3}(x) - K_{\frac{1}{3}}(z_q) \frac{\omega}{\varepsilon} \bigl( \bm{\eta} \cdot \bm{b} \bigr), \\
\underline{\bm{S}} &= \Bigl[ 2 K_{\frac{2}{3}}(z_q) - \int^\infty_{z_q} dx K_\frac{1}{3}(x) \Bigr] \bm{\eta} + \Bigl[ K_{\frac{2}{3}}(z_q)  - \int^\infty_{z_q} dx K_\frac{1}{3}(x)\Bigr]\frac{\omega^2}{\varepsilon \varepsilon^\prime}(\bm{\eta}\cdot \hat{\bm{v}}) \hat{\bm{v}} - K_{\frac{1}{3}}(z_q) \frac{\omega}{\varepsilon^\prime} \bm{b}, \label{pe_pol_spin_res_0} \\
\underline h^1 &= K_\frac{1}{3}(z_q) \frac{\omega}{\varepsilon^\prime} \bigl( \bm{\eta} \cdot \bm{s} \bigr), \\
\underline h^2 &= \Bigl[ K_\frac{2}{3}(z_q) \frac{\varepsilon^2 -\varepsilon^{\prime 2}}{ \varepsilon \varepsilon^\prime} - \frac{\omega}{ \varepsilon} \int^\infty_{z_q} dx K_\frac{1}{3}(x)  \Bigr] \bigl( \hat{\bm{v}} \cdot \bm{\eta} \bigr), \\
\underline h^3 &= K_\frac{2}{3}(z_q) - K_\frac{1}{3}(z_q) \frac{\omega}{\varepsilon^\prime} \bigl( \bm{\eta} \cdot \bm{b} \bigr), \\
\underline{P}^{1j} &= - K_\frac{2}{3}(z_q)\frac{\varepsilon^2 -\varepsilon^{\prime 2}}{2 \varepsilon \varepsilon^\prime} (\hat{\bm{v}} \times \bm{\eta} ) - \frac{\omega^2}{2 \varepsilon \varepsilon^\prime} \bigl[(\bm{\eta}\cdot \bm{s}) \bm{b} + (\bm{\eta} \cdot \bm{b}) \bm{s} \bigr] \int^\infty_{z_q} dx K_\frac{1}{3}(x) + K_\frac{1}{3}(z_q) \frac{\omega}{\varepsilon} \bm{s}, \label{pe_pol_spin_res_1} \\
\underline{P}^{2j} &= \Bigl[ K_\frac{2}{3}(z_q) \frac{\varepsilon^2 -\varepsilon^{\prime 2}}{ \varepsilon \varepsilon^\prime} - \frac{\omega}{\varepsilon^\prime} \int^\infty_{z_q} dx K_\frac{1}{3}(x)  \Bigr] \hat{\bm{v}} + K_\frac{1}{3}(z_q) \Bigl\{ \frac{\varepsilon^2 -\varepsilon^{\prime 2}}{2 \varepsilon \varepsilon^\prime} ( \bm{s} \times \bm{\eta} ) - \frac{\omega^2}{2\varepsilon\varepsilon^\prime}\bigl[(\bm{\eta}\cdot \hat{\bm{v}}) \bm{b} + (\bm{\eta} \cdot \bm{b}) \hat{\bm{v}} \bigr] \Bigr\}, \label{pe_pol_spin_res_2} \\
\underline{P}^{3j} &= K_\frac{2}{3}(z_q) \Bigl[\frac{\varepsilon^2 + \varepsilon^{\prime 2}}{2\varepsilon\varepsilon^\prime} \bm{\eta} - \frac{\omega^2}{2\varepsilon \varepsilon^\prime}(\bm{\eta}\cdot \hat{\bm{v}}) \hat{\bm{v}} \Bigr] - \frac{\omega^2}{2\varepsilon\varepsilon^\prime} \bigl[(\bm{\eta} \cdot \bm{s}) \bm{s} - (\bm{\eta} \cdot \bm{b}) \bm{b} \bigr] \int^\infty_{z_q} dx K_\frac{1}{3}(x) - K_\frac{1}{3}(z_q) \frac{\omega}{\varepsilon} \bm{b}, \label{pe_pol_spin_res_3}
\end{align}
where
\begin{equation}
z_q = \frac{2 \omega}{3 (\varepsilon - \omega) \chi_e}.
\end{equation}
%


\section{\label{CCFint}Angle- and spin-resolved differential distribution in a constant crossed field}

This Appendix provides derivation details and explicit expressions for all quantities entering Eq.~\eqref{matrix_element_squared_final}, as used in Sec.~\ref{subsec:sign}. We also show that, for an arbitrary initial lepton state specified by momentum $\bm{p}$ and rest-frame spin $\bm{\eta}$, the ultrarelativistic limit of the spin- and polarization-summed result (i.e., after summing over the outgoing electron spin and photon polarization) reduces to the corresponding quasiclassical-method expression.

\subsection*{1. Reduction to a single phase integral in a CCF}

We start from Eq.~\eqref{squared}. Specializing to the CCF vector potential introduced in Sec.~\ref{subsec:sign} and defining
\begin{align}
\bar z(\phi_+) &\equiv 2u\bigl(\mu + 2\beta \phi_+ + 3\kappa \phi_+^{2}\bigr),
\label{eq:zbar}
\\
u &\equiv \frac{1}{2}\left(\frac{4}{3\kappa}\right)^{1/3},
\label{eq:u_def}
\end{align}
where
\begin{equation}
\mu = p^\prime_+ + k_+ - p_+, \qquad \beta = \frac{e}{2} \Bigl( \frac{p^\prime_\mu}{p^\prime_-} - \frac{p_\mu}{p_-} \Bigr) a^\mu \qquad \text{and} \qquad \kappa = - \frac{e^2}{6} a^2 \Bigl( \frac{1}{p^\prime_-} - \frac{1}{p_-} \Bigr),  
\end{equation}
we proceed as follows: (i) change integration variables according to
\begin{equation}
\phi=\phi_+ + u\tilde\phi,
\qquad
\phi^\prime=\phi_+ - u\tilde\phi;
\label{change of vars}
\end{equation}
(ii) impose the gauge conditions in Eqs.~\eqref{gauge_fixing_0}--\eqref{gauge_fixing}; and (iii) use the polarization completeness relation~\eqref{completeness}. This yields
\begin{align}
dP^{(\zeta,\zeta^\prime)}_{\mathrm{NCS}} &=
\frac{d^{3}k}{8\pi}\,
\frac{\alpha}{p_- p_-^\prime \omega}
\int_{-\infty}^{+\infty} d\phi_+
\int_{-\infty}^{+\infty}\frac{d\tilde\phi}{2\pi}\,
\exp\!\left[i \bar z(\phi_+)\tilde\phi + \frac{i}{3}\tilde\phi^{3}\right]
\left(-g_{\mu\nu}+\frac{k_\mu d_\nu + d_\mu k_\nu}{k_-}\right)
\notag\\
&\quad\times
\frac{u}{2}\,
\mathrm{Tr} \Biggl\{
\Biggl[
\gamma^\mu
+\frac{e}{2}\Biggl(
\frac{\slashed{a}\slashed{d}}{p_-^\prime}\gamma^\mu
+\gamma^\mu\frac{\slashed{d}\slashed{a}}{p_-}
\Biggr)\bigl(\phi_+ + u\tilde\phi\bigr)
\Biggr]
(\slashed{p}+m)(1+\gamma^{5}\slashed{\zeta})
\notag\\
&\qquad\qquad\qquad\qquad\times
\Biggl[
\gamma^\nu
+\frac{e}{2}\Biggl(
\frac{\slashed{a}\slashed{d}}{p_-}\gamma^\nu
+\gamma^\nu\frac{\slashed{d}\slashed{a}}{p_-^\prime}
\Biggr)\bigl(\phi_+ - u\tilde\phi\bigr)
\Biggr]
(\slashed{p}^\prime+m)(1+\gamma^{5}\slashed{\zeta}^\prime)
\Biggr\}.
\label{matrix_element_squared_3}
\end{align}
The integral over $\tilde\phi$ is performed using the Airy-function representation in Appendix~\ref{sec:airy}, which leads to
\begin{align}
dP^{(\zeta,\zeta^\prime)}_{\mathrm{NCS}} &=
\frac{d^{3}k}{8\pi}\,
\frac{\alpha}{p_- p_-^\prime \omega}
\int_{-\infty}^{+\infty} d\phi_+\,
\left(-g_{\mu\nu}+\frac{k_\mu d_\nu + d_\mu k_\nu}{k_-}\right)
\frac{u}{2}\,
\mathrm{Tr} \Biggl\{
\Biggl[
\gamma^\mu (\slashed{p}+m)(1+\gamma^{5}\slashed{\zeta})\gamma^\nu \Ai(\bar z)
\notag\\
&\qquad
+ \frac{e}{2}\gamma^\mu(\slashed{p}+m)(1+\gamma^{5}\slashed{\zeta})
\Biggl(\frac{\slashed{a}\slashed{d}}{p_-}\gamma^\nu + \gamma^\nu\frac{\slashed{d}\slashed{a}}{p_-^\prime}\Biggr)
\bigl(\phi_+\Ai(\bar z)+ i u\,\Ai^\prime(\bar z)\bigr)
\notag\\
&\qquad
+ \frac{e}{2}\Biggl(\frac{\slashed{a}\slashed{d}}{p_-^\prime}\gamma^\mu + \gamma^\mu\frac{\slashed{d}\slashed{a}}{p_-}\Biggr)
(\slashed{p}+m)(1+\gamma^{5}\slashed{\zeta})\gamma^\nu
\bigl(\phi_+\Ai(\bar z)- i u\,\Ai^\prime(\bar z)\bigr)
\notag\\
&\qquad
+ \frac{e^{2}}{4}\Biggl(\frac{\slashed{a}\slashed{d}}{p_-^\prime}\gamma^\mu + \gamma^\mu\frac{\slashed{d}\slashed{a}}{p_-}\Biggr)
(\slashed{p}+m)(1+\gamma^{5}\slashed{\zeta})
\Biggl(\frac{\slashed{a}\slashed{d}}{p_-}\gamma^\nu + \gamma^\nu\frac{\slashed{d}\slashed{a}}{p_-^\prime}\Biggr)
\bigl(\phi_+^{2}\Ai(\bar z)+ u^{2}\bar z\,\Ai(\bar z)\bigr)
\Biggr]
(\slashed{p}^\prime+m)(1+\gamma^{5}\slashed{\zeta}^\prime)
\Biggr\}.
\label{matrix_element_squared_3_trace_integrated}
\end{align}
After evaluating the Dirac trace and contracting the indices $\mu$ and $\nu$, Eq.~\eqref{matrix_element_squared_3_trace_integrated} reduces to Eq.~\eqref{matrix_element_squared_final}, where $\bar I(\phi_+)$ is the spin-independent contribution and $\bar l^\mu(\phi_+)$ is the spin-coupling four-vector. Their explicit forms are
\begin{align}
\bar I(\phi_+) &=
u\Biggl[
- 2 e^2 (a  a)
\left(\frac{p_-^\prime}{p_-}+\frac{p_-}{p_-^\prime}\right)
\bigl(u^{2}\bar z+\phi_+^{2}\bigr)\Ai(\bar z)
-\frac{4}{k_-} e \phi_+ \Ai(\bar z)
\Bigl(p_-(a  k)+p_-^\prime (a  k)-\frac{k_- p_-}{p_-^\prime}(a  p^\prime)\Bigr)
\notag \\
&+\frac{4}{k_-}\Ai(\bar z)\bigl(-k_- m^{2}+p_-^\prime (k  p)+p_-(k  p^\prime)\bigr)
+ 4 e m u\,\Ai^\prime(\bar z)\,\epsilon^{\alpha\beta\rho\sigma}\zeta_\alpha a_\beta
\Bigl(\frac{1}{p_-}k_\rho d_\sigma-\frac{1}{p_-}d_\rho p_\sigma+\frac{1}{p_-^\prime}d_\rho p^\prime_\sigma\Bigr)
\Biggr],
\label{Ibar}
\\[2mm]
\bar l^\mu(\phi_+)
&=
\bar A\,\zeta^\mu + \bar B\,a^\mu + \bar C\,k^\mu + \bar D\,d^\mu + \bar E\,p^\mu
+ \bar F\,\epsilon^{\mu\nu\rho\sigma}a_\nu k_\rho d_\sigma
+ \bar G\,\epsilon^{\mu\nu\rho\sigma}a_\nu d_\rho p_\sigma
+ \bar H\,\epsilon^{\mu\nu\rho\sigma}a_\nu d_\rho p^\prime_\sigma.
\label{final_lS_bar}
\end{align}
The coefficients in Eq.~\eqref{final_lS_bar} depend on the phase $\phi_+$ and read
\begin{align}
\bar A(\phi_+) = u \Bigl[ & 4 (a  a) e^{2} \Ai(\bar z) ( \phi_+^{2} + u^{2} \bar z ) - \frac{4}{k_-} e \phi_+ \Ai(\bar z) \Bigl( - p_- (a  k) - p^\prime_- (a  k) + k_- (a  p^\prime{}) \Bigr) \notag \\
& \,			+ \frac{4}{k_-} \Ai(\bar z) \Bigl( k_- (p  p^\prime{}) - p^\prime_- (k  p) - p_- (k  p^\prime{}) \Bigr) \Bigr], \\
\bar B(\phi_+) = u \Bigl[ & - \frac{4}{k_-} e \phi_+ \Ai(\bar z) \Bigl( - p_- (\zeta  k) + p^\prime_- (\zeta  k) - k_- (\zeta  p^\prime{}) + \zeta_- (k  p) \notag \\
& \,         + \frac{\zeta_- k_-}{p_-} (p  p^\prime{}) - \frac{\zeta_- k_-}{ p^\prime_-} m^{2} - \frac{\zeta_- p^\prime_-}{p_-} (k  p) \Bigl) \Bigr], \\
\bar C(\phi_+) = u \Bigl[ & \frac{4}{k_-} \Ai(\bar z) \Bigl( - e \phi_+ (\zeta  a) (p_- - p^\prime_-) - \zeta_- (p  p^\prime{}) + \zeta_- m^{2} + p_- (\zeta  p^\prime{}) \Bigr) \Bigr], \\
\bar D(\phi_+) = u \Bigl[ & 4 (\zeta  a) e \phi_+ \Ai(\bar z) \Bigl( \frac{m^{2}}{p_-} - \frac{(p  p^\prime{})}{p^\prime_-} \Bigr) + \frac{4}{k_-} e \phi_+ \Ai(\bar z) \Bigl( \frac{p_-}{p^\prime_-} - 1 \Bigr) \bigl( (\zeta  a) (k  p^\prime{}) - (\zeta  k) (a  p^\prime{}) \bigr) \notag \\
& \,          - \frac{4}{k_-} (\zeta  k) \Ai(\bar z) \bigl( (p  p^\prime{}) - m^{2} \bigr) - 2 \zeta_- (a  a) e^{2} m^{2} \Ai(\bar z) ( \phi_+^{2} + u^{2} \bar z) \Bigl( \frac{1}{p_-^2} + \frac{1}{p^\prime_-{}^2} \Bigr) \notag \\
& \,          + 4 \frac{\zeta_-}{p_- p^\prime_-} (a  a) (p  p^\prime{}) e^{2} \Ai(\bar z) ( \phi_+^{2} + u^{2} \bar z) + \frac{4}{k_-} \zeta_- (a  k) e \phi_+ \Ai(\bar z) \Bigl( \frac{1}{p_-} + \frac{1}{p^\prime_-} \Bigr) \bigl( (p  p^\prime{}) - m^{2} \bigr) \notag \\
& \,          - \frac{4}{k_-} \zeta_- (a  p^\prime{}) (k  p) e \phi_+ \Ai(\bar z) \Bigl( \frac{1}{p_-} - \frac{1}{p^\prime_-} \Bigr) - 4 (\zeta  p^\prime{}) (a  a) \frac{1}{p^\prime_-} e^{2} \Ai(\bar z) ( \phi_+^{2} + u^{2} \bar z ) \notag \\
& \,          - \frac{4}{k_-} (\zeta  p^\prime{}) (a  k) e \phi_+ \Ai(\bar z) \Bigl( \frac{p_-}{p^\prime_-} + 1 \Bigr) + \frac{4}{k_-} (\zeta  p^\prime{}) (k  p) \Ai(\bar z)\Bigr], \\
\bar E(\phi_+) = u \Bigl[ & 4 (\zeta  a) e \phi_+ \Ai(\bar z) - \frac{4}{k_-} \zeta_- e \phi_+ \Ai(\bar z) \Bigl( \frac{p^\prime_-}{p_-} (a  k) + (a  k) - \frac{k_-}{p_-} (a  p^\prime{}) \Bigr) \notag \\
 & \,           - 4 \frac{\zeta_-}{p_-} (a  a) e^{2} \Ai(\bar z)( \phi_+^{2} + u^{2} \bar z ) + \frac{4}{k_-} \Ai(\bar z) \bigl( \zeta_- (k  p^\prime{}) - k_- (\zeta  p^\prime{}) + p^\prime_- (\zeta  k) \bigr) \Bigr], \\
\bar F(\phi_+) = - & \frac{4}{k_-} e m u^2 \Ai^\prime(\bar z) \Bigl( 1 - \frac{p_-}{p^\prime_-} \Bigr), \\
\bar G(\phi_+) = - & 4 e m u^2 \Ai^\prime(\bar z) \frac{1}{p_-}, \\
\bar H(\phi_+) = + & 4 e m u^2 \Ai^\prime(\bar z) \frac{1}{p^\prime_-} .
\end{align}

\subsection*{2. Ultrarelativistic limit and connection to the quasiclassical method}

In the ultrarelativistic limit the integrand in Eq.~\eqref{matrix_element_squared_final} can be mapped directly onto the quasiclassical method differential in Eq.~\eqref{spa_res_2_sum_pola}. For brevity, we illustrate this explicitly for the case where the outgoing electron spin is summed. From Eq.~\eqref{matrix_element_squared_final} one has
\begin{equation}
\sum_{\zeta^\prime}\frac{dP^{(\zeta,\zeta^\prime)}_{\mathrm{NCS}}}{d\phi_+\, d\omega\, d\theta\, d\varphi} = \frac{\alpha}{4\pi}\, \frac{\omega}{p_- p_-^\prime}\, \sin\theta\, \bar I(\phi_+),
\label{matrix_element_squared_simpl_3_bar_sum_spin}
\end{equation}
while from Eq.~\eqref{spa_res_2_sum_pola} one obtains, after summing over the final spin,
\begin{equation}
\sum_{\eta^\prime}\frac{dP^{(\eta\eta^\prime)}_{\mathrm{NCS}}}{dt\, d\omega\, d\theta\, d\varphi} = \frac{\alpha \omega}{\pi^{2}}\sqrt{\frac{2}{3}}\, \lambda^{1/2}\frac{\gamma^{3}}{\chi_e\ \varepsilon^\prime}\, \sin\theta\, w,
\label{spa_res_2_sum_pola_spin}
\end{equation}
To demonstrate that Eq.~\eqref{matrix_element_squared_simpl_3_bar_sum_spin} reduces to the QM result in Eq.~\eqref{spa_res_2_sum_pola_spin} in the ultrarelativistic limit, we adopt the same CCF configuration as in Sec.~\ref{subsec:sign} and parametrize the photon momentum as
\begin{equation}
\bm{k} = \omega\bigl(\cos\theta\,\hat{\bm{v}} + \sin\theta\cos\varphi\,\bm{s} + \sin\theta\sin\varphi\,\bm{b}\bigr),
\end{equation}
in the local instantaneous basis $(\bm{s}, \bm{b}, \hat{\bm{v}})$ (see Fig.~\ref{angles_pic}).
In this limit the combination in Eq.~\eqref{eq:zbar} yields
\begin{align}
\mu + 2 \beta \phi_+ + 3 \kappa \phi_+^2 & \approx \left( \frac{\varepsilon\omega}{2(\varepsilon-\omega)} \right)\Biggl[ \frac{1}{2}\Bigl(\theta^2+\frac{m^2}{\varepsilon^2}\Bigr) - t \omega_c \theta \cos\varphi + \frac{\omega_c^2}{2}t^2 \Biggr],
\label{lambda riconosco}
\end{align}
where $\omega_c \equiv 2|e|E/\varepsilon$ and $E\approx m^3\chi_e/(2\varepsilon|e|)$.
For an ultrarelativistic electron, the square bracket in Eq.~\eqref{lambda riconosco} coincides with the kinematic factor $\lambda(t)$ in Eq.~\eqref{lambda_def}
\begin{equation}
\lambda(t) = \frac{1}{2} \Bigl( \theta^2 + \frac{m^2}{\varepsilon^2}\Bigr) - t \omega_c \theta \cos\varphi + \frac{\omega_c^2}{2} t^2.
\end{equation}
Using the Airy--Bessel relations in Eqs.~\eqref{Airy to Bessel 1}--\eqref{Airy to Bessel 2}, and expressing the spin four-vector $\zeta^\mu$ in terms of the rest-frame polarization $\bm{\eta}$ as $\zeta^{0} = (|\bm{p}|/m)\,\eta_{\parallel}$, $\bm{\zeta}_{\perp}=\bm{\eta}_{\perp}$ and $\zeta_{\parallel}= (\varepsilon/m)\,\eta_{\parallel}$ (with $\parallel$ and $\perp$ denoting components parallel and perpendicular to $\bm{p}$), the leading contributions in Eq.~\eqref{Ibar} reduce to 
\begin{align}
\frac{4}{3 \kappa} & \approx \frac{32}{m^6} \frac{\varepsilon^4}{\chi_e^2} \frac{2(\varepsilon - \omega)}{\varepsilon \omega}, \label{useful var} \\
\Bigl( \frac{4}{3 \kappa} \Bigr)^\frac{1}{3} \Ai(\bar z) &\approx 4 \sqrt{\frac{2 \lambda}{3}} \frac{\varepsilon^2}{m^3 \pi \chi_e}  K_{\frac{1}{3}}\Bigl(\frac{2}{3} \bar z^\frac{3}{2}\Bigr),\label{r123prefactor} \\
\Bigl( \frac{4}{3 \kappa} \Bigr)^\frac{2}{3} \Ai^\prime(\bar z) &\approx - \frac{32 \lambda}{\sqrt{3}} \frac{\varepsilon^4}{m^6 \pi \chi_e^2} K_{\frac{2}{3}}\Bigl(\frac{2}{3} \bar z^\frac{3}{2} \Bigr), \label{r4prefactor} \\
- 2 e^{2} (a  a) \Bigl( \frac{p^\prime_-}{p_-} + \frac{p_-}{p^\prime_-} \Bigr) u^2 \bar z &\approx 4 \varepsilon^2 \lambda \frac{\varepsilon^2 + \varepsilon^\prime{}^2}{\varepsilon \varepsilon^\prime}, \label{r1t1} \\
- 2 e^{2} (a  a) \Bigl( \frac{p^\prime_-}{p_-} + \frac{p_-}{p^\prime_-} \Bigr) \phi_+^{2} &\approx 4 \varepsilon^2 \frac{\omega_c^2 t^2}{2} \frac{\varepsilon^2 + \varepsilon^\prime{}^2}{\varepsilon \varepsilon^\prime}, \label{r1t2} \\
- \frac{4}{k_-} e \phi_+ \bigl( p_- (a  k) + p^\prime_- (a  k) - \frac{k_- p_-}{p^\prime_-} (a  p^\prime) \bigr) &\approx - 4 \varepsilon^2 \omega_c t \theta \cos\varphi \frac{\varepsilon^2 + \varepsilon^\prime{}^2}{\varepsilon \varepsilon^\prime}, \label{r2} \\
\frac{4}{k_-} \bigl( - k_-m^{2} + p^\prime_- (k  p) + p_- (k  p^\prime) \bigr) &\approx 4 \varepsilon^2 \Bigl[ \frac{1}{2} \Bigl( \theta^2 + \frac{m^2}{\varepsilon^2}\Bigr)\frac{\varepsilon^2 + \varepsilon^\prime{}^2}{\varepsilon \varepsilon^\prime} - \frac{1}{\gamma^2} \Bigr], \label{r3} \\
2 e m \varepsilon^{\mu\nu\rho\sigma} \zeta_\mu a_\nu \Bigl( \frac{1}{p_-} k_\rho d_\sigma - \frac{1}{p_-} d_\rho p_\sigma + \frac{1}{p^\prime_-} d_\rho p^\prime_\sigma \Bigr) &\approx m \chi_e \Bigl( - \theta \sin\varphi \frac{\varepsilon^2 - \varepsilon^\prime{}^2}{\varepsilon \varepsilon^\prime} \hat{\bm{v}} + \frac{\omega}{\varepsilon \gamma} \bm{b} \Bigr) \cdot \bm{\eta}.
\label{r4}
\end{align}
Using the relation $2\bar z^{3/2}/3\approx \xi_e$ [see Eq.~\eqref{eq:zbar} and Eq.~\eqref{xi_e_def}] and collecting terms yields
\begin{align}
\sum_{\zeta^\prime}\frac{dP^{(\zeta,\zeta^\prime)}_{\mathrm{NCS}}}{d\phi_+\, d\omega\, d\theta\, d\varphi}
& \approx
\frac{1}{2} \sum_{\eta^\prime}\frac{dP^{(\eta\eta^\prime)}_{\mathrm{NCS}}}{dt\, d\omega\, d\theta\, d\varphi} =
\frac{1}{2}\frac{\alpha \omega}{\pi^{2}}\sqrt{\frac{2}{3}}\,
\lambda^{1/2}\frac{\gamma^{3}}{\chi_e\,\varepsilon^\prime}\,
\sin\theta
\Biggl[
K_{1/3}(\xi_e)
\Biggl(2\lambda\frac{\varepsilon^{2}+\varepsilon^{\prime 2}}{\varepsilon\varepsilon^\prime}
-\frac{1}{\gamma^{2}}\Biggr)
\notag \\
& + K_{2/3}(\xi_e)\sqrt{2\lambda}\,
\Biggl(\theta\sin\varphi\frac{\varepsilon^{2}-\varepsilon^{\prime 2}}{\varepsilon\varepsilon^\prime}\hat{\bm{v}}
-\frac{\omega}{\varepsilon\gamma}\bm{b}\Biggr)\!\cdot\bm{\eta}
\Biggr],
\label{all limit}
\end{align}
The factor of $1/2$ originates from the different evolution variables: Eq.~\eqref{matrix_element_squared_simpl_3_bar_sum_spin} is differential in the phase $\phi_+$, whereas Eq.~\eqref{spa_res_2_sum_pola_spin} is differential in laboratory time $t = t_+$ (see Sec.~\ref{subsec:QM}). In the present CCF configuration $\phi_+\approx 2t$.


\section{\label{app-calc}Phase-integrated distributions}

In Sec.~\ref{sec:newMC} we derived the phase-integrated NCS probability density in a CCF. Here we report the explicit expressions of the outgoing spin- and polarization-independent term $\underline{I}$ and of the spin detector four-vector $\underline{l}^\mu$ appearing in Eq.~\eqref{matrix_element_squared_simpl_3}.

\subsection*{1. Spin-dependent structure}

The four-vector $\underline{l}^\mu$ can be decomposed as
\begin{align}
\underline{l}^\mu = & \underline{A}\,\zeta^\mu + \underline{B}\,a^\mu + \underline{C}\,k^\mu + \underline{D}\,d^\mu + \underline{E}\,p^\mu + \epsilon^{\mu\nu\rho\sigma} \big(\underline{F}\,a_\nu k_\rho d_\sigma + \underline{G}\,a_\nu d_\rho p_\sigma +\underline{H}\,a_\nu d_\rho p^\prime_\sigma\big),
\label{final_lS}
\end{align}
where the coefficients $\underline{A},\ldots,\underline{H}$ depend only on the emission kinematics and on the background-field parameters.

The explicit expressions for $\underline{I}$ and for the coefficients in Eq.~\eqref{final_lS} are
\begin{align}
\underline{A}
&=
2 (a  a) e^{2}
\left[
\frac{\Ai^\prime(z)^{2}}{(3\kappa)^{2/3}}
+\phi_0^{2}\Ai(z)^{2}
\right]
+\frac{2}{k_-}e\phi_0\,\Ai(z)^{2}\Bigl(-p_-(a  k)-p_-^\prime(a  k)+k_-(a  p^\prime)\Bigr)
\notag\\
&\quad
+\frac{2}{k_-}\Ai(z)^{2}\Bigl(k_-(p  p^\prime)-p_-^\prime(k  p)-p_-(k  p^\prime)\Bigr),
\\
\underline{B}
&=
\frac{2}{k_-}e\phi_0\,\Ai(z)^{2}
\Bigl(
- p_- (\zeta  k) + p_-^\prime (\zeta  k) - k_- (\zeta  p^\prime)
+ \zeta_- (k  p)
+ \frac{\zeta_- k_-}{p_-}(p  p^\prime)
- \frac{\zeta_- k_-}{p_-^\prime}m^{2}
- \frac{\zeta_- p_-^\prime}{p_-}(k  p)
\Bigr),
\\
\underline{C}
&=
\frac{2}{k_-}e\phi_0(\zeta  a)\Ai(z)^{2}\,(p_- - p_-^\prime)
+\frac{2}{k_-}\Ai(z)^{2}\Bigl(-\zeta_-(p  p^\prime)+\zeta_-m^{2}+p_-(\zeta  p^\prime)\Bigr),
\\
\underline{D}
&=
-2(\zeta  a)e\phi_0\Ai(z)^{2}\left(\frac{m^{2}}{p_-}-\frac{(p  p^\prime)}{p_-^\prime}\right)
-\frac{2}{k_-}e\phi_0\Ai(z)^{2}\left(\frac{p_-}{p_-^\prime}-1\right)
\Bigl((\zeta  a)(k  p^\prime)-(\zeta  k)(a  p^\prime)\Bigr)
\notag\\
&\quad
-\frac{2}{k_-}(\zeta  k)\Ai(z)^{2}\bigl((p  p^\prime)-m^{2}\bigr)
-\zeta_-(a  a)e^{2}m^{2}
\left[
\frac{\Ai^\prime(z)^{2}}{(3\kappa)^{2/3}}
+\phi_0^{2}\Ai(z)^{2}
\right]
\left(\frac{1}{p_-^{2}}+\frac{1}{p_-^{\prime 2}}\right)
\notag\\
&\quad
+2\frac{\zeta_-}{p_-p_-^\prime}(a  a)(p  p^\prime)e^{2}
\left[
\frac{\Ai^\prime(z)^{2}}{(3\kappa)^{2/3}}
+\phi_0^{2}\Ai(z)^{2}
\right]
-2\frac{\zeta_-}{k_-}(a  k)e\phi_0\Ai(z)^{2}
\left(\frac{1}{p_-}+\frac{1}{p_-^\prime}\right)\bigl((p  p^\prime)-m^{2}\bigr)
\notag\\
&\quad
+2\frac{\zeta_-}{k_-}(a  p^\prime)(k  p)e\phi_0\Ai(z)^{2}\left(\frac{1}{p_-}-\frac{1}{p_-^\prime}\right)
-2(\zeta  p^\prime)(a  a)\frac{1}{p_-^\prime}e^{2}
\left[
\frac{\Ai^\prime(z)^{2}}{(3\kappa)^{2/3}}
+\phi_0^{2}\Ai(z)^{2}
\right]
\notag\\
&\quad
+\frac{2}{k_-}(\zeta  p^\prime)(a  k)e\phi_0\Ai(z)^{2}\left(\frac{p_-}{p_-^\prime}+1\right)
+\frac{2}{k_-}(\zeta  p^\prime)(k  p)\Ai(z)^{2},
\\
\underline{E}
&=
-2(\zeta  a)e\phi_0\Ai(z)^{2}
-2\frac{\zeta_-}{p_-}(a  a)e^{2}
\left[
\frac{\Ai^\prime(z)^{2}}{(3\kappa)^{2/3}}
+\phi_0^{2}\Ai(z)^{2}
\right]
+2\frac{\zeta_-}{k_-}e\phi_0\Ai(z)^{2}
\Bigl(\frac{p_-^\prime}{p_-}(a  k)+(a  k)-\frac{k_-}{p_-}(a  p^\prime)\Bigr)
\notag\\
&\quad
+\frac{2}{k_-}\Ai(z)^{2}\Bigl(\zeta_-(k  p^\prime)-k_-(\zeta  p^\prime)+p_-^\prime(\zeta  k)\Bigr),
\\
\underline{F}
&=
-\frac{2}{k_-}em\,\Ai(z)\frac{\Ai^\prime(z)}{(3\kappa)^{1/3}}
\left(1-\frac{p_-}{p_-^\prime}\right),
\\
\underline{G}
&=
-2em\,\Ai(z)\frac{\Ai^\prime(z)}{(3\kappa)^{1/3}}\frac{1}{p_-},
\\
\underline{H}
&=
+2em\,\Ai(z)\frac{\Ai^\prime(z)}{(3\kappa)^{1/3}}\frac{1}{p_-^\prime},
\\
\underline{I}
&=
- e^{2}(a  a)\left(\frac{p_-^\prime}{p_-}+\frac{p_-}{p_-^\prime}\right)
\left[
\frac{\Ai^\prime(z)^{2}}{(3\kappa)^{2/3}}
+\phi_0^{2}\Ai(z)^{2}
\right]
+\frac{2}{k_-}e\phi_0\Ai(z)^{2}
\Bigl(p_-(a  k)+p_-^\prime(a  k)-\frac{k_- p_-}{p_-^\prime}(a  p^\prime)\Bigr)
\notag\\
&\quad
+\frac{2}{k_-}\Ai(z)^{2}\Bigl(-k_-m^{2}+p_-^\prime(k  p)+p_-(k  p^\prime)\Bigr)
+2em\,\Ai(z)\frac{\Ai^\prime(z)}{(3\kappa)^{1/3}}
\epsilon^{\mu\nu\rho\sigma}\zeta_\mu a_\nu
\Bigl(\frac{1}{p_-}k_\rho d_\sigma-\frac{1}{p_-}d_\rho p_\sigma+\frac{1}{p_-^\prime}d_\rho p^\prime_\sigma\Bigr).
\label{underl I}
\end{align}
As shown in the main text, applying the relations in Eq.~\eqref{4dto3drest} recasts Eqs.~\eqref{final_lS}--\eqref{underl I} into the form of Eq.~\eqref{matrix_element_squared_simpl_4}, thereby identifying the vector $\bm{\delta}$ [Eq.~\eqref{delta def}]. The mean outgoing spin vector then follows from Eq.~\eqref{legitimate_SPA}.

\subsection*{2. Polarization-dependent structure}

For the photon polarization, Eq.~\eqref{matrix_element_squared_simpl_5} involves the polarization-coupling vector $\bm{y}=(y^{1},y^{2},y^{3})$, from which the mean Stokes vector can be calculated [Eq.~\eqref{legitimate_PPA}]. The explicit expressions for $y^{1}$, $y^{2}$, and $y^{3}$ are
\begin{align}
y^1 &=  4 (e_1{}  a) (e_2{}  a) e^{2} \Bigl[ \frac{\Ai^\prime(z)^2}{(3\kappa)^\frac{2}{3}} + \phi_0^{2} \Ai(z)^{2} \Bigr] + 2 \Ai(z)^{2} \Bigl( (e_1{}  p) (e_2{}  p^\prime{}) + (e_1{}  p^\prime{}) (e_2{}  p) \Bigr) \notag \\
& \quad          + 2 e \phi_0 \Ai(z)^{2} \Bigl[ (e_1{}  a) \Bigl( (e_2{}  p) + (e_2{}  p^\prime{}) \Bigr) + (e_2{}  a) \Bigl( (e_1{}  p) + (e_1{}  p^\prime{}) \Bigr) \Bigr] \notag \\ 
& \quad          + e m \Ai(z) \frac{\Ai^\prime(z)}{(3\kappa)^\frac{1}{3}} \epsilon^{\alpha\beta\rho\sigma} \Bigl\{ \frac{1}{p_-} e_1{}_\alpha e_2{}_\beta \zeta_\rho a_\sigma \Bigl( p^\prime_{-} - \frac{p_{-}}{2} \Bigr) + \frac{1}{p_-} e_1{}_\alpha e_2{}_\beta a_\rho \Bigl( (\zeta  p^\prime{}) d_\sigma - \frac{\zeta_-}{2} p_\sigma \Bigr) \notag \\ 
& \qquad \qquad \qquad \qquad \qquad           +  \frac{1}{p_-} (\zeta  a) e_1{}_\alpha e_2{}_\beta d_\rho \Bigl( p^\prime{}_\sigma - \frac{1}{2} p_\sigma \Bigr) \notag \\
& \qquad \qquad \qquad \qquad \qquad           +  e_1{}_\alpha \zeta_\beta a_\rho d_\sigma \Bigl[ \Bigl( \frac{1}{p^\prime_{-}} - \frac{1}{2 p_-} \Bigr) (e_2{}  p) - \frac{2}{p_-} (e_2{}  p^\prime{}) \Bigr] + (e_1{}  p) e_2{}_\alpha \zeta_\beta a_\rho d_\sigma \Bigl( \frac{1}{p^\prime_{-}} + \frac{1}{2 p_-} \Bigr) \notag \\ 
& \qquad \qquad \qquad \qquad \qquad           - \frac{(e_2{}  a)}{p^\prime_{-}} e_1{}_\alpha \zeta_\beta d_\rho p_\sigma + (e_2{}  \zeta) e_1{}_\alpha a_\beta d_\rho \Bigl[ \Bigl( \frac{2}{p^\prime_{-}} - \frac{1}{p_-} \Bigr) p^\prime{}_\sigma - \Bigl( \frac{1}{p^\prime_{-}} - \frac{1}{2 p_-} \Bigr) p_\sigma \Bigr] \notag \\
& \qquad \qquad \qquad \qquad \qquad           + (e_1{}  a) e_2{}_\alpha \zeta_\beta d_\rho \Bigl( \frac{2}{p_-} p^\prime{}_\sigma - \frac{1}{p^\prime_{-}} p_\sigma \Bigr) + (e_1{}  \zeta) e_2{}_\alpha a_\beta d_\rho \Bigl[ \Bigl( \frac{2}{p^\prime_{-}} - \frac{1}{p_-} \Bigr) p^\prime{}_\sigma - \Bigl( \frac{1}{p^\prime_{-}} + \frac{1}{2 p_-} \Bigr) p_\sigma \Bigr] \Bigr\}, \\
y^2 &= 2 e \Ai(z) \frac{\Ai^\prime(z)}{(3\kappa)^\frac{1}{3}} \Bigl[ \frac{p^\prime_{-}}{p_-} \Bigl( - (e_1{}  a) (e_2{}  p) + (e_1{}  p) (e_2{}  a) \Bigr) + \frac{p_-}{p^\prime_{-}} \Bigl( - (e_1{}  a) (e_2{}  p^\prime{}) + (e_1{}  p^\prime{}) (e_2{}  a) \Bigr) \Bigr] \notag \\
& \quad          + 2 m \Ai(z)^{2} \epsilon^{\alpha\beta\rho\sigma}e_1{}_\alpha e_2{}_\beta \zeta_\rho \Bigl( p^\prime{}_\sigma - p_\sigma \Bigr) \notag \\
& \quad          +  \frac{1}{p^\prime_{-}} e^{2} m \Bigl[ \frac{\Ai^\prime(z)^2}{(3\kappa)^\frac{2}{3}} + \phi_0^{2} \Ai(z)^{2} \Bigr] \epsilon^{\alpha\beta\rho\sigma} \Bigl[ 2 \frac{(e_2{}  a)}{p_-} e_1{}_\alpha a_\beta d_\rho \Bigl( p^\prime_{-} \zeta_\sigma - \zeta_- p^\prime{}_\sigma \Bigr) - (a  a) e_1{}_\alpha e_2{}_\beta d_\rho \Bigl( \frac{\zeta_- p^\prime_{-}}{p_-^2} p^\prime{}_\sigma - \zeta_\sigma \Bigr) \Bigr] \notag \\
& \quad          + e m \phi_0 \Ai(z)^{2} \frac{1}{p_-} \epsilon^{\alpha\beta\rho\sigma} \Bigl\{ e_1{}_\alpha e_2{}_\beta a_\rho \Bigl[ (\zeta  p^\prime{}) d_\sigma - \frac{\zeta_-}{2} p_\sigma + \Bigl( p^\prime_{-} - \frac{3}{2} p_- \Bigr) \zeta_\sigma \Bigr] \notag \\
& \qquad \qquad \qquad \qquad \qquad \qquad \quad         + e_1{}_\alpha e_2{}_\beta d_\rho \Bigl[ 2 \frac{(a  p^\prime{})}{p^\prime_{-}} p_- \zeta_\sigma - (\zeta  a) \Bigl( p^\prime{}_\sigma - \frac{3}{2} p_\sigma \Bigr) \Bigr] \notag \\ 
& \qquad \qquad \qquad \qquad \qquad \qquad \quad         + \frac{1}{2} \Bigl( (e_1{}  p) e_2{}_\alpha - (e_2{}  p) e_1{}_\alpha \Bigr) \zeta_\beta a_\rho d_\sigma + \Bigl( (e_1{}  \zeta) e_2{}_\alpha - (e_2{}  \zeta) e_1{}_\alpha \Bigr) a_\beta d_\rho \Bigl( p^\prime{}_\sigma - \frac{1}{2} p_\sigma \Bigr) \Bigr\}, \\
y^3  &= 2 e \phi_0 \Ai(z)^{2} \Bigl[ (e_1{}  a) \Bigl( (e_1{}  p) + (e_1{}  p^\prime{}) \Bigr) - (e_2{}  a) \Bigl( (e_2{}  p) + (e_2{}  p^\prime{}) \Bigr) \Bigr] \notag \\
& \quad          + 2 e^{2} \Bigl[ \frac{\Ai^\prime(z)^2}{(3\kappa)^\frac{2}{3}} + \phi_0^{2} \Ai(z)^{2} \Bigr] \Bigl( (e_1{}  a)^{2} - (e_2{}  a)^{2} \Bigr) + 2 \Ai(z)^{2} \Bigl( (e_1{}  p) (e_1{}  p^\prime{}) - (e_2{}  p) (e_2{}  p^\prime{}) \Bigr) \notag \\
& \quad          + \frac{2}{p^\prime_{-}} e m \Ai(z) \frac{\Ai^\prime(z)}{(3\kappa)^\frac{1}{3}} \epsilon^{\alpha\beta\rho\sigma} \Bigl\{ \Bigl[ \Bigl( (e_1{}  p) - \frac{p^\prime_{-}}{p_-} (e_1{}  p^\prime{}) \Bigr) e_1{}_\alpha - \Bigl( (e_2{}  p) - \frac{p^\prime_{-}}{p_-} (e_2{}  p^\prime{}) \Bigr) e_2{}_\alpha \Bigr] \zeta_\beta a_\rho d_\sigma \notag \\
& \quad \qquad \qquad \qquad \qquad \qquad \qquad  + \Bigl( (e_1{}  \zeta) e_1{}_\alpha - (e_2{}  \zeta) e_2{}_\alpha \Bigr) a_\beta d_\rho \Bigl( p^\prime{}_\sigma - p_\sigma \Bigr) \Bigr\}.
\end{align}
\end{widetext}
Finally, we emphasize that the coefficients $(y^{1},y^{2},y^{3})$, and hence the Stokes vector $\langle \bm{\xi} \rangle=\bm{y}/\underline{I}$, are defined with respect to the event-dependent polarization basis specified in Eq.~\eqref{double_gauge_fixed_pol} (see also Sec.~\ref{stokes_param}).


\bibliography{biblio}

\end{document}